\documentclass[twocolumn]{aastex61}
\pdfoutput=1
\usepackage[latin1]{inputenc}
\usepackage{amsmath}
\usepackage{amsfonts}
\usepackage{amssymb}
\usepackage{chngcntr}

\usepackage{graphicx}
\usepackage[varg]{txfonts}
\usepackage{mwe}  
\usepackage{url}
\usepackage{longtable}
\usepackage{float}

\accepted{by ApJ on December 2018} 

\shorttitle{CXB and Population Synthesis Model}
\shortauthors{Ananna et al.}

\begin{document}
\title{The Accretion History of AGN I: Supermassive Black Hole Population Synthesis Model}
	
\author{Tonima Tasnim Ananna}
\affiliation{Department of Physics, Yale University, PO BOX 201820, New Haven, CT 06520-8120}
\affiliation{Yale Center for Astronomy and Astrophysics, P.O. Box 208120, New Haven, CT 06520, USA 0000-0002-5554-8896}
\email{tonimatasnim.ananna@yale.edu}

\author{Ezequiel Treister}
\affiliation{Instituto de Astrof\'{\i}sica, Facultad de F\'{\i}sica, Pontificia Universidad Cat\'{o}lica de Chile, Casilla 306, Santiago 22, Chile}

\author{C. Megan Urry}
\affiliation{Department of Physics, Yale University, PO BOX 201820, New Haven, CT 06520-8120}
\affiliation{Yale Center for Astronomy and Astrophysics, P.O. Box 208120, New Haven, CT 06520, USA 0000-0002-5554-8896}
\affiliation{Department of Astronomy, Yale University, P.O. Box 208101, New Haven, CT 06520, USA}

\author{C. Ricci}
\affiliation{N\'ucleo de Astronom\'ia de la Facultad de Ingenier\'ia, Universidad Diego Portales, Av. Ej\'ercito Libertador 441, Santiago, Chile}
\affiliation{Kavli Institute for Astronomy and Astrophysics, Peking University, Beijing 100871, China}
\affiliation{Chinese Academy of Sciences
South America Center for Astronomy and China-Chile Joint Center for Astronomy,
Camino El Observatorio 1515, Las Condes, Santiago, Chile}

\author{Allison Kirkpatrick}
\affiliation{Department of Physics, Yale University, PO BOX 201820, New Haven, CT 06520-8120}
\affiliation{Yale Center for Astronomy and Astrophysics, P.O. Box 208120, New Haven, CT 06520, USA 0000-0002-5554-8896}

\author{Stephanie LaMassa}
\affil{Space Telescope Science Institute, 3700 San Martin Dr, Baltimore, MD 21218}

\author{Johannes Buchner}
\affiliation{Instituto de Astrof\'{\i}sica, Facultad de F\'{\i}sica, Pontificia Universidad Cat\'{o}lica de Chile, Casilla 306, Santiago 22, Chile}
\affiliation{Millenium Institute of Astrophysics, Vicu$\tilde{n}$a. MacKenna 4860, 7820436 Macul, Santiago, Chile}

\author{Francesca Civano}
\affiliation{Harvard-Smithsonian Center for Astrophysics, 60 Garden Street, Cambridge, MA 02138, USA}

\author{Michael Tremmel}
\affiliation{Department of Physics, Yale University, PO BOX 201820, New Haven, CT 06520-8120}
\affiliation{Yale Center for Astronomy and Astrophysics, P.O. Box 208120, New Haven, CT 06520, USA 0000-0002-5554-8896}

\author{Stefano Marchesi}
\affiliation{Department of Physics and Astronomy, Clemson University,  Kinard Lab of Physics, Clemson, SC 29634, USA}

\begin{abstract}
	As matter accretes onto the central supermassive black holes in active galactic nuclei (AGN), X-rays are emitted. We present a population synthesis model that accounts for the summed X-ray emission from growing black holes;  modulo the efficiency of converting mass to X-rays, this is effectively a record of the accreted mass. We need this population synthesis model to reproduce observed constraints from X-ray surveys: the X-ray number counts, the observed fraction of Compton-thick AGN [$\log$ (N$_{\rm H}$/cm$^{\rm -2}$) $>$ 24] and the spectrum of the cosmic X-ray background (CXB), after accounting for selection biases. Over the past decade, X-ray surveys by {\it XMM-Newton}, {\it Chandra}, \textit{NuSTAR} and \textit{Swift}-BAT have provided greatly improved observational constraints. We find that no existing X-ray luminosity function (XLF) consistently reproduces all these observations. We take the uncertainty in AGN spectra into account, and use a neural network to compute an XLF that fits all observed constraints, including observed Compton-thick number counts and fractions. This new population synthesis model suggests that, intrinsically, 50$\pm$9\% (56$\pm$7\%) of all AGN within z $\simeq$ 0.1 (1.0) are Compton-thick.

\end{abstract}

\section{Introduction}\label{sec:intro}

Supermassive black holes (SMBH) are found at the cores of most galaxies and their masses correlate closely with the host bulge mass, velocity dispersion and luminosity (\citealp{richstone1998,magorrian1998,gebhardt2000,kormendy2001,merrit2001,ferrarese2005,kormendyho2013}). This suggests SMBH may regulate star formation rates, e.g., through molecular and ionized wind mass outflows (\citealp{ferrarese2000,gebhardt2000,matteo2005, merloni2010,fiore2017,martinnavarro2018}). 
If so, the accretion history of SMBH has important implications for the evolution of galaxies.

The growth of SMBH over cosmic time can be traced through the light emitted during rapid growth phases, when the galaxy appears as an active galactic nuclei (AGN).
A population synthesis model describes the number density of AGN as a function of their luminosity and redshift (X-ray luminosity function); 
together with the spectral energy distributions, this model describes all the radiation produced by SMBH growth throughout the Universe.
 
High-energy X-rays are a prime tracer of AGN because they are produced close to the black hole and they can penetrate all but the thickest columns of absorbing material \citep{brandthasinger2005,Cardamone2008,donley2012,mendez2013,kirkpatrick2015,delmoro2016}. Additionally, X-ray surveys detect mostly active galaxies rather than inactive galaxies. Therefore X-rays have the advantage of both sensitivity to obscured AGN and efficiency of detecting AGN. Hard X-ray bands are especially important for heavily obscured Compton-thick objects (with column densities N$_{\rm H} > 10^{24}$~cm$^{-2}$), as well as obscured Compton-thin sources (N$_{\rm H} = 10^{22} - 10^{24}$~cm$^{-2}$). Indeed, the number density of heavily obscured objects was one of the most uncertain parts of early population synthesis models because the first X-ray surveys were fairly soft (\citealp{maccacaro1991,boyle1993,comastri1995,jones1997,page1997,miyaji2000,gilli2001}).  As higher energy X-ray data ($>$ 3 keV) became available (\citealp{boyle1998,cowie2003,ueda2003,gilli2007,treister2009,ueda2014,aird2015xlf,buchner2015}), more obscured AGN were included. 

At this point, there exists a large ensemble of broad-band X-ray surveys with different combinations of depth and volume,  collectively spanning an extensive range in luminosity and redshift (which in any one flux-limited survey are strongly correlated).
In particular, the \textit{Chandra} X-ray observatory \citep{weisskopf2002} has contributed the \textit{Chandra} Deep Field South (CDFS) 7 Ms catalog \citep{luo2017} which reaches the faintest fluxes at E $<$ 10 keV;
extended CDFS \citep{xue2012,lehmer2012,luo2017};
COSMOS  \citep{elvis2009ccosmos};
ChaMP \citep{kim2007}; and
Stripe 82X \citep{lamassa2013a,lamassa2013b,lamassa2016,ananna2017}. 
The \textit{X-ray Multi-Mirror Mission (XMM)} - Newton observatory has a slightly harder response function than \textit{Chandra} and has also carried out both deep and wide surveys, including 
{\it XMM}-CDFS \citep{ranalli2013};
{\it XMM}-COSMOS \citep{nicocosmos2007};
2XMMi \citep{mateos2008};
Stripe~82X  \citep{lamassa2013b,lamassa2016} and 
\textit{XMM}-XXL \citep{pierre2016}. 

At still higher X-ray energies, both the \textit{Neil Gehrels Swift Observatory} Burst Alert Telescope (BAT; 14$-$195 keV; \citealp{gehrels2004,barthelmy2005})  and \textit{NuSTAR} (3$-$79 keV; \citealp{harrison2013}) have contributed the most unbiased surveys to date. \textit{Swift}-BAT and \textit{NuSTAR} are particularly sensitive to heavily obscured AGN as higher energy X-ray photons are less susceptible to absorption. \textit{Swift}-BAT is a non-focusing X-ray observatory that images the sky in five bands between 14$-$195 keV \citep{barthelmy2005}. \textit{NuSTAR} observes in a lower energy band (3$-$79 keV), and is the first orbiting telescope that focuses X-ray light above 10 keV, increasing its sensitivity by two orders of magnitude.

In this work, we show that existing X-ray luminosity functions (XLF) cannot explain the X-ray data observed in all these new surveys. We explore the uncertainty in AGN X-ray spectra, and using a neural network, find an XLF which satisfy all observed constraints. Those constraints include the integrated spectrum of the cosmic X-ray background (CXB), 
the overall X-ray number counts (i.e., the number of AGNs observed per unit area of the sky as a function of flux),
and the Compton-thick AGN number counts and fraction in each survey.

Our new population synthesis model is presented as follows: the X-ray spectra of AGN is discussed in \S~\ref{sec:spectral_model}. The most recent XLFs are described in \S~\ref{sec:xlfs}. The observational constraints from X-ray surveys are discussed in \S~\ref{sec:constraints}. Our approach of formulating a new population synthesis model is described in \S~\ref{sec:new_model}.  Our results are presented  in \S~\ref{sec:results}. The conclusions and summary of this work are presented in \S~\ref{sec:conclusion} and  \S~\ref{sec:summary}, respectively.

\section{AGN X-ray spectra}\label{sec:spectral_model}

\begin{figure*}[th]
	\centering
	\includegraphics[width=0.8\linewidth]{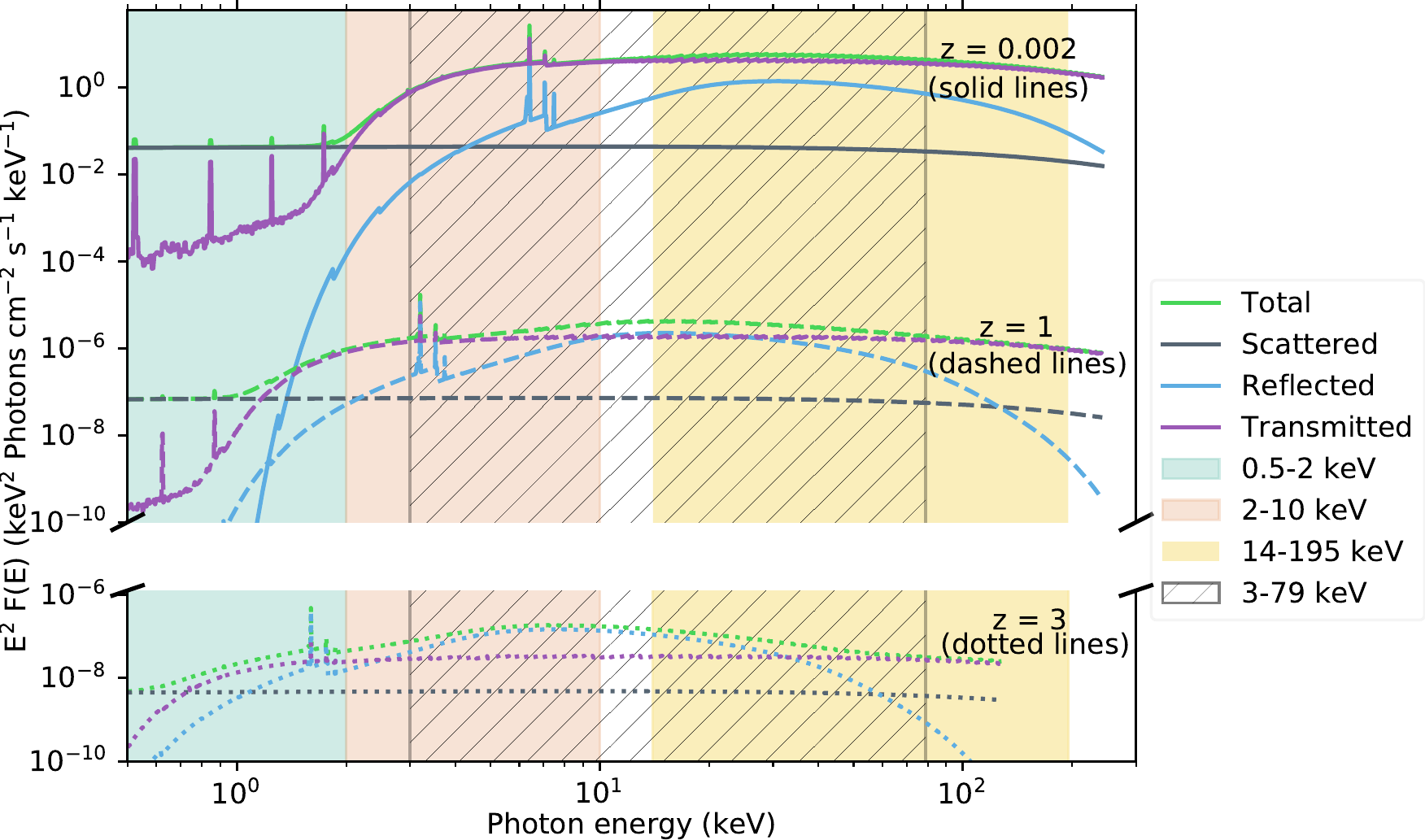}
	\caption{Observed-frame spectra with intrinsic L$_{\rm 2-10}$ = 10$^{\rm 44}$ erg/s, $\log$ (N$_{\rm H}$/cm$^{\rm -2}$) = 23 at z = 0.002 (\textit{solid lines}), z = 1 (\textit{dashed lines}) in the \textit{top panel} and z = 3 (\textit{dotted lines}) in the \textit{bottom panel}. The plot shows  unabsorbed component scattered by gas outside the torus region \textit{(dark gray lines)}, reflection component (\textit{light blue lines}) from the accretion disk, transmitted emission (\textit{purple lines}) from the torus, the total AGN spectra (\textit{light green lines}), and window of observation in typical energy bands. \textit{Chandra} observes in 0.5$-$8 keV band, and \textit{XMM-Newton} observes in 0.5$-$10 keV band, \textit{Swift}-BAT in 14$-$195 keV band and \textit{NuSTAR} in 3$-$79 keV band at this redshift. At z$=$1.0, the observed flux is six orders of magnitude lower than in the local Universe due to distance, while redshifting allows the instruments to probe higher rest-frame energy bands. The \textit{Chandra}, \textit{XMM} hard bands and to a larger extent, \textit{NuSTAR} plays a vital role in quantifying the strength of the reflection component.} 
	\label{fig:buchner_spectra_z_variation} 
\end{figure*}

In this work, we focus on the light emitted in X-ray bands as SMBHs grow. Understanding AGN X-ray spectra is necessary to interpret observed X-ray samples, and to constrain the population synthesis model. Figure~\ref{fig:buchner_spectra_z_variation} shows the X-ray spectra of a moderately obscured AGN at three different redshifts while keeping all other spectral parameters constant. It also shows energy windows of \textit{Chandra} and \textit{XMM}-Newton ($<$ 10 keV), \textit{NuSTAR} (8$-$24 keV) and \textit{Swift}-BAT (14$-$195 keV) X-ray instruments.

\begin{figure}[th]
	\centering
	\includegraphics[width=1.0\linewidth]{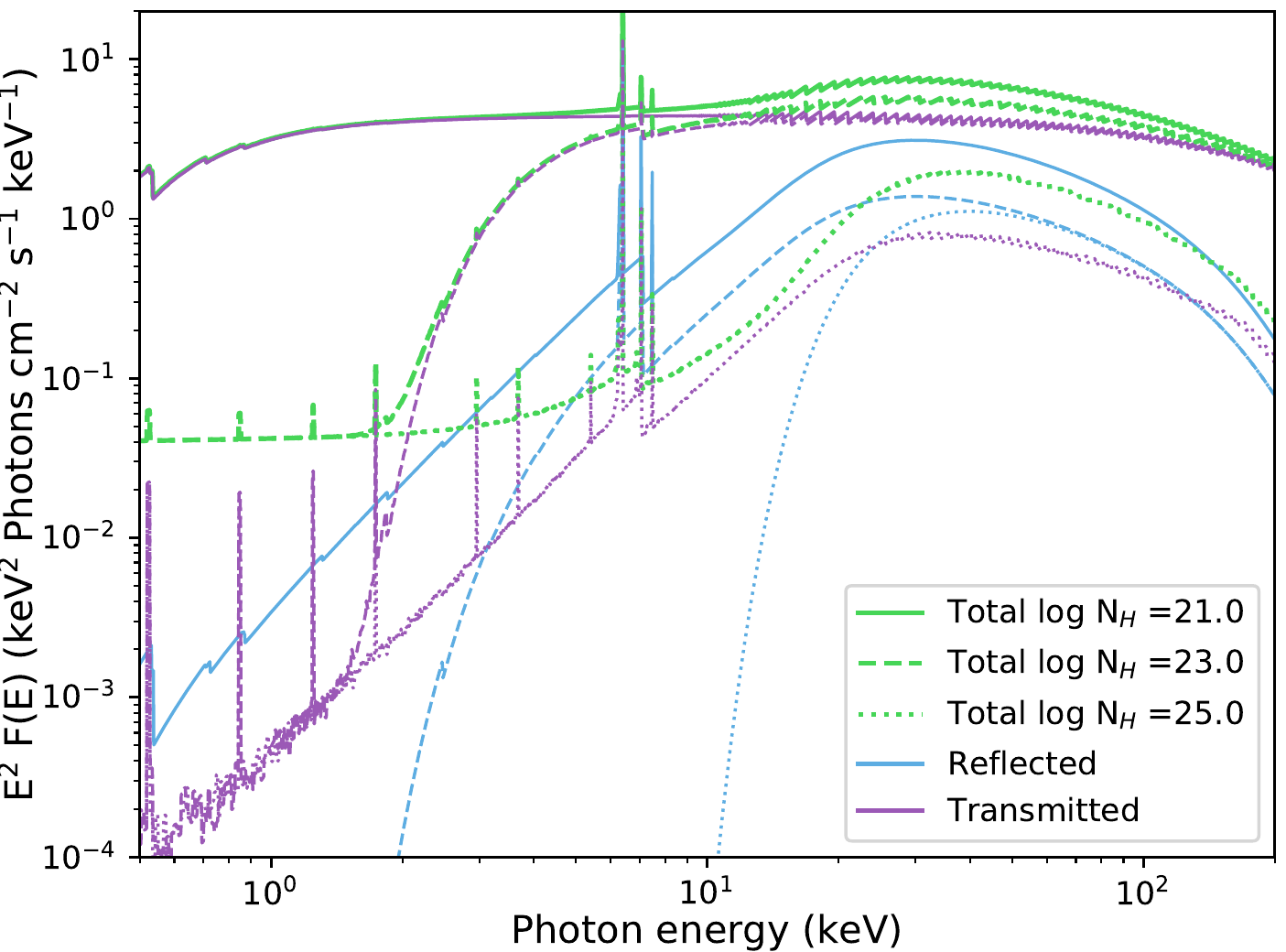}
	\caption{Variation in X-ray spectra with absorbing column density $\log$ (N$_{\rm H}/{\rm cm}^{\rm -2}$), where column density varies from $\log$ (N$_{\rm H}/{\rm cm}^{\rm -2}$) = 21 (\textit{solid lines}) to $\log$ (N$_{\rm H}/{\rm cm}^{\rm -2}$) = 23 (\textit{dashed lines}) to $\log$ (N$_{\rm H}/{\rm cm}^{\rm -2}$) = 25 (\textit{dotted lines}). The components are reflection (\textit{blue lines}) from the accretion disk, reprocessed emission (\textit{purple lines}) from the torus, and the sum of all components (\textit{green lines}). The scattered component is not shown for clarity. Spectral parameters other than absorption are fixed at $\Gamma$ = 1.96, R $= 0.83$ for unabsorbed, and R $= 0.37$ for absorbed sources, E$_{\rm C}$ = 200 keV and f$_{\rm scatt}$=1\%.} 
	\label{fig:absoprtion_variation_spectra} 
\end{figure}

\begin{figure}[th]
	\centering
	\includegraphics[width=1.0\linewidth]{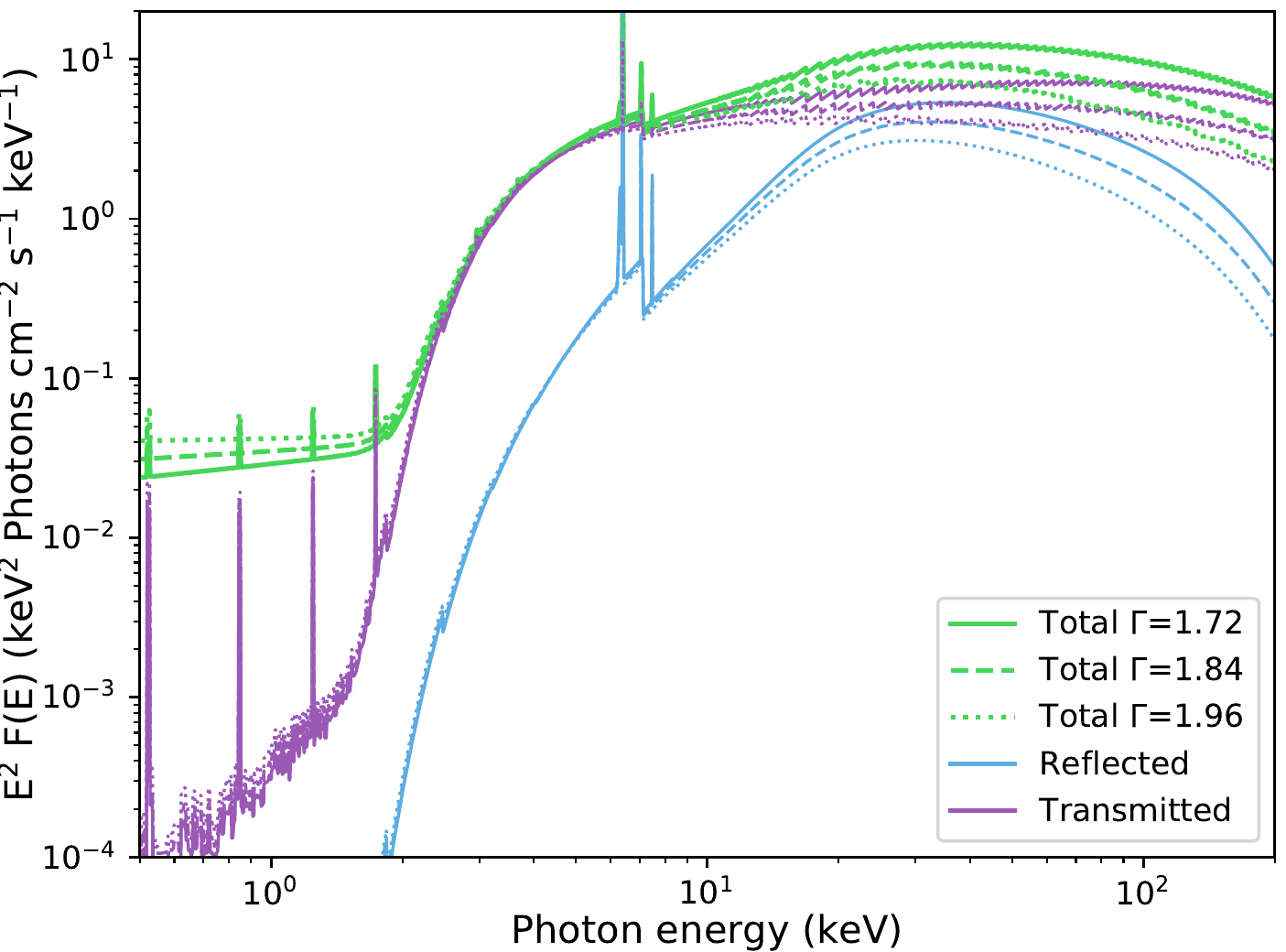}
	\caption{Variation in X-ray spectra with photon index $\Gamma$, where $\Gamma$ is varied from 1.72 (\textit{solid lines}) to 1.84 (\textit{dashed lines}) to 1.96 (\textit{dotted lines}). The components are reflection (\textit{blue lines}) from the accretion disk, reprocessed emission (\textit{purple lines}) from the torus, and the sum of all components (\textit{green lines}). The scattered component is not shown for clarity. Spectral parameters other than $\Gamma$ are fixed at a constant scattering fraction (f$_{\rm scatt}$=1\%), cutoff energy (E$_{\rm C}$ = 200 keV), absorbing column density $\log$ (N$_{\rm H}/{\rm cm}^{\rm -2}$) = 23 and reflection scaling factor (R = 0.83).} 
	\label{fig:gamma_variation_spectra} 
\end{figure}

\begin{figure}[th]
	\centering
	\includegraphics[width=1.0\linewidth]{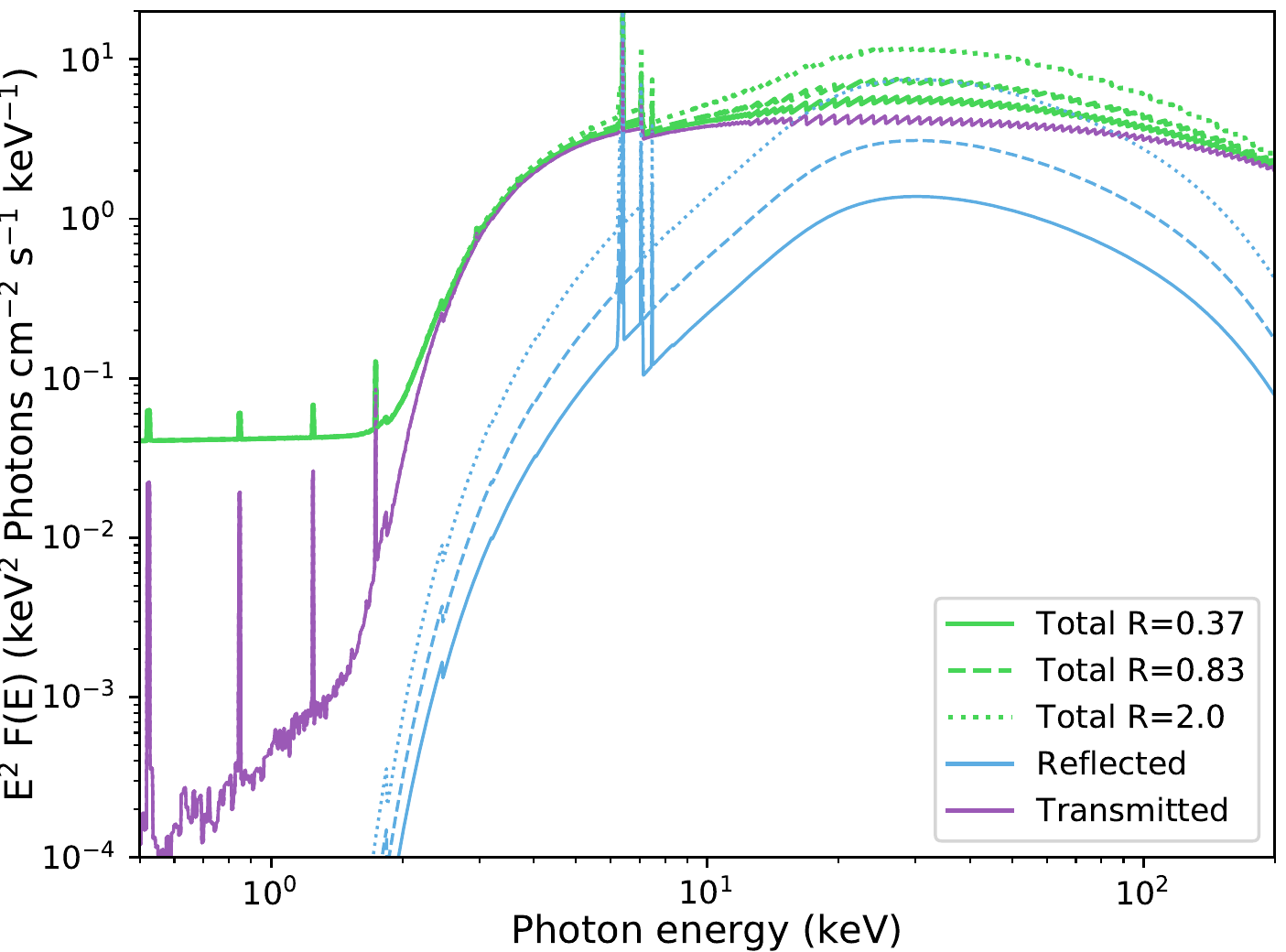}
	\caption{Variation in X-ray spectra with reflection scaling factor R, where R is varied from 0.37 (\textit{solid lines}), 0.83 (\textit{dashed lines}) to 2 (\textit{dotted lines}). The components are reflection (\textit{blue lines}) from the accretion disk, reprocessed emission (\textit{purple lines}) from the torus, and the sum of all components (\textit{green lines}). The scattered component is not shown for clarity. Spectral parameters other than reflection scaling factor are fixed at $\Gamma$ = 1.96, cutoff energy (E$_{\rm C}$ = 200 keV), $\log$ (N$_{\rm H}/{\rm cm}^{\rm -2}$) = 23 and f$_{\rm scatt}$=1\%.} 
	\label{fig:R_variation_spectra} 
\end{figure}

The AGN X-ray spectrum affects the conversion between number counts and flux, as well as the sensitive area of each survey. The origin of the X-ray spectra (shown in Figures~\ref{fig:absoprtion_variation_spectra}, \ref{fig:gamma_variation_spectra}, \ref{fig:R_variation_spectra}) is the hot corona around the accretion disk, and the shape of the spectrum is a power-law, with a photon index in the range $\Gamma \simeq$ 1.4 - 2.1 \citep{nandra1994,ueda2014,claudio2017bat} and an exponential cutoff energy, i.e., F(E) $\propto E^{-\Gamma} \exp(-E/E_{\rm cutoff})$. This emission is reflected by the accretion disk, which is $< 1$ pc from the central SMBH  \citep{nenkova2002,accretionsize}, and reprocessed by a torus-like distribution of obscuring material $\simeq$ 10-100 pc from the AGN \citep{nenkova2002}. The torus absorbs optical, ultraviolet and X-ray photons and re-emits it in infrared. The unabsorbed continuum can be scattered by gas outside the torus region. Fe K-$\alpha$ emission line at around 6.4~keV is prominent in AGN spectra and is thought to have originated either in the outer regions of the accretion disk or the inner region of the torus \citep{nandra2006}. 

Each of these components has to be modeled in order to calculate observable quantities, such as CXB. For this work, each component was modeled using \textsc{xspec} \citep{xspec}. The Compton reflection from the accretion disk is modeled using \textsc{pexrav} \citep{pexrav} or \textsc{pexmon} \citep{pexmon} model, where the latter updates the former with self-consistent Fe K-$\alpha$ emission lines relative to the power-law. Torus models such as \textsc{BNTorus} \citep{bntorus2011} self-consistently account for transmitted power-law, reflection and fluorescence lines from metals. The Thomson scattered component, from ionized material within the torus opening angle, has the same shape as the power-law, and some fraction (f$_{\rm scatt}$) of its magnitude. This component dominates at E $<$ 2 keV, as shown in Figure~\ref{fig:buchner_spectra_z_variation}.

\citet{buchner2015} shows that the sum of torus, \textsc{pexmon} and scattering is currently the best prescription to model AGN spectra. A second-order effect that could slightly modify the spectra is a Compton scattering of the \textsc{pexmon} component, which itself is a neutral Compton reflection from the accretion disk, by the compact torus. A multiplicative \textsc{xspec} model which down-scatters photons for a Compton-thick obscurer is not currently available. However, this effect is likely to be marginal compared to the uncertainty due to the range of reflection scaling factors used in the literature.

Figure~\ref{fig:absoprtion_variation_spectra} demonstrates the effects of obscuration on AGN spectra by varying levels of equivalent Hydrogen atom column density (N$_{\rm H}/{\rm cm}^{\rm -2}$). We can see that at E $<$ 10 keV, a Compton-thin obscuring column density of log (N$_{\rm H}/{\rm cm}^{\rm -2}$) = 23 can significantly decrease the observed fluxes, especially in the soft band where the total E$^{\rm 2}$ F(E) drops by almost two orders of magnitude. For this reason, heavily obscured Compton-thin and Compton-thick objects are difficult to observe in the local Universe using E $<$ 10 keV bands, and the higher energy bands from \textit{NuSTAR} and \textit{Swift}-BAT are required at z $<$ 1. At higher redshifts, the spectrum gets shifted to lower energy bands (as shown in Figure~\ref{fig:buchner_spectra_z_variation}) and can be detected at observed frame E $ < $ 10 keV. 

The contribution to the CXB by objects in each N$_{\rm H}$ bin distinctly shapes the overall CXB spectrum because of the way obscuration affects AGN spectra, as shown in Figure~\ref{fig:absoprtion_variation_spectra}. Even with a high reflection scaling factor of R = 0.83, the unabsorbed spectrum is relatively flat. If the CXB were dominated by unabsorbed objects with log (N$_{\rm H}/{\rm cm}^{\rm -2}$) $<$ 22, it should have a shape similar to an unabsorbed AGN spectra. Similarly,  a  Compton-thin dominated CXB should be low at $<$ 2 keV but rise and become approximately flat until 60$-$70 keV, depending on the cutoff energy of the intrinsic power-law. A substantial contribution from Compton-thick objects will produce the characteristic peak (Compton hump) at 20$-$30 keV that we observe in the CXB, similar to the spectrum of a Compton-thick object.

Figures~\ref{fig:gamma_variation_spectra} and \ref{fig:R_variation_spectra} provide some insight into how the spectrum varies due to variation in photon index and reflection scaling factor, which in turn helps us understand how this affects the CXB. These figures show the reflected and the transmitted components of the AGN spectra (the scattered component does not vary greatly, so it is removed from the figures for clarity), and the sum of all components. Figure~\ref{fig:gamma_variation_spectra} shows that higher $\Gamma$ causes steeper decline at E $>$ 10 keV.  Figure~\ref{fig:R_variation_spectra} shows how the reflection component changes with reflection scaling factor R, with a more prominent bump for stronger reflection. Thus a high R value, rather than a large number of obscured sources, can also cause the prominent bump; however spectral fitting of \textit{Swift}-BAT and \textit{NuSTAR} sources show that R generally lies below 1 \citep{claudio2017bat,zappacosta2018}. 

\subsection{Observed AGN Spectra}\label{sec:observed_spectra}

\citet{claudio2017bat} presented a detailed X-ray spectral analysis of AGNs of the local Universe. This analysis was carried out using \textit{Swift}-XRT \citep{swiftxrt,swiftxrt2}, \textit{XMM}-Newton and \textit{Chandra} data for E $<$ 10 keV and \textit{Swift}-BAT data in 14$-$150 keV for 836 sources in the local Universe (85\% of the sources are at z $<$ 0.1). The wide wavelength coverage, large sample size and relatively unbiased data makes the parameter distribution from this sample a robust empirical measure of X-ray spectral parameters. \citet{claudio2017bat} report that a Kolmogorov-Smirnov (KS) test between the distributions of photon indices of unobscured and obscured sources show that the two distributions are significantly different. Fitting a Gaussian to the photon indices of unobscured AGN yields $\langle \Gamma \rangle \simeq$ 1.8 and $\sigma_{\Gamma}$ = 0.24, higher than that of obscured AGN: $\langle \Gamma\rangle \simeq$ 1.72 and $\sigma_{\Gamma}$ = 0.31. \citet{tueller2008,burlon2011,ueda2014,z2000} also find a higher photon index for unobscured sources compared to obscured sources. 

\citet{claudio2017bat} reports that the reflection scaling factor R varies significantly based on obscuration: R$_{\rm median}$ = 0.83 $\pm$ 0.14 for unobscured sources and R$_{\rm median}$ = 0.37 $\pm$ 0.11 for obscured sources. We performed a KS test on the cutoff energy parameter of the AGN power-law for obscured and unobscured objects, but did not find a statistically significant difference between the two distributions (p-value = 0.42). The cutoff energy was well constrained for 161 sources. The median cutoff energy found by \textit{Swift}-BAT for these 161 objects is 76 keV, and most of these energies are below 100 keV. The overall distributions of these three parameters are shown in Figure~\ref{fig:swift_bat_distributions}.

 \begin{deluxetable*}{lcccc}[th]
	\tablewidth{0pt}
	\tablecaption{\label{tab:spectral} \textsc{Summary of X-ray spectral parameters in recent population synthesis models and observations.}}
	\tablehead{\colhead{\textsc{Model/Observation}} &\colhead{\textsc{Photon Index  $\langle\Gamma\rangle$}} & \colhead{\textsc{Refl. Scaling Factor (R)}}&\colhead{\textsc{E$_{\rm cutoff}$ (keV)}}&\colhead{\textsc{f$_{\rm scatt}$}} }
	\startdata
	\citealp{claudio2017bat}\tablenotemark{1}  & 1.72 (obscur), 1.8 (unobsc) & 0.37 (obscur), 0.83 (unobsc) & 76\tablenotemark{2} & $\simeq$ 1\% \\
	\citealp{ueda2014}\tablenotemark{3} &  1.84 (obsc),  1.94 (unobsc) & 0.5 & 300 & $\simeq$ 1\% \\
	\citealp{aird2015xlf}\tablenotemark{3} &1.9  & 0  - 2. (uniform) & 300  & $\simeq$ 1\% \\
	\citealp{buchner2015}\tablenotemark{3} & 1.95 & 0.1  - 2. (log uniform) & No cutoff  & $\simeq$ 1\% \\
	\enddata	
    \tablenotetext{1}{Observed parameters determined by detailed X-ray spectral fittings to \textit{Swift}-BAT 70-month survey sources.}
    \tablenotetext{2}{The cutoff energies measured in \textit{Swift}-BAT can only be adequately constrained when the value is lower than 100 keV.}
    \tablenotetext{3}{Parameter values assumed to model X-ray spectra for each respective X-ray luminosity function.}
\end{deluxetable*}	

\textit{Swift}-BAT covers the widest waveband, but the BAT sample is at low redshifts: 85\% of the sample is below z $=$ 0.1. Spectral fitting is also susceptible to biases: \citet{claudio2017bat} reports that cutoff energies in the \textit{Swift}-BAT 70-month survey data can only be constrained for sources where this value lies below 100 keV, and reflection parameters are easier to constrain when this value is large. Even though the median of all \textit{observed} cutoff energy values is 76 keV, a Kaplan-Meier estimator on these energy values, including all the lower limits,  yields a median value of 200 $\pm$ 29 keV. Additionally, spectral parameters may be coupled. \citet{z99,petrucci2001} report a correlation between reflection parameter and photon index, \citet{matt2001} report a positive correlation between photon index and cutoff energies. These correlations may occur due to the intrinsic nature of the spectra, or due to the fact that they are strongly related in the fitting procedure. Similarly, the difference of observed parameters between obscured and unobscured sources might be intrinsic but can also arise due to imperfections in the modeling of the obscurer \citep{mislav2018}.

\citet{ueda2014} fit 14$-$195 keV spectra of \textit{Swift}-BAT 9-month catalog sources with a power-law model, with fixed reflection parameter (R=0.5) and cutoff energy (300 keV), and derived photon indices of $\langle \Gamma \rangle$ = 1.84 for obscured sources and $\langle \Gamma \rangle$ = 1.94 for unobscured sources. \citet{nandra1994} reported a $\Gamma$ = 1.9$-$2.0, assuming a power-law spectra without  a cutoff energy. \citet{claudio2017bat} also finds a $\langle \Gamma \rangle$ value consistent with these results by fitting 14$-$195 keV data with a simple power-law model ($ \langle \Gamma_{\rm BAT} \rangle$ = 1.96), even though the $\langle \Gamma \rangle$ for overall broadband spectral fitting is lower (i.e., $\langle \Gamma \rangle$ = 1.72 for obscured sources and $\langle \Gamma \rangle$ = 1.80 for unobscured sources). 

The predicted CXB and number counts from any XLF varies depending on the assumed spectra. The consequence of spectral parameter uncertainties on observed constraints are explored further in \S~\ref{sec:new_model} and \S~\ref{sec:results}.

\subsection{Modeled AGN Spectra in Existing Population Synthesis Models}\label{sec:further}

\begin{figure*}[th]
	\centering
	\includegraphics[width=0.33\linewidth]{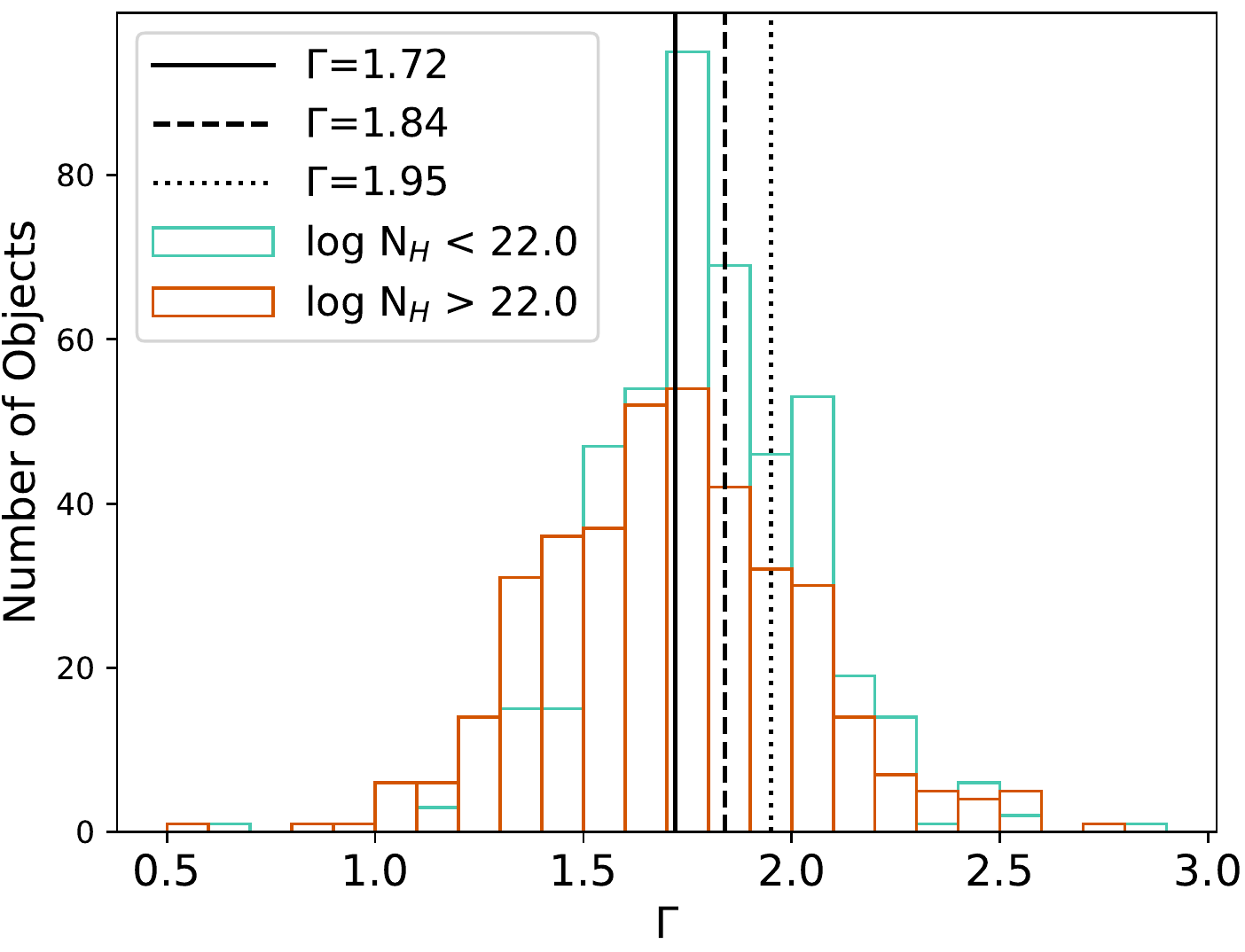}~
	\includegraphics[width=0.33\linewidth]{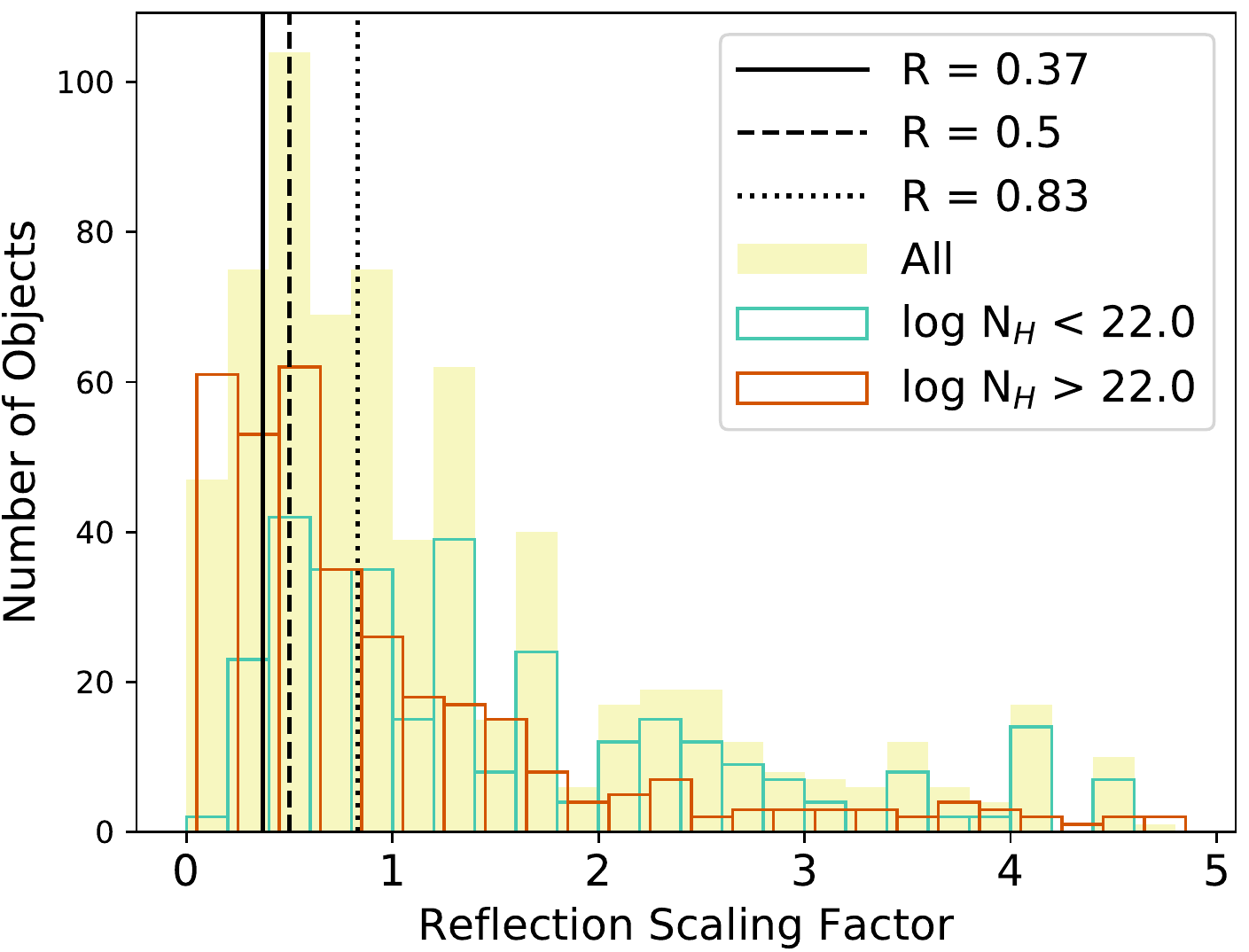}~
	\includegraphics[width=0.33\linewidth]{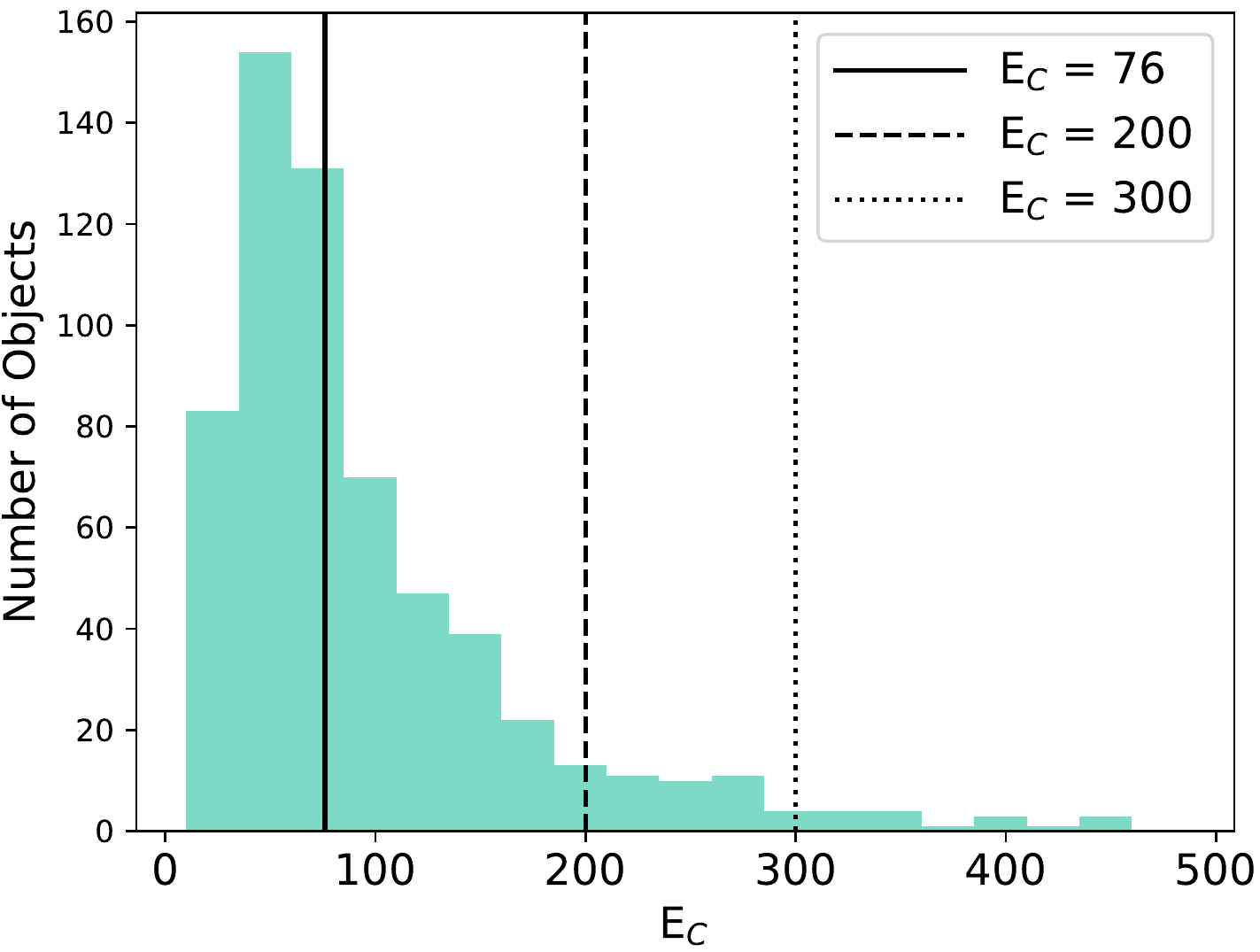}
	\caption{AGN power-law spectral photon index (\textit{left}), reflection scaling factor (\textit{middle}) and cutoff energy (\textit{right}) distributions for \textit{Swift}-BAT 70-month sample \citep{claudio2017bat}. In the \textit{left} and \textit{middle panels}, the unabsorbed and absorbed population $\Gamma$ distributions are plotted in two distinct histograms, log (N$_{\rm H}/{\rm cm}^{\rm -2}$) $<$ 22 (\textit{light green}) and   log (N$_{\rm H}/{\rm cm}^{\rm -2}$) $>$ 22 (\textit{orange line}). Most of the reflection scaling factor values are upper limits, whereas most of the cutoff energy values are lower limits. \cite{ueda2014} parameter values are indicated by \textit{black solid and dashed vertical lines} in the \textit{left} and \textit{middle panels}. In the \textit{right panel}, U14 cutoff energy, E$_{\rm C}=$ 300 keV, is shown with dotted line, and \textit{Swift}-BAT 70-month observed median (E$_{\rm C}=$ 76 keV) and bias-corrected median (E$_{\rm C}=$ 200 keV) are shown with \textit{solid} and \textit{dashed lines}, respectively.} 
	\label{fig:swift_bat_distributions} 
\end{figure*} 

Here we describe the AGN spectra of the three population synthesis models examined in this paper. The main spectral parameters are summarized in Table~\ref{tab:spectral}. The XLFs of these models are discussed in more detail in \S~\ref{sec:xlfs}. \citeauthor{ueda2014} (\citeyear{ueda2014}; henceforth U14) assumes constant $\Gamma \simeq$ 1.84 and 1.94 for obscured and unobscured sources respectively, a reflection scaling factor of R = 0.5, based on averaged reflection strength of local Seyfert galaxies, modeled using \textsc{pexrav}. For the torus component, U14 uses \textsc{BNTorus}, where the opening angle of the torus is related to the fraction of absorbed AGNs. The scattering component is dependent on the torus opening angle $\propto$ (1 - $\cos \theta_{\rm OA}$). The cutoff energies of these spectra are assumed to be 300 keV.

The same three spectral components are used in the \citeauthor{aird2015xlf} (\citeyear{aird2015xlf}; henceforth A15) AGN template spectra as well. The photon index is modeled by a normal distribution with $\langle\Gamma\rangle $ = 1.9 and $\sigma_{\Gamma} = 0.2$. The reflection scaling factor is drawn from a uniform distribution between 0 and 2.0, and the scattering fraction is of the order of 1\%, drawn from a lognormal distribution. The cutoff energy is assumed to be 300 keV as well.

\cite{buchner2014spectra} presented a Bayesian analysis of spectra of $\simeq$ 350 AGN from the 4 Ms CDFS. The most probable model is similar to the U14 model, in that it also has scattering, reflection and torus components. A power-law is assumed as the intrinsic spectrum, but without any cutoff energies. The reflected component is modeled with \textsc{pexmon} instead of \textsc{pexrav} (which was used by U14) and the transmitted component was modeled using \textsc{BNTorus}. \citeauthor{buchner2015} (\citeyear{buchner2015}; henceforth B15) use this spectrum in their population synthesis model. 
The scattering component is independent of opening angle and is allowed to vary uniformly between 0.0001 and 0.1 in log space. The photon index can vary within a Gaussian distribution with $\langle\Gamma\rangle=1.95$ and $\sigma_{\Gamma} = 0.15$, and the reflection scaling factor R is drawn from a log uniform distribution in the range 0.1-2. 

We compared the distributions used in previous population synthesis models with the parameters observed in \textit{Swift}-BAT 70-month survey spectra. Figure~\ref{fig:swift_bat_distributions} shows the observed distribution plotted against suggested values for U14 spectra (A15 and B15 uses a distribution for these parameters with $\langle\Gamma\rangle$ close to the dotted line). The $\Gamma$ is slightly higher than observed values for all the models, but the most noticeable difference is the cutoff energy which is much higher for the models. Note that if observational biases are taken into account, the cutoff energy is estimated to be higher (200 keV; \citealp{claudio2017bat}).

\section{A brief review of X-ray luminosity functions}\label{sec:xlfs}

Early XLFs were based on soft X-ray bands (\citealp{maccacaro1991,boyle1993,jones1997,page1997,miyaji2000}). One of the first hard XLFs, \citet{ueda2003}, introduced an ``absorption function" as part of the XLF, which takes into account what fraction of objects at each luminosity and redshift falls in each N$_{\rm H}$ bin. This addition meant that the spectrum of an AGN can be corrected in the rest frame for absorption effects. This is important in hard XLFs which include more absorbed AGNs. The population synthesis model of \citet{ueda2003} defined the following three components: i) an AGN template spectrum, the shape and normalization of which varies with N$_{\rm H}$ and intrinsic rest-frame luminosity (L$_{\rm 2-10}$) respectively, as described in \S~\ref{sec:spectral_model}; ii) A distribution of how space density varies with L$_{\rm 2-10}$ and z; iii) An absorption function of how this space density is distributed in N$_{\rm H}$ bins. The second component, the space density per co-moving Mpc$^{3}$, follows a double power-law relationship as a function of luminosity (as shown in Equation~11 in \citealp{ueda2003} and Equation~14 in \citealp{ueda2014}). The second and third components of population synthesis models can be dependent on each other, so the most general XLFs give space densities based on all three parameters (z, L$_{\rm 2-10}$, N$_{\rm H}$).  

Most recent population synthesis models provide the same three components (spectrum, XLF and N$_{\rm H}$ distribution), although the form of the function is sometimes different \citep{aird2010,aird2015b} or the AGN spectrum has different components or different distribution of spectral parameters.  We discuss three of the most recent XLF in this section, which were formulated using the most recent surveys with the most representative samples of AGN.  We consider these three models when fitting all the latest observed constraints.

\subsection{Ueda et al 2014}

U14 used a maximum likelihood method to fit a double power-law luminosity function for the local Universe, and a redshift evolution function, which together follows a complex Luminosity Dependent Density Evolution (LDDE) relationship. The AGN samples used to derive the XLF are selected in 0.5$-$195 keV in X-ray bands, and have high identification completeness ($\geq 90\%$). 

Even though all available samples are used to formulate the XLF, to construct a robust absorption function, U14 only uses samples with the highest photon counts: the \textit{Swift}-BAT 9-month survey \citep{tueller2008}, \textit{ASCA} Medium Sensitivity Survey (AMSS; \citealp{ueda2001amss,akiyama2003}, and Subaru/\textit{XMM-Newton} Deep Survey (SXDS; \citealp{ueda2008sxds}) data. \textit{Swift}-BAT data were used to quantify the local absorption function and AMSS, SXDS data were used to formulate the redshift/luminosity evolution. \textit{XMM-Newton} and \textit{Chandra} sources were not used to constrain the absorption function because the faint flux limits result in too few photons to construct a reliable X-ray spectrum. U14 constrains the absorption function separately from the XLF to avoid strong parameter coupling.

\subsection{Buchner et al 2015}

B15 used a non-parametric approach on $\simeq$ 2000 AGN selected in 2$-$7 keV band to derive an XLF that does not impose any form on the luminosity function. The final product of this approach are  3-dimensional matrices of space densities in z, L$_{\rm 2-10}$ and $\log$ (N$_{\rm H}/{\rm cm}^{\rm -2}$) bins.  A thousand equally likely Markov Chain Monte Carlo (MCMC) samples/matrices are generated based on the uncertainties imposed by the data. The Bayesian prior in this approach are two types of smoothness assumptions about how space densities vary from one bin to another: i) constant value prior and ii) constant slope prior. The constant value prior requires that the space density from one bin to the next stays constant unless  constraints are imposed by the data. The value of a bin scatters around its neighbor's density value following a normal distribution with an allowed correlation width for luminosity and redshift axes. The constant slope prior is only applied to L$_{\rm 2-10}$ and log (N$_{\rm H}/{\rm cm}^{\rm -2}$), and requires that the space density follows a constant power-law slope from one bin to the next unless constraints are imposed by the data. The slope scatters around the neighbor's slope following a normal distribution. Each of the two assumptions provides 500 samples, resulting in a total of 1000. 

\subsection{Aird et al 2015}

The A15 XLF was formulated using a parametric Bayesian approach. A15 derived XLFs for 0.5$-$2 keV and 2$-$7 keV X-ray samples separately, then incorporated absorption effects and modeled the unobscured and obscured samples separately. Consequently, there are two components of the A15 XLF, with different sets of parameters, one for absorbed and one for unabsorbed AGNs. Unlike U14 and B15, A15 does not calculate N$_{\rm H}$ for individual sources. However, their approach statistically predicts an N$_{\rm H}$ distribution by global comparisons between the soft and hard band samples. A15 account for contribution from star-forming galaxies to CXB by formulating a galaxy luminosity function. This contribution should not be ignored when synthesizing CXB using AGN XLFs.

We briefly summarize these XLFs again when we discuss our approach in this work in \S~\ref{sec:new_model}.

\section{The Observed Constraints}\label{sec:constraints}

In Table~\ref{tab:observed_constraints}, we list all the observed constraints considered in this work, and we explain each of these constraints in this section. Along with CXB, we consider AGN number counts, and observed Compton-thick fractions. The AGN number counts, i.e., the number of AGNs observed per square degree of the sky at a given flux limit in an X-ray band, should be reproduced by a complete population synthesis model, for surveys of all depths, volumes and energy ranges. 

Every X-ray survey probes some region of L$_X$, z and N$_{\rm H}$ space. Typically, large volume X-ray surveys, such as Stripe~82X (\citealp{lamassa2013a,lamassa2013b,lamassa2016,ananna2017}) and \textit{XMM}-XXL \citep{pierre2016}, sample more rare, luminous quasars, whereas deep pencil beam surveys, such as \textit{Chandra} Deep Fields South (CDFS; \citealp{cdfs2002,lehmer2012,luo2017}), are sensitive down to very low fluxes but are limited to finding low to moderate luminosity AGN.  The \textit{Chandra} Deep Field South (CDFS) 7 Ms \citep{luo2017} catalog is the deepest of X-ray surveys, covering a total area of 484.2 arcmin$^2$, with 1008 sources detected in  0.5-7 keV energy range. It has been previously shown that existing luminosity functions reasonably reproduces number counts down to 10$^{-15}$ $\rm erg\,cm^{-2}\,s^{-1}$ in this energy range \citep{ballantyne2011}. The deeper CDFS 7 Ms catalog allows comparison to even fainter fluxes, and our new results are presented in \S~\ref{sec:results}. 

\begin{turnpage}
	\begin{deluxetable*}{lcccccc}[th]
		\tablewidth{0pt}
		\tablecaption{\label{tab:observed_constraints} \textsc{Observed constraints on AGN population in X-ray band.}}
		\tablehead{\colhead{\textsc{Constraint}} &\colhead{\textsc{Survey}} & \colhead{\textsc{Band (keV)}} & \colhead{\textsc{Area (deg$^{\rm 2}$)}} & \colhead{\textsc{Depth ($\rm erg\,cm^{-2}\,s^{-1}$)}} & \colhead{\textsc{Num of Sources}} & \colhead{\textsc{Reference}}}
		\startdata
		Integrated X-ray Background & \textit{Swift}-BAT & 14 - 195 & All sky &  & & \citealp{ajello2008}  \\
		& \textit{Chandra} COSMOS  & 0.3 - 7 & 2.15 &   & & \citealp{nico2017} \\
		& \textit{\textit{RXTE}} & 3 - 20 & 22600 &  & & \citealp{rxte} \\
		& \textit{ASCA} SIS & 2 - 10 & 0.14 &  & & \citealp{ascasisxrb} \\
		Number Counts 	& \textit{NuSTAR} Extragal. overall number counts: & 8 - 24 &  &  & 124 & \citealp{harrison2016} \\
		& 1) \textit{NuSTAR} COSMOS & 8 - 24 & 1.7 & 1.3 $\times 10^{-13}$ & 91 & \citealp{civano2015nustarcosmos} \\
		& 2) \textit{NuSTAR} ECDFS & 8 - 24 & 0.3 & 2.5 $\times 10^{-14}$  & 19 & \citealp{mullaney2015}\\
		& 3) \textit{NuSTAR} EGS & 8 - 24 & 0.23 & 2.5 $\times 10^{-14}$ &  & J. Aird in prep\\
		& 4)  \textit{NuSTAR} Serendipitous Survey & 8 - 24 & 13 & 2-10 $\times 10^{-14}$ & 24 & \citealp{lansbury2017}\\
		& \textit{NuSTAR} Ser. Compton-thick counts and fraction & 8 - 24 &  & & 4 & \citealp{lansbury2017Ctk} \\
		& \textit{NuSTAR}-COSMOS Compton-thick fraction & 8 - 24 & &  & 2 & \citealp{civano2015nustarcosmos} \\
		& \textit{NuSTAR} UDS Compton-thick fraction & 8 - 24 & 0.6 &2.7 $\times 10^{-14}$ & 6.8 $\pm$ 1.2 & \citealp{masini2018} \\
		& \textit{Swift}-BAT 70-month all source counts & 14-195 & All sky &  & 838 &\citealp{claudio2017bat} \\
		& \textit{Chandra} Deep Fields South 7 Ms & 0.5 - 7 & 0.1345 & 2.7 $\times 10^{-17}$ & 1008 & \citealp{luo2017} \\
		& 2\textit{XMM}i & 0.5 - 10 & 132.3 & $\times 10^{-14}$ & 30,000 & \citealp{mateos2008} \\
		&\textit{XMM}-COSMOS & 0.5 - 10 & 2.13 & 7 $\times 10^{-16}$ & 1416 & \citealp{nicocosmos2007} \\
		& \textit{Chandra}-COSMOS & 0.5 - 10 & 0.5 & 5.7 $\times 10^{-16}$ &  1655 & \citealp{elvis2009ccosmos} \\
		& ChaMP & 0.5 - 8 & 10 & 9 $\times 10^{-16}$ & 6800 & \citealp{kim2007} \\
		& Stripe~82X & 0.5 -10 & 31.3 & 2.1 $\times 10^{-15}$ & 6181 & \citealp{lamassa2013a,lamassa2013b,lamassa2016} \\
		& \textit{XMM}-CDFS & 2 - 10 & 0.1345 & 6.6 $\times 10^{-16}$ & 339 & \citealp{ranalli2013} \\
		& {Extended CDFS} & 0.5 - 8 & 0.3 & 6.7 $\times 10^{-16}$ & 915  &\citealp{lehmer2005} \\
		& \textit{Chandra} COSMOS Leg. Comp.-thick num counts& 2 - 8 &  &  &  41.9 & \citealp{lanzuisi2018cosmosctk} \\
		X-ray \textbf{luminosity function}s &  \citealp{ueda2014} & 0.5 - 195 &  &  & \\
		& \citealp{buchner2015} & 2$-$7 &  &  &  \\
		& \citealp{aird2015xlf} & 0.5$-$7 &  &  & \\
		\enddata		
	\end{deluxetable*}	
	\clearpage
\end{turnpage}

We also include number counts in the 0.5$-$2 and 2$-$10 keV bands from 2\textit{XMM}i \citep{mateos2008}, \textit{XMM}-COSMOS \citep{nicocosmos2007}, \textit{Chandra}-COSMOS \citep{elvis2009ccosmos}, \textit{XMM}-CDFS \citep{ranalli2013} and Stripe~82X (\citealp{lamassa2013a,lamassa2013b,lamassa2016,ananna2017}), and 0.5$-$2 keV and 2$-$8 keV number counts from Extended \textit{Chandra} Deep Field Survey (ECDFS; \citealp{lehmer2005}) and \textit{Chandra} Multi-wavelength Project (ChaMP; \citealp{kim2007}). 2\textit{XMM}i is a \textit{XMM}-Newton Serendipitous Survey covering 132.3~deg$^2$, contains more than 30,000 objects down to flux limits of 10$^{-15}$ $\rm erg\,cm^{-2}\,s^{-1}$ in the 0.5$-$2 keV bin and 10$^{-14}$ $\rm erg\,cm^{-2}\,s^{-1}$ above 2 keV. \textit{XMM}-COSMOS is a 2.13~deg$^{\rm 2}$ survey with a total exposure time of $\sim$1.5 Ms, reaching similar flux levels as 2\textit{XMM}i homogeneously for 90\% of the total area. \textit{Chandra}-COSMOS covers a smaller area in the COSMOS-Legacy field (0.9 deg$^2$) but with twice the effective exposure time as \textit{XMM}-COSMOS, reaches nearly  10$^{-16}$ $\rm erg\,cm^{-2}\,s^{-1}$ flux levels in both soft and hard bands. ChaMP covers a $\sim$10 deg$^2$ area, with the deepest 0.5-8 keV levels reaching 9 $\times$ 10$^{-16}$ $\rm erg\,cm^{-2}\,s^{-1}$, and has a range of exposure times of 0.9-124 Ks. For ease of comparison, we convert all the hard band \textit{Chandra} and \textit{XMM} surveys to the 2$-$7 keV band. To convert the harder band fluxes (2$-$10 and 2$-$8 keV) to 2$-$7 keV, we use the photon indices adopted by each survey: $\Gamma = 1.6$ for 2\textit{XMM}i, $\Gamma = 1.7$ for \textit{XMM}-CDFS, \textit{XMM}-COSMOS, Stripe~82X, and $\Gamma = 1.4$ for ECDFS, \textit{Chandra}-COSMOS and ChaMP.

Heavily obscured Compton-thick sources are one of the biggest remaining uncertainties in population synthesis models.  \citet{lanzuisi2018cosmosctk} provides a careful analysis of \textit{Chandra} COSMOS Legacy survey spectra to produce Compton-thick number counts in the 0.04 $<$ z $<$ 3.5 range. We compare these Compton-thick number counts with existing models in \S~\ref{sec:results}.

At E $>$ 10 keV bands, the \textit{Swift} BAT 70-month catalog provides a hard X-ray selected sample in the nearby Universe (z $<$ 0.1) and \textit{NuSTAR} Extragalactic Surveys provide an equivalent sample up to z $\sim 1$ \citep{aird2015b,harrison2016,lansbury2017}. We calculate overall number counts of the \textit{Swift}-BAT 70-month catalog presented in \citet{claudio2017bat}, in the 14$-$195 keV band. We compare the \textit{Swift}-BAT overall counts, as well as \textit{NuSTAR} Extragalactic Survey overall number counts from \citet{harrison2016} with existing models. The \textit{NuSTAR} Extragalactic Surveys are a wedding cake survey, and we look at three tiers in this work: UKIDSS Ultra Deep Survey (UDS; \citealp{masini2018}),  COSMOS \citep{civano2015nustarcosmos} and Serendipitous Survey \citep{alexander2013,lansbury2017Ctk}.
 The details of each of these surveys are given in Table~\ref{tab:observed_constraints}. \citet{masini2018} finds an observed Compton-thick fraction in the UDS field - 11.5 $\pm$ 2.0\%. \citet{civano2015nustarcosmos} calculates Compton-thick fraction using two objects (out of 91) from the \textit{NuSTAR} COSMOS field. The observed fraction of Compton-thick objects found in this work is between 13-20\%.  

\citet{lansbury2017Ctk} calculated \textit{NuSTAR} band ratios from the Serendipitous Survey to select eight (out of 497) heavily obscured objects with the hardest X-ray spectra. They present an analysis of the soft and hard X-ray properties of these sources (the soft bands provided by \textit{Chandra}, \textit{Swift}-XRT or \textit{XMM-Newton}) as well as multi-wavelength properties. Of these Compton-thick objects (in this case, N$_{\rm H} >$ 1.5 $\times$ 10$^{\rm 24}$ cm$^{\rm -2}$) three are at low redshifts (z  $\simeq$ 0.036, 0.034, 0.069) and one Compton-thick AGN is at a relatively higher redshift (z $\sim$ 0.16). Using these four objects, \citet{lansbury2017Ctk} calculated a low-redshift bias Compton-thick number counts, an upper limit in Compton-thick number counts without any bias, and a Compton-thick fraction. They report a Compton-thick fraction of $\simeq$ 30\% in the local Universe ($z < 0.07$). 

In Table~\ref{tab:observed_constraints}, we list the three luminosity functions we examine in this work as observed constraints as well. Since these luminosity functions were derived from data, they should reasonably agree with each other, as well as any new luminosity function. Each survey and model is a step towards converging on the correct solution, and so we compare luminosity functions against each other to verify whether they are in reasonable agreement. The results of these comparisons are presented in \S~\ref{sec:results}.

\section{New Model}\label{sec:new_model}

Existing population synthesis models do not reproduce all observed constraints (shown in \S~\ref{sec:results}), so we attempted to update the existing XLF which is best fit to CXB --- U14 --- using a newer absorption function by \citeauthor{ricci2015} (\citeyear{ricci2015}; henceforth, R15). The R15 absorption function is based on \textit{Swift}-BAT 70-month sample whereas the U14 function was based on \textit{Swift}-BAT 9-month sample.

R15 reports a completeness-corrected absorption function for the \textit{Swift}-BAT 70-month survey, based on assumptions about the geometry of the torus component. Specifically, the corrections to the absorption functions are calculated by assuming two different opening angles of the torus component of the AGN, 35$^{\circ}$ and 60$^{\circ}$. These corrections are calculated  in two luminosity bins: $\log$  (${\rm L}_{\rm 14-195}$/$\text{erg s}^{-1}$) = 40$-$43.7 and $\log$ (${\rm L}_{\rm 14-195}$/$\text{erg s}^{-1}$) = 43.7$-$46. The corrected fractions are shown in Figure~4 of R15 and also in \S~\ref{sec:results} of this work.

Both U14 and R15 absorption functions are normalized in $\log$ (N$_{\rm H}$/cm$^{2}$) = 20-24 range. Using \textit{Swift}-BAT 9-month data, U14 quantified the fraction of objects in four equally spaced $\log$ (N$_{\rm H}$/cm$^{2}$) bins in the 20-24 range at a fixed luminosity: $\log$  (${\rm L}_{\rm 2-10}$/$\text{erg s}^{-1}$) = 43.75. Then luminosity dependence and redshift dependence were added to these fractions based on observed relationships. U14 assumes that number of Compton-thin objects are equal to number of Compton-thick objects at any redshift and luminosity bin, and evenly divided over $\log$ (N$_{\rm H}$/cm$^{2}$) = 24-26 bin. 

As U14 underestimates Compton-thick number counts, we updated it with the R15 absorption function with the higher Compton-thick fractions: the correction which assumes a torus opening angle of 60$^{\circ}$.  As \textit{Swift}-BAT is a local sample, we assume the R15 distribution locally and add the same redshift evolution as U14. The details of the update are explained in Appendix~\ref{sec:ricci}. We found that this update still leaves the Compton-thick counts largely underestimated. Therefore we used a neural network to modify the XLF further. In this section, we describe this neural network.

\subsection{Neural Network to Optimize XLF}

\begin{figure*}[t]
	\centering
	\includegraphics[width=1.\linewidth]{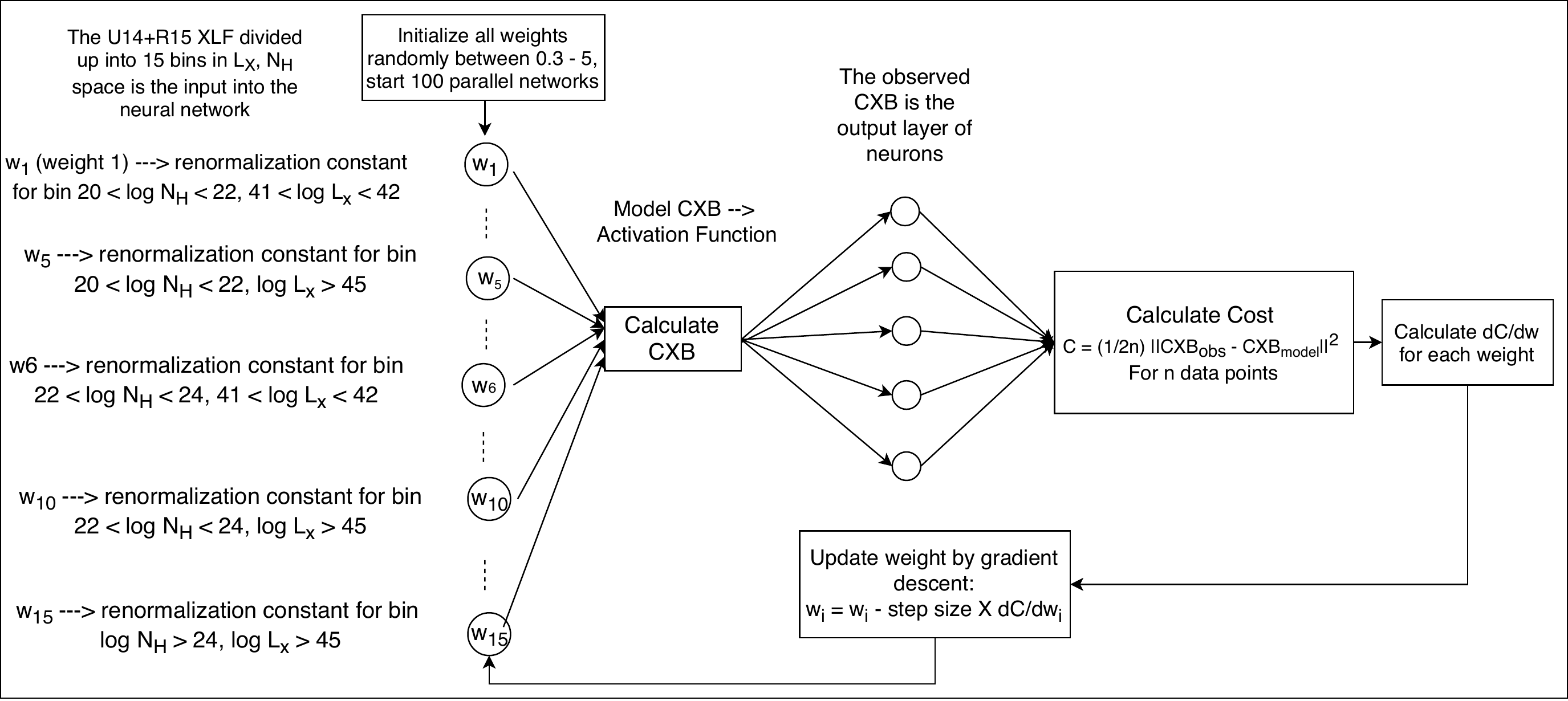}
	\caption{Summary of our neural network to find space densities that reproduce the X-ray background. The result is then validated using number counts and Compton-thick fractions from \textit{Chandra}, \textit{XMM}, \textit{Swift}-BAT and \textit{NuSTAR} survey.}
	\label{fig:neuralnetwork} 
\end{figure*}

\begin{deluxetable*}{lcc}[th]
	\tablewidth{0pt}
	\tablecaption{\label{tab:xlf_summary} \textsc{Summary of approaches taken to formulate XLF in recent works\tablenotemark{1}.}}
	\tablehead{\colhead{\textsc{Model}} &\colhead{\textsc{Approach}} & \colhead{\textsc{Results}}}
	\startdata
	\citealp{ueda2014} & Maximum Likelihood methods on  survey data & Parametric function \\
	&  to formulate XLF and N$_{\rm H}$ distribution & \\
	\citealp{aird2015xlf}  & Bayesian analysis of survey data & Parametric function \\
	\citealp{buchner2015}  & Bayesian analysis of survey data  & Non-parametric space density 3D grid \\
	\textbf{This work} & Neural Network fitting to X-ray background, cross-validation & Non-parametric space density 3D grid \\
	& using number counts/fractions from surveys & \\
	\enddata
    \tablenotetext{1}{The details of the spectra for each model is given in Table~\ref{tab:spectral}, where this work uses observed parameter distributions from \textit{Swift}-BAT, explained in \S~\ref{sec:spectral_param_desc}.}
\end{deluxetable*}	

In order to reasonably modify the luminosity function to find a solution that fit all the observed constraints, we used a neural network. This neural network finds all the XLFs that fits the CXB given a set of input spectra. The spectral parameter distributions of this input spectra can be luminosity, redshift and/or N$_{\rm H}$ dependent. 

The distinct contribution to the CXB from each N$_{\rm H}$ bin is discussed in \S~\ref{sec:spectral_model}. However, different proportions of AGN within the same absorption range, but with different luminosities, can produce the same CXB. To break the degeneracy within luminosity bins, all available number counts can be used for cross-validation.

Therefore, the neural network modifies the space densities to find all the solutions that fit the CXB. In this way, the X-ray background acts as a training set. 
 The rest of the number counts and Compton-thick fractions act as a test set to verify the accuracy of the output models.

We carried out the changes as follows. We convert the U14+R15 XLF described in Appendix~\ref{sec:ricci} into a 3D matrix of space densities rather than a parametric function because it provides more flexibility to apply changes. This is a 3D matrix with dimensions z, L$_{\rm 2-10}$ and N$_{\rm H}$, and a simple linear interpolator will provide space densities at any (z, L$_{\rm 2-10}$ and N$_{\rm H}$) coordinate. It differs from B15 final product as there is no binning involved. B15 space densities are flat over the width of each 3D bin, whereas the space densities in our matrix vary continuously, similar to U14 and A15.

A neural network is used to tune this matrix so that it produces an increasingly better fit to the CXB. This neural network employs back propagation and gradient descent algorithms, which are described in Appendix~\ref{sec:backpropagation}.

We summarize our approach to deriving a new XLF in Figure~\ref{fig:neuralnetwork}. After optimizing the neural network for best performance, the best configuration was as follows: we divided the matrix re-weighting into 15 blocks, i.e., 3 N$_{\rm H}$ bins (unabsorbed, Compton-thin and Compton-thick) times 5 luminosity bins [$\log$  (${\rm L}_{\rm 2-10}$/$\text{erg s}^{-1}$): 41-42, 42-43, 43-44, 44-45, 45-47]. The neural network has 15 input neurons, and each block is input into each neuron. The weights associated with each input neuron are the factors by which all space densities in each of these blocks are renormalized. After renormalization using these weights, CXB is calculated using the pre-defined spectra and this modified XLF. The neural network then calculates the cost function, which is the sum of squares of the difference between observed CXB and the model prediction, divided by 2 $\times$ the number of observed data points. We use CXB observed data points from \textit{Chandra} COSMOS \citep{nico2017}, \textit{RXTE} \citep{rxte} and \textit{Swift}-BAT \citep{ajello2008}, as these are the most updated estimates of the CXB. After calculating costs, the neural network then updates the space densities in these 15 bins simultaneously by calculating derivatives of the cost function with respect to the weights.

We initialize weights for the 15 neurons randomly between 0.3 and 5, and run 100 neural networks in parallel to find all the solutions that converge. Some parallel networks sometimes get trapped in local minimas or diverge. We consider all branches that converge to costs $\leq$ 5.0 as it roughly corresponds to a reduced $\chi^2 \leq 2$ and should not be ruled out without cross-validation. We used the CXB as a training set, and then cross-validated the resulting XLFs on the rest of the observed constraints: the number counts and the Compton-thick fractions. If an XLF fits the CXB with a reduced $\chi^2 <$ 2.0, it has contributions from the three absorption bins (unabsorbed, Compton-thin and Compton-thick) in correct proportions. However different distribution in luminosity bins can produce the same CXB. The degeneracy in distribution in luminosity bins is broken by choosing the solution that minimizes reduced $\chi^2$ with respect to all the observed number counts/Compton-thick fractions.

The number of parallel networks is limited by computational power. Increasing the number of parallel networks increases the probability of a quick convergence as the number of steps needed to reach a minimum via gradient descent is sensitive to initial position. However, each point in the CXB requires a numerical integration over redshift, L$_{\rm X}$, N$_{\rm H}$ and energy, using Monte Carlo sampling. Therefore, the number of multiple processes that can be run is limited. 

We present a summary of the approaches taken to formulate U14, A15, B15 and this work in Table~\ref{tab:xlf_summary}. 

\subsection{A Discussion on Spectral Parameters}\label{sec:spectral_param_desc}

To calculate the expected CXB and number counts from the XLF, we have to integrate the AGN spectra over a range of energies for all AGN. Because of the uncertainty in spectral parameters due to various biases and parameter couplings, as explained in \S~\ref{sec:observed_spectra}, we tested our neural network with different combinations of $\Gamma$, R and E$_{\rm cutoff}$. The combinations of spectral parameters used are determined by observed values that exist in the literature. A representative sample is listed in Table~\ref{tab:combination_spectra}. 
The \textsc{xspec} syntax of the spectral model is:

\textsc{fscatt $\times$ cutoffpl} + \textsc{wabs $\times $pexmon} + \textsc{bntorus $\times$ highecut}

For all five sets of spectra, the scattered component is proportional to the initial cutoff power-law, with a scattering fraction (f$_{\rm scatt}$) of 0.01. The torus is modeled using \textsc{BNTorus}. The value of the half-opening angle is drawn from a uniform distribution between 55$^{\circ}$ and 61$^{\circ}$ (\S~5.3.1 of \citealp{claudio2017bat}). For unabsorbed sources, the physical inclination angle should be pole-on \citep{masini2016}. \textsc{BNTorus} assumes line-of-sight column density for all angles, so the change in observed spectra with respect to inclination angle is very small. However, to be physically consistent, we fix the inclination angle for $\log (\text{N}_{\rm H}/\text{cm}^{-2}) < 21$ objects at 25$^{\circ}$, smaller than the opening angle of the torus. The inclination angles for $\log (\text{N}_{\rm H}/\text{cm}^{-2})$ = 21-24 objects are larger than the opening angle by 15$^{\circ}$, and the inclination angle of $\log (\text{N}_{\rm H}/\text{cm}^{-2}) > 24$ is fixed at maximum possible value of 87$^{\circ}$. The reflection component is modeled using self-consistent \textsc{pexmon} model, with an inclination angle of 30$^{\circ}$ (all parameters typical of \textit{Swift}-BAT AGN).

Spectral Set 1 in Table~\ref{tab:combination_spectra} is the observed spectral parameter distributions from \textit{Swift}-BAT 70-month catalog. For the intrinsic cutoff power-law, we draw $\Gamma$ from two normal distributions:
$\langle \Gamma \rangle \simeq$ 1.8 and $\sigma_{\Gamma}$ = 0.24 for unobscured AGN, and $\langle \Gamma\rangle \simeq$ 1.72 and $\sigma_{\Gamma}$ = 0.31
for obscured AGN. We choose cutoff energy values of the 161 objects for which this value was properly constrained, and draw from that distribution to produce AGN spectra. The reflection scaling factor is drawn from two Gaussian distributions: ($\langle R \rangle$ = 0.83, $\sigma_R$ = 0.14) and ($\langle R \rangle$ = 0.37,  $\sigma_R$ = 0.11) for unobscured and obscured AGN respectively. These $\langle R \rangle$ were calculated by \citet{claudio2017bat} by taking the upper and lower limits in R into account to produce a representative median for obscured and unobscured sources. The lower $\langle R \rangle$ value for obscured sources could arise because the reflection component from accretion disk is higher for objects which are observed pole-on rather than the ones which are observed edge-on. 

\begin{figure*}[th]
	\centering
	\includegraphics[width=0.5\linewidth]{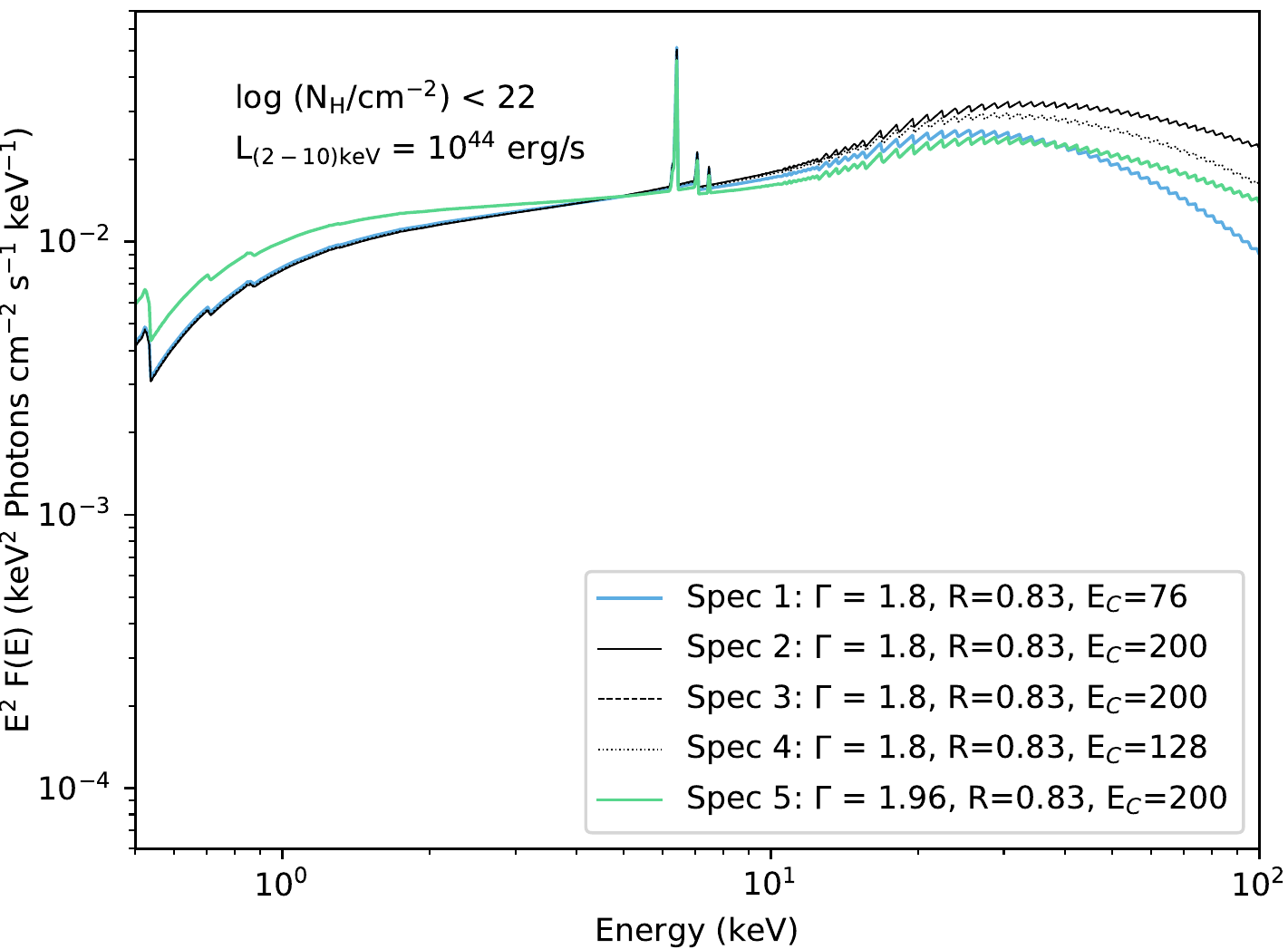}~
	\includegraphics[width=0.5\linewidth]{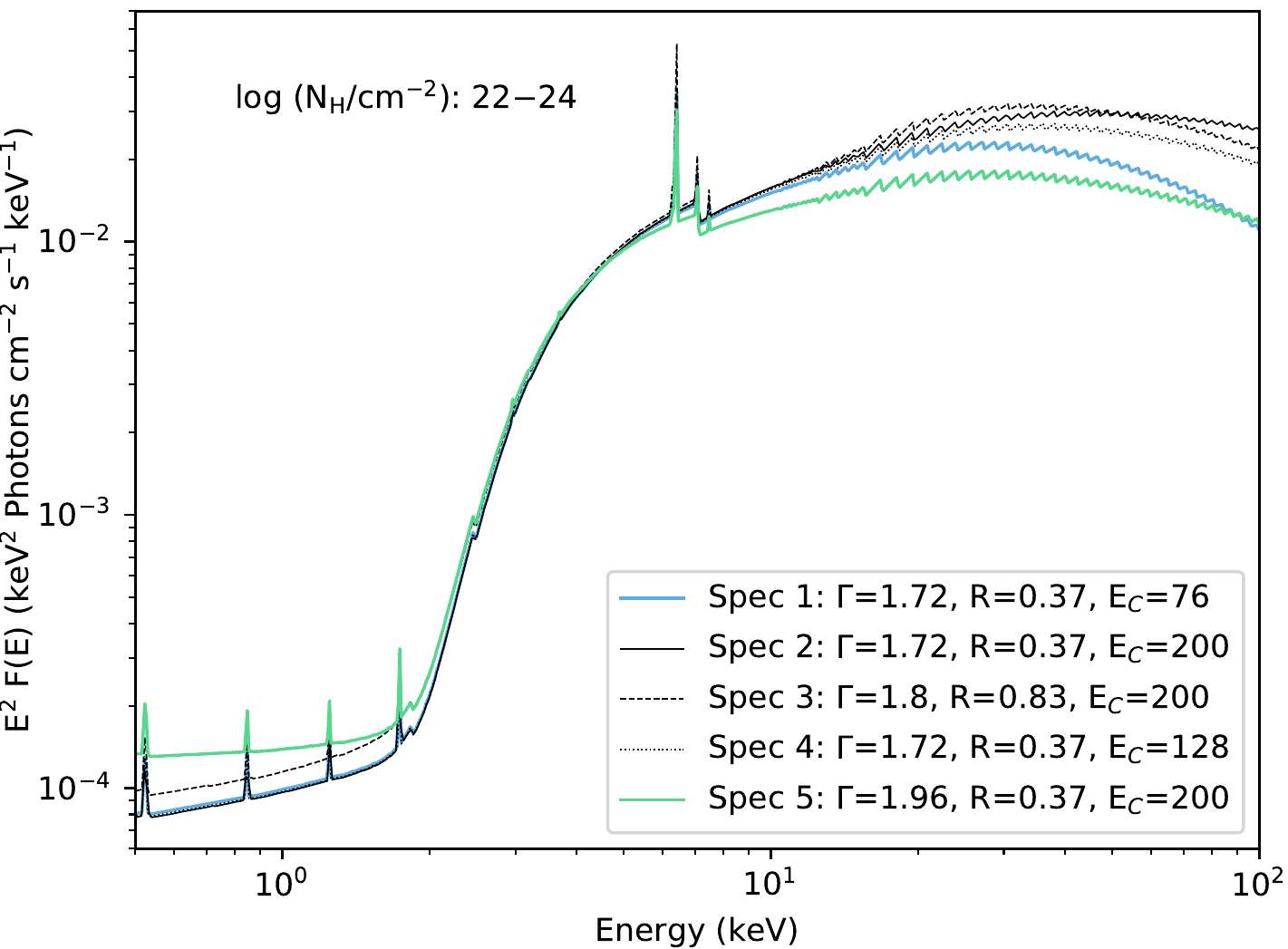}
	\includegraphics[width=0.5\linewidth]{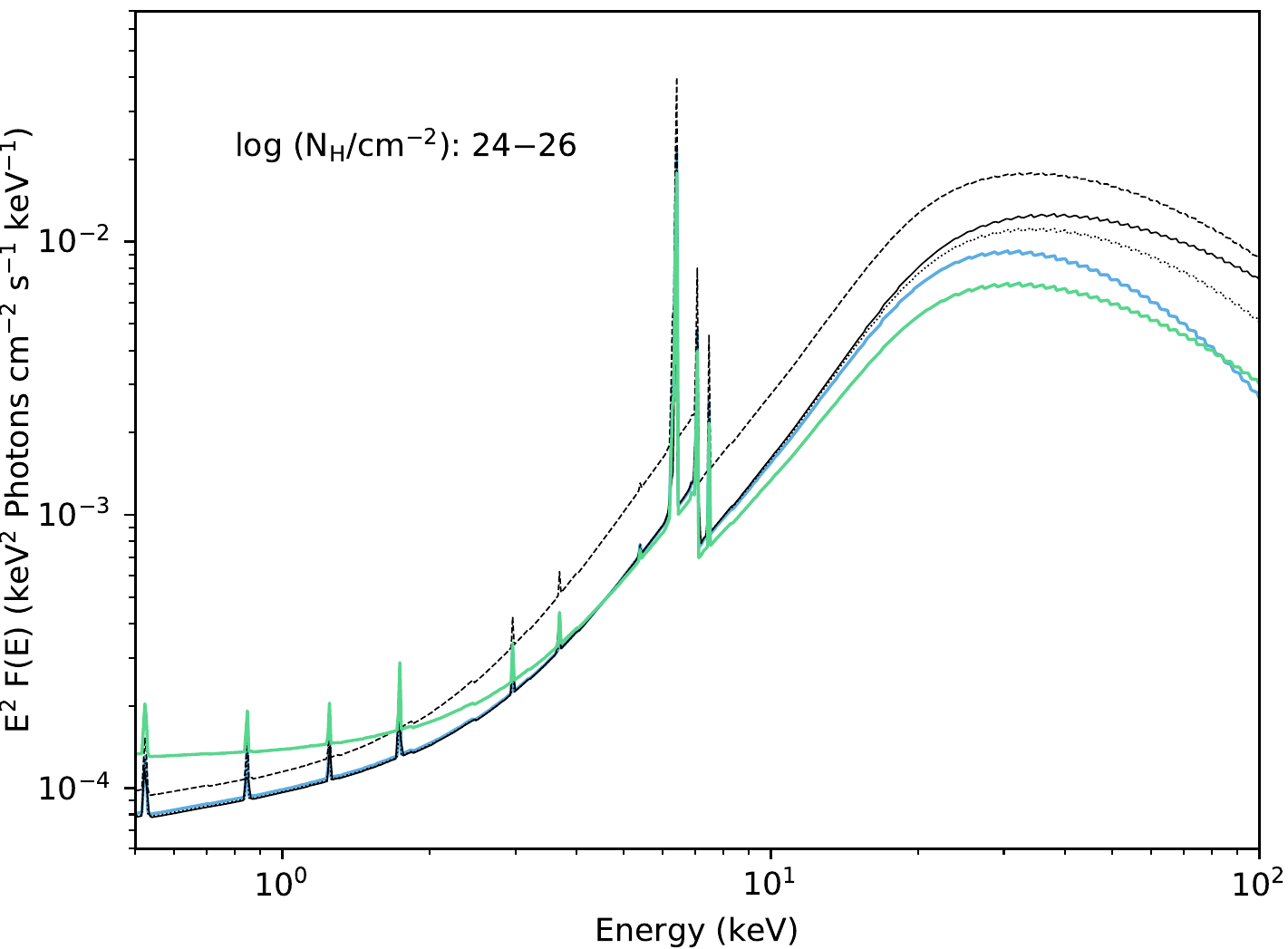}
	\caption{The resultant spectra from the five spectral parameter distributions explored in this work to find XLF that fits all observed constraints. \textit{Top left:} Spectra of an unabsorbed, \textit{top right:} Compton-thin and \textit{bottom:} Compton-thick object. The same parameters are used for both Compton-thin and Compton-thick objects. All the spectra have an intrinsic X-ray 2$-$10 keV luminosity of 10$^{\rm 44}$ erg/s. The values of the parameters come from \textit{Swift}-BAT 70-month survey \citep{claudio2017bat} and other spectral fittings \citep{nandra1994,malizia2014,ueda2014}.} 
    \label{fig:spectral_combinations} 
\end{figure*}

To account for the fact that the observed cutoff energies are biased against high values (i.e., E $>$ 100 keV), we attempted Spectral Set 2, where the observed values of $\Gamma$ and R are unchanged, and cutoff energy is drawn from a Gaussian distribution of $\langle E_{\rm cutoff} \rangle$ = 200 keV and $E_{{\rm cutoff},\sigma}$ = 29 keV, as found using the Kaplan-Meier estimator on observed values. For Spectral Set 3, we use $\langle \Gamma \rangle$ and $\langle R \rangle$ for unobscured \textit{Swift}-BAT sources and a cutoff energy of 200 keV, as there is a possibility that the difference in spectral parameters between obscured and unobscured objects arise due to imperfections in the modeling of the obscurer. For Spectral Set 4, we adopted a Gaussian cutoff energy distribution with $\langle E_{\rm cutoff} \rangle$ = 128 keV and $E_{{\rm cutoff},\sigma}$ = 46 keV, as reported in \citet{malizia2014}
keeping all other parameters identical to Spectral Set 1. In Spectral Set 5, we adopt 
 $\langle E_{\rm cutoff} \rangle$ = 200 keV and $E_{\sigma}$ = 29 keV along with $\langle \Gamma \rangle$ = 1.96, which is the median $\Gamma_{\rm BAT}$ reported in \citet{claudio2017bat}, and consistent with high $\Gamma$ values assumed in earlier models and observed values in \citet{nandra1994}, \citet{gilli2007} and U14. The five sets of spectra in each N$_{\rm H}$ bin are shown in Figure~\ref{fig:spectral_combinations}. The results for these spectral analysis are discussed in \S~\ref{sec:results}.

 \begin{deluxetable*}{lcccc}[th]
	\tablewidth{0pt}
	\tablecaption{\label{tab:combination_spectra} \textsc{Spectral parameters used to construct spectra in this work.}}
	\tablehead{\colhead{\textsc{Spectral Set}} &\colhead{\textsc{Photon Index  $\langle\Gamma\rangle$}} & \colhead{\textsc{Refl. Scaling Factor (R)}}&\colhead{\textsc{E$_{\rm cutoff}$ (keV)}}&\colhead{\textsc{f$_{\rm scatt}$}} }
	\startdata
	1 \tablenotemark{1}  & 1.72 (obscur), 1.8 (unobsc) & 0.37 (obscur), 0.83 (unobsc) & 76 & $\simeq$ 1\% \\
    2 \tablenotemark{2}  & 1.72 (obscur), 1.8 (unobsc) & 0.37 (obscur), 0.83 (unobsc) & 200 $\pm$ 29 & $\simeq$ 1\% \\
    3 \tablenotemark{3}  &  1.8  &  0.83 & 200 $\pm$ 29 & $\simeq$ 1\% \\
    4 \tablenotemark{4}  & 1.72 (obscur), 1.8 (unobsc) & 0.37 (obscur), 0.83 (unobsc) & 128 $\pm$ 46 & $\simeq$ 1\% \\
    \textbf{5} \tablenotemark{5} & 1.96 & 0.37 (obscur), 0.83 (unobsc) & 200 $\pm$ 29 & $\simeq$ 1\% \\
	\enddata	
    \tablenotetext{1}{Observed parameters determined by detailed X-ray spectral fittings to \textit{Swift}-BAT 70-month survey sources by \citealp{claudio2017bat}.}
    \tablenotetext{2}{The cutoff energies measured in \textit{Swift}-BAT can only be adequately constrained when the value is lower than 100 keV, so using a distribution that takes the lower limits into account using a Kaplan-Meier estimator, the true median is found to be 200 keV.}
	\tablenotetext{3}{$\Gamma$ and R for unobscured sources from \textit{Swift}-BAT, and E$_{\rm cutoff}$ = 200 keV.}
	\tablenotetext{4}{Same parameter distribution as Spectrum 1, with E$_{\rm cutoff}$ = 128 $\pm$ 46 keV Gaussian distribution. This cutoff energy distribution is reported by \citet{malizia2014}.}
    \tablenotetext{5}{Spectrum 2 parameter distribution, with $\Gamma$ = 1.96 $\pm$ 0.1, median  $\Gamma_{\rm BAT}$  for non-blazar AGN as observed in \textit{Swift}-BAT 70-month sample, consistent with \citet{nandra1994}, \citet{gilli2007} and U14.}
\end{deluxetable*}	

\section{Results and Discussion}\label{sec:results}

\begin{figure*}[t]
	\centering
	\includegraphics[width=0.5\linewidth]{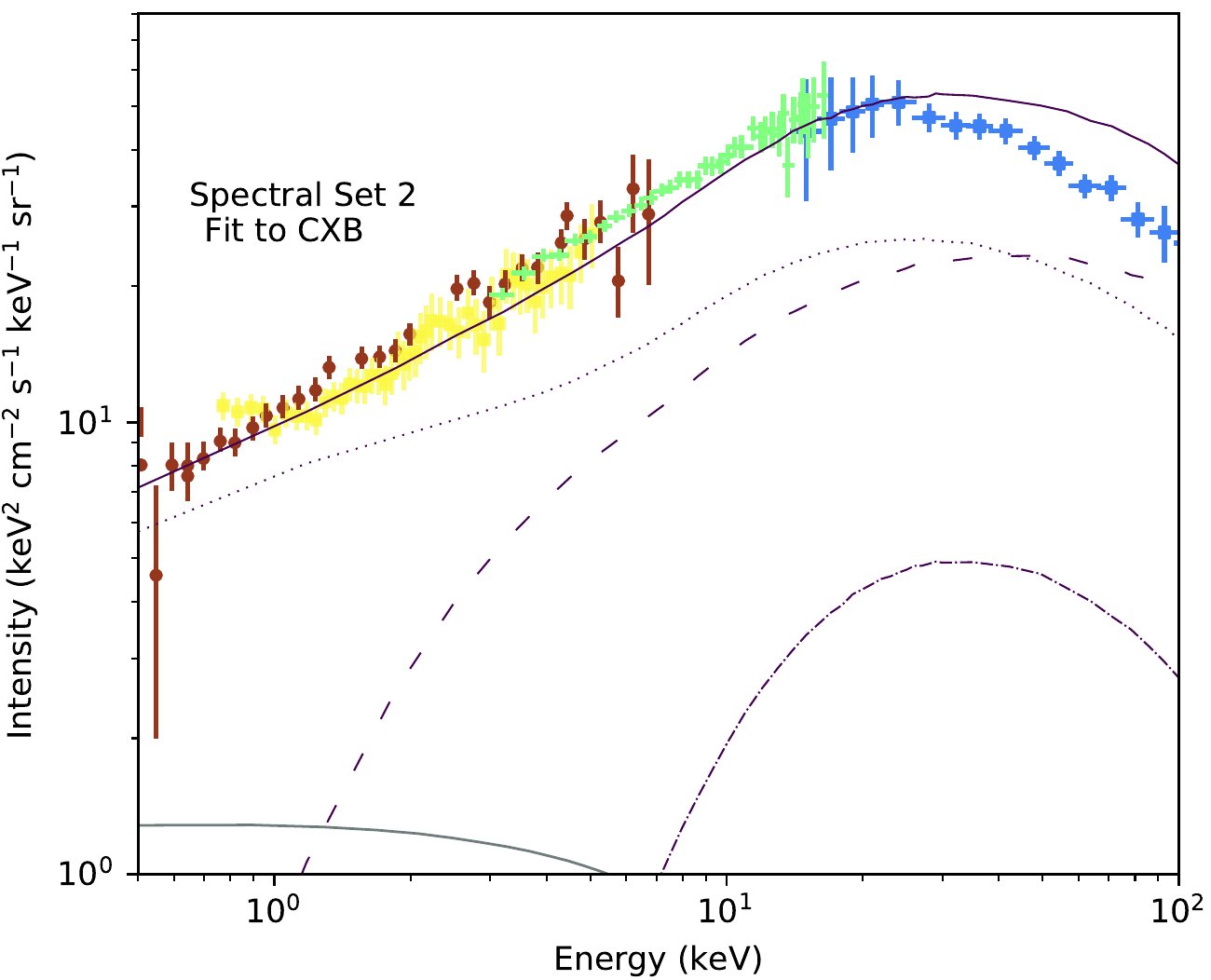}~
	\includegraphics[width=0.5\linewidth]{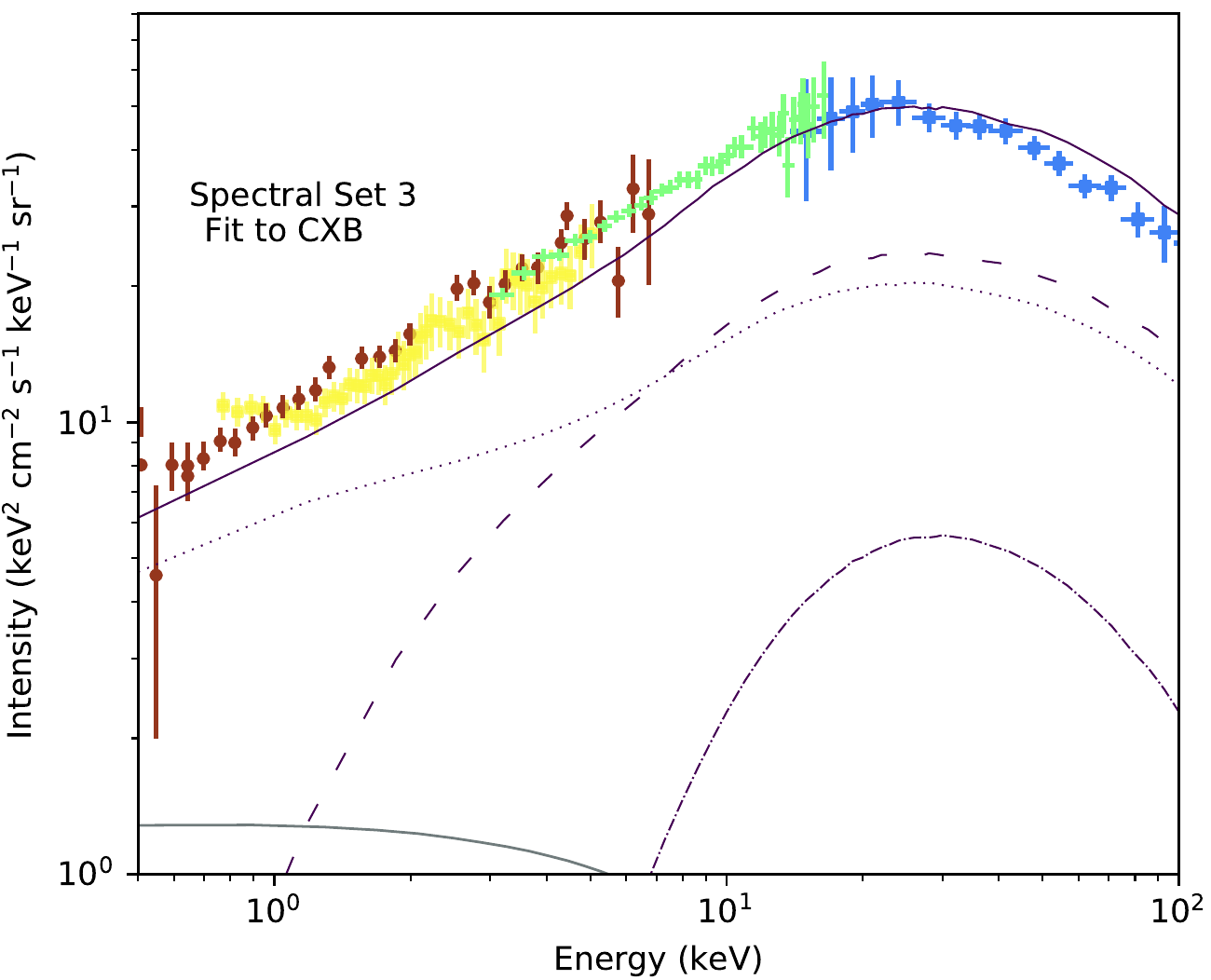}
    \includegraphics[width=0.5 \linewidth]{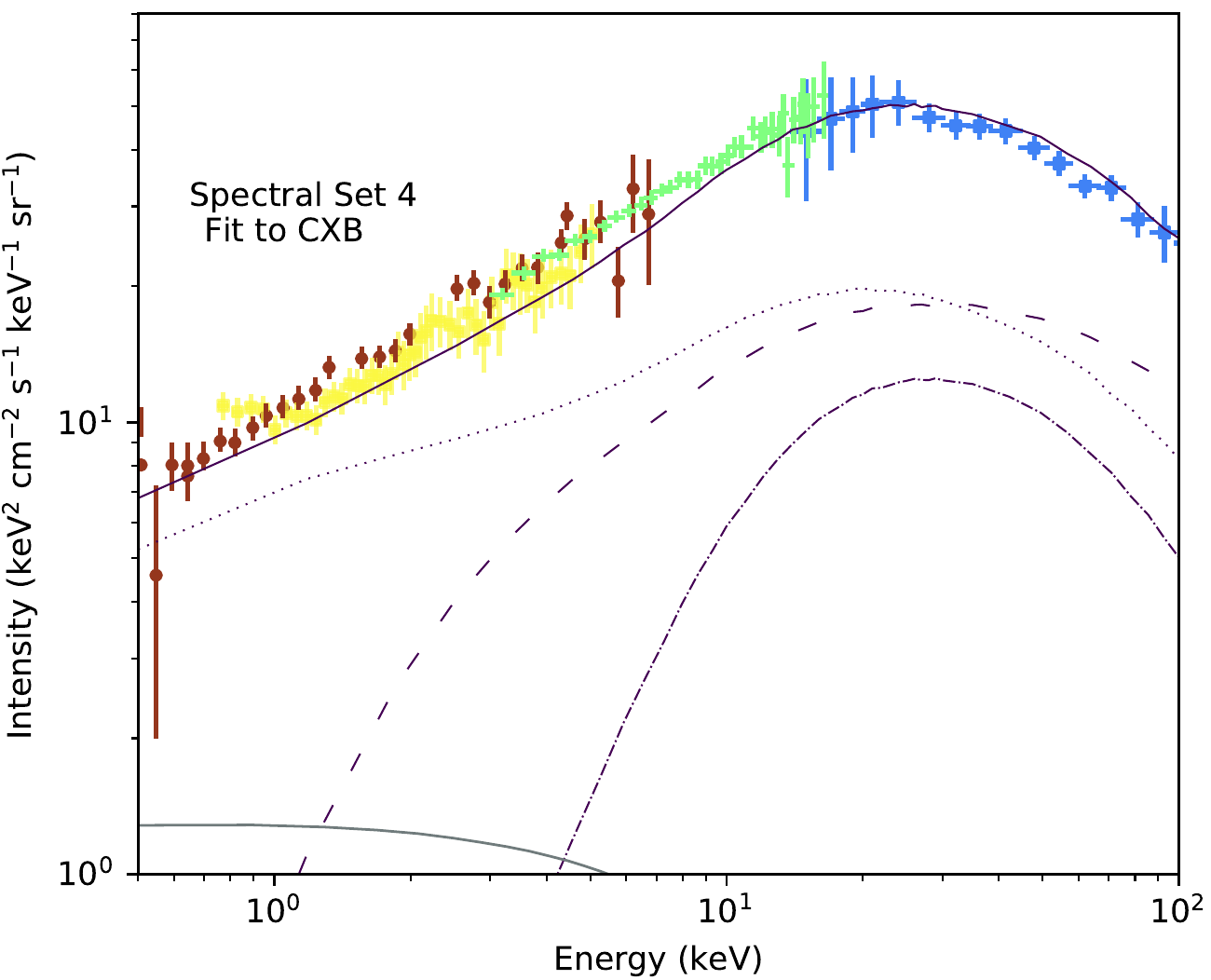}~
    \includegraphics[width=0.5 \linewidth]{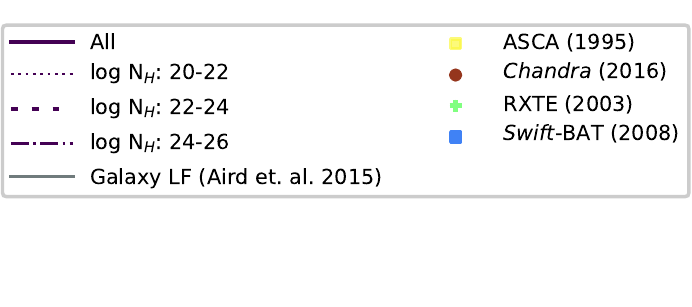}
	\caption{CXB data points fitted with three different assumed AGN spectra. \textit{Top left panel:} The assumed spectra are from Spectral Set 2 from Table~\ref{tab:combination_spectra}. Unabsorbed AGNs (\textit{dotted line}) dominate the E $<$ 1 keV region, so we fix the unabsorbed fraction such that this region is well fitted. Changing unabsorbed contribution will over- or underproduce CXB in this region. The Compton-thin objects (\textit{dashed line}) contribute heavily in the 3$-$10 keV region, where CXB is underestimated by $\geq$ 3$\sigma$ with respect to \textit{RXTE} data. Increasing Compton-thin contribution will improve fit in this region, but overestimate CXB at $>$ 30 keV, even with negligible contribution from Compton-thick objects (\textit{dash-dotted line}). Increasing Compton-thick fraction will contribute minimally in the 3$-$10 keV region but will lead to greater overestimation at higher energies, as Compton-thick contribution peaks at 20$-$60 keV. The \textit{black solid lines} in all three panels is the galaxy contribution to CXB, calculated using A15 galaxy LF. \textit{Top right panel:} Spectral Set 3 adopts a higher photon index (1.8) and reflection coefficient (0.87) for absorbed objects, equal to those of unabsorbed objects. This model reaches better agreement with CXB than Spectral Set 2 at E $>$ 30 keV, but continues to overestimate it, and it does not match all observed constraints. \textit{Bottom left panel:} Spectral Set 4 adopts the same reflection parameter and photon index distributions as Spectral Set 2, but a lower cutoff energy of 128 $\pm$ 46 keV. It produces improved (but not perfect) fits to CXB compared to Spectral Sets 2 and 3, but heavily underestimates Compton-thick number counts and fractions as higher Compton-thick contribution will overproduce CXB at E $>$ 30 keV.} 
	\label{fig:combinations} 
\end{figure*}

We find a modified XLF that satisfies all observed constraints assuming the observed \textit{Swift}-BAT 70-month spectral parameter distributions (Spectral Set 1 in Table~\ref{tab:combination_spectra}). The results for Spectral Sets 2, 3 and 4 are shown in Figure~\ref{fig:combinations}. For Spectral Set 2, where the observed values of $\Gamma$ and R are unchanged, but the cutoff energy is drawn from a Gaussian distribution of $\langle E_{\rm cutoff} \rangle$ = 200 keV and $E_{\sigma}$ = 29 keV, the CXB at E $>$ 30 keV is generally overestimated. We demonstrate the cause of this overestimation in Figure~\ref{fig:combinations}. In the top left panel of the figure, the unabsorbed contribution to CXB is fixed to perfectly reproduce observations at E $<$ 2 keV. Generally, the Compton-thin contribution becomes more significant at E $>$ 2 keV, and contributes to reproducing the slope of the CXB between 3$-$10 keV. However, in the figure, this region of the CXB is underestimated, whereas CXB at E $>$ 30 keV is overestimated. The Compton-thick contribution to 3$-$10 keV is much smaller than at E $>$ 20 keV, so increasing Compton-thick contribution will improve fits at 3$-$10 keV minimally but increase overestimation at E $>$ 30 keV.

Similarly, for Spectral Sets 3 and 4, CXB is either overestimated at high energies (E $>$ 30 keV) or the slope cannot be matched at lower energies (E $<$ 10 keV). Although Spectral Set 4 provides much more improved fits to CXB, the Compton-thick contribution is very low and does not match observed number counts.

We obtain the best fit to all constraints using Spectral Set 5, where we adopt 
 $\langle E_{\rm cutoff} \rangle$ = 200 keV and $E_{\sigma}$ = 29 keV along with $\langle \Gamma \rangle$ = 1.96. Our analysis indicates that to fit the CXB, the effective mean of $\Gamma$ and $E_{\rm cutoff}$ distributions has to be such that if $\langle E_{\rm cutoff} \rangle < 100$ keV then $\langle \Gamma \rangle \simeq 1.7-1.8$, or if $\langle E_{\rm cutoff} \rangle \geq 200$ keV then $\langle \Gamma \rangle > 1.8$. This is possibly a consequence of the correlation between the two parameters. Since the cutoff energies and $\Gamma$ values in \textit{Swift}-BAT are constrained together, the results of the spectral fit for these two parameters are not independent. However, we do find a closer fit to the observed constraints using a higher $\Gamma$ and the unbiased $\langle E_{\rm cutoff} \rangle$ value of 200 keV.
 
\begin{figure*}[th]
	\centering
\includegraphics[width=0.48\linewidth]{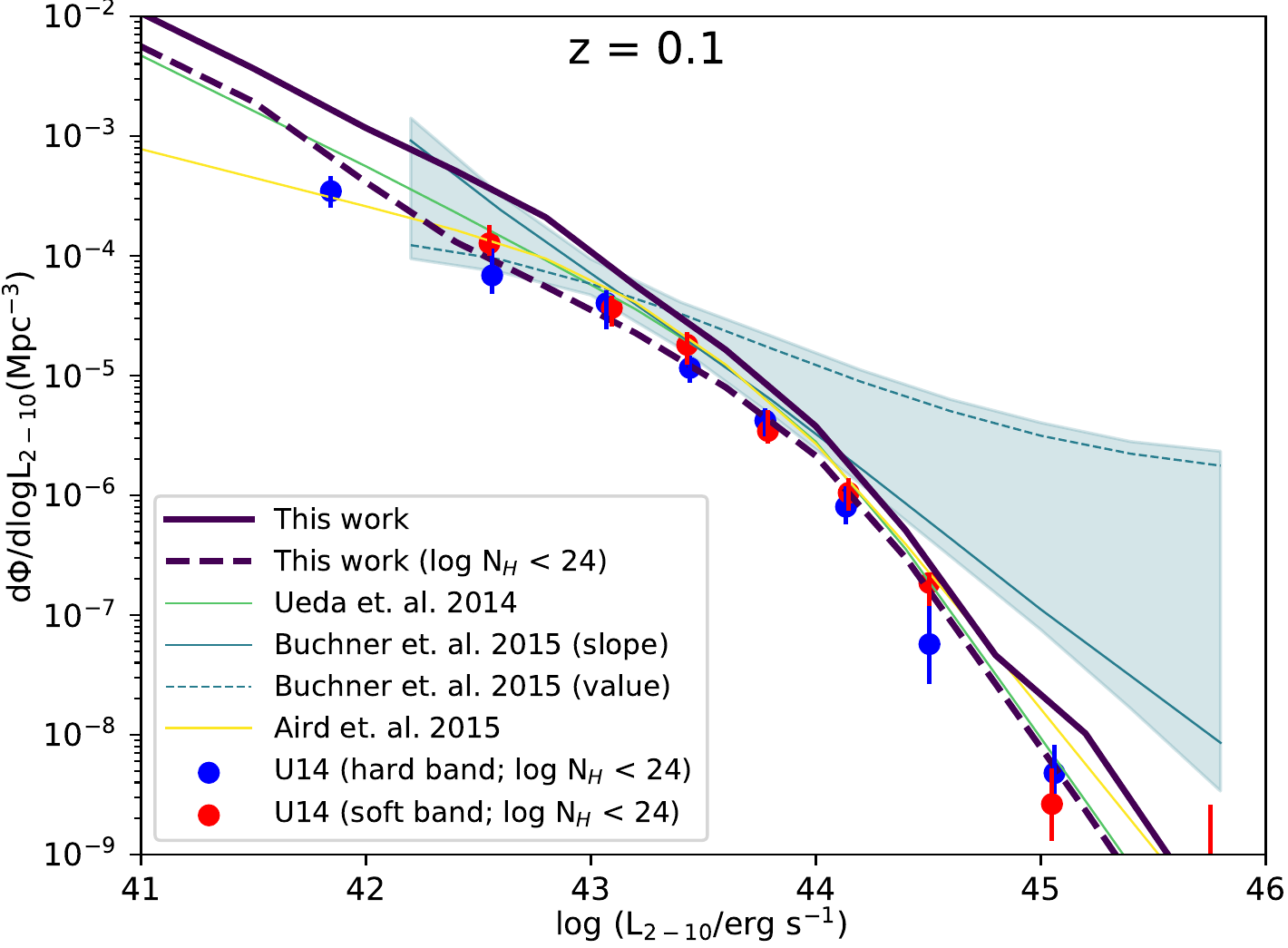}~
\includegraphics[width=0.48\linewidth]{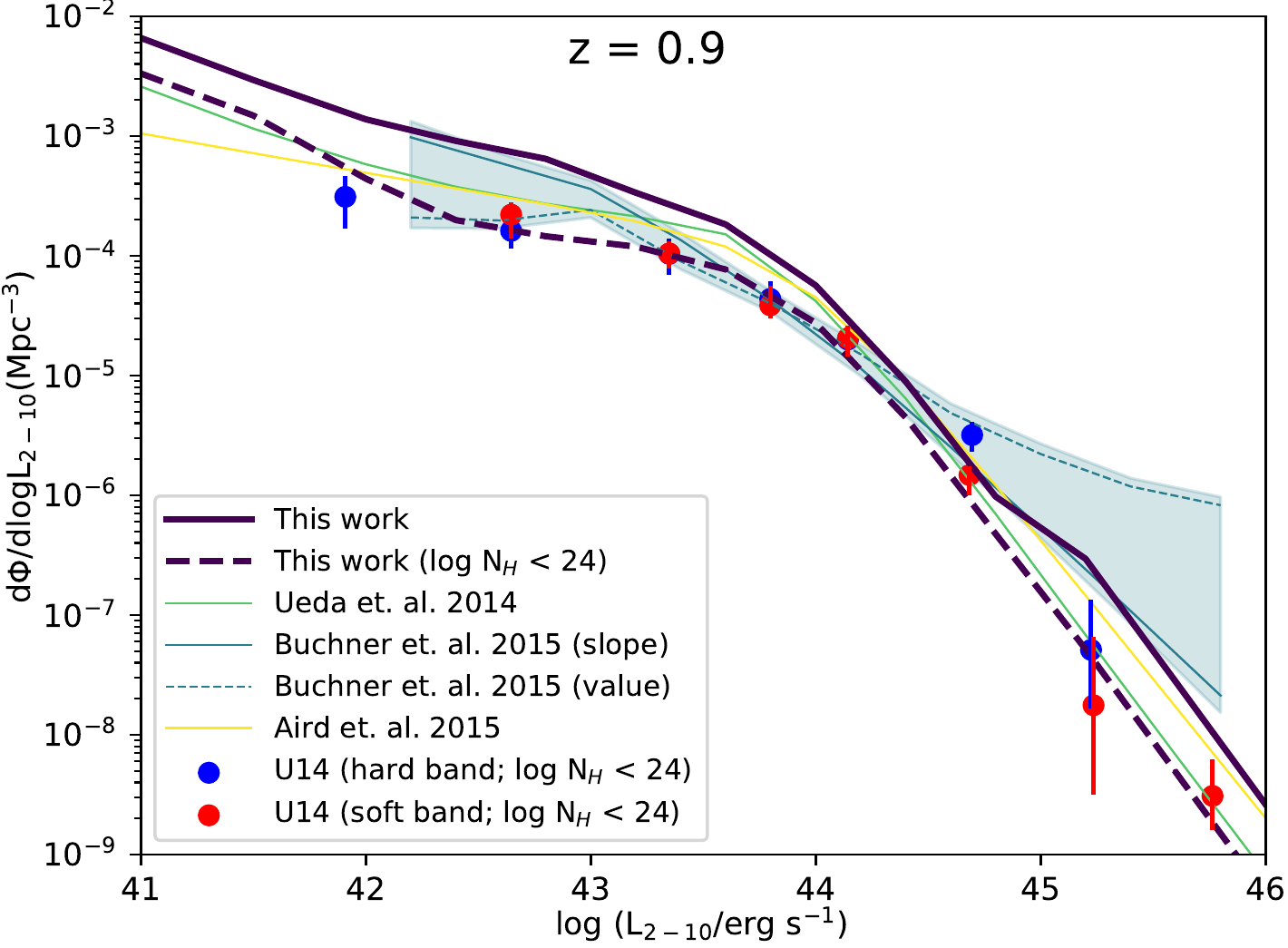}
\includegraphics[width=0.48\linewidth]{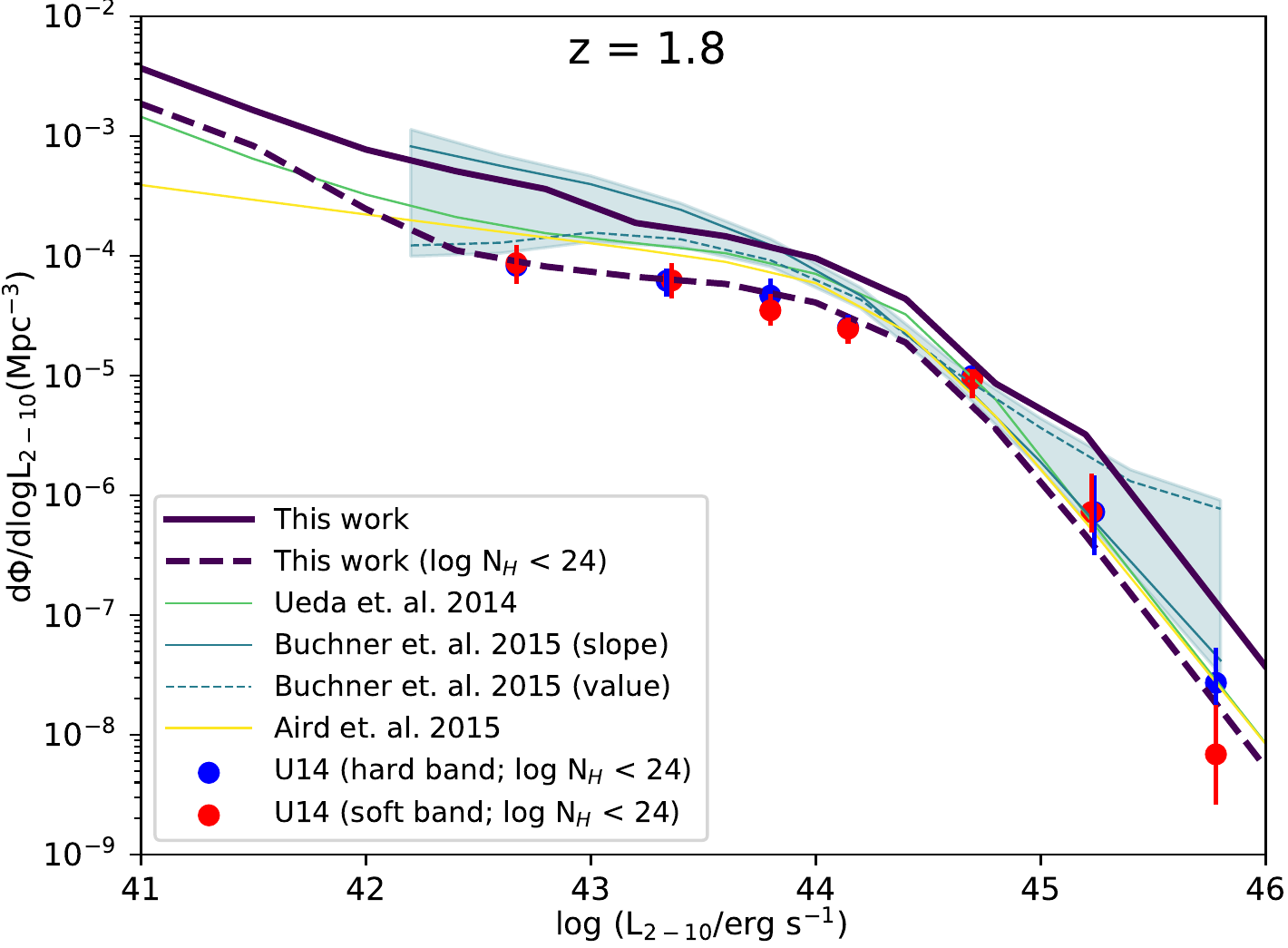}~
\includegraphics[width=0.48\linewidth]{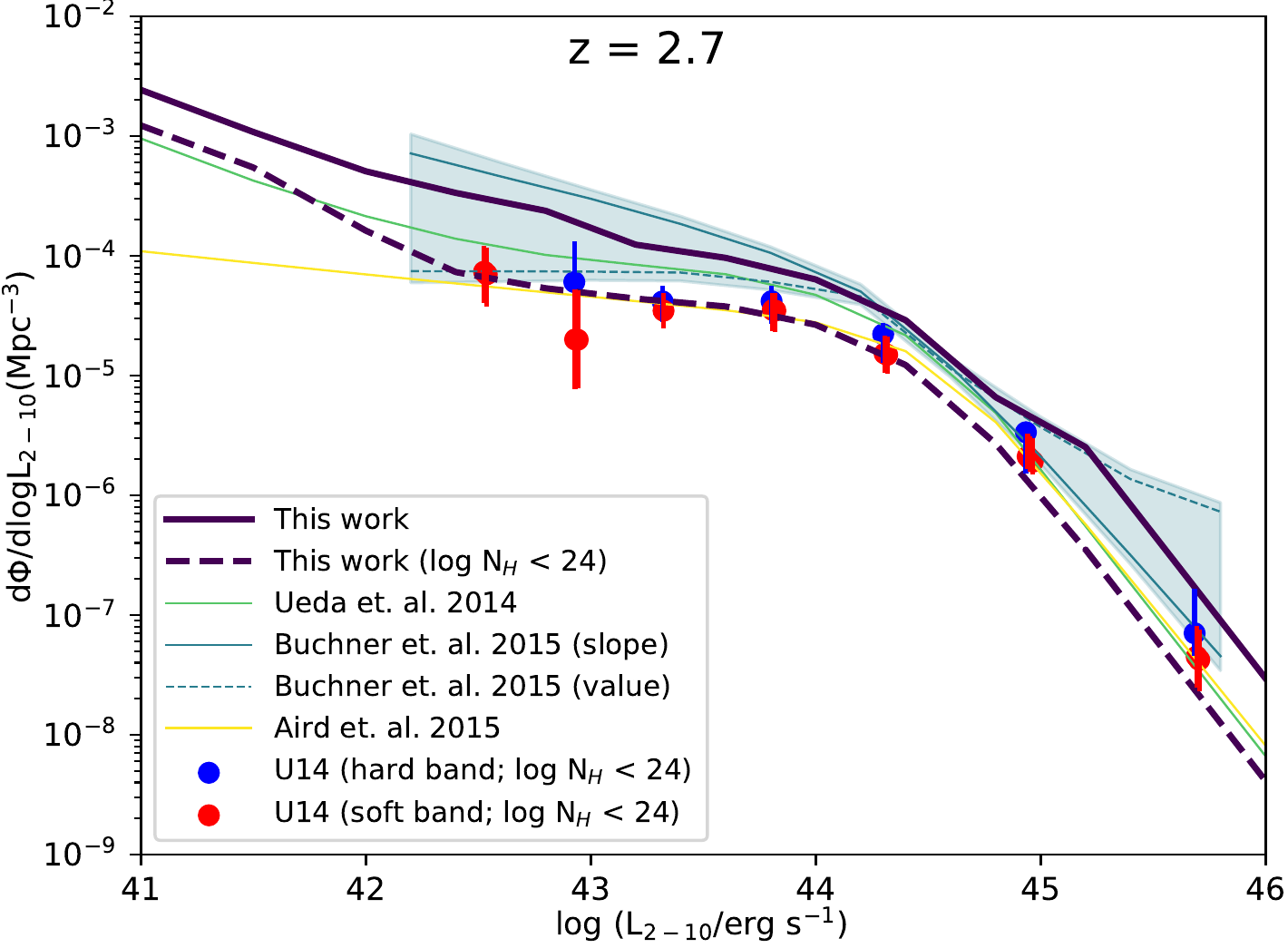}
	\caption{Number of objects per comoving Mpc$^{\rm 3}$ against log L$_{\rm x}$ for  z = 0.1 (\textit{top left}), z = 0.9 (\textit{top right}), z = 1.8 (\textit{bottom left}) and z = 2.7 (\textit{bottom right}), summed over all absorption bins, including Compton-thick objects (\textit{solid lines}), and summed in the $\log$ ${\rm N}_{\rm H} = 20-24$ bins (\textit{dashed lines}). The models are this work (\textit{purple lines}), U14 (\textit{light green line}), A15 (\textit{yellow lines}), B15 median of constant slope prior (\textit{gray solid lines}) and median of constant value prior (\textit{gray dashed lines}). The shaded gray region is B15 uncertainty (1-99th percentile). The \textit{blue and red data points} are binned counts in $\log$ ${\rm N}_{\rm H} = 20-24$ range from \citet{ueda2014}, in the hard and soft X-ray selected bands, respectively. The overall XLF in the $\log$ ${\rm N}_{\rm H} = 20-24$ for this work (\textit{dashed purple line}) is also given for comparison with data points.} 
	\label{fig:dphi} 
\end{figure*}

\begin{figure*}[th]
	\centering
	\includegraphics[width=0.5\linewidth]{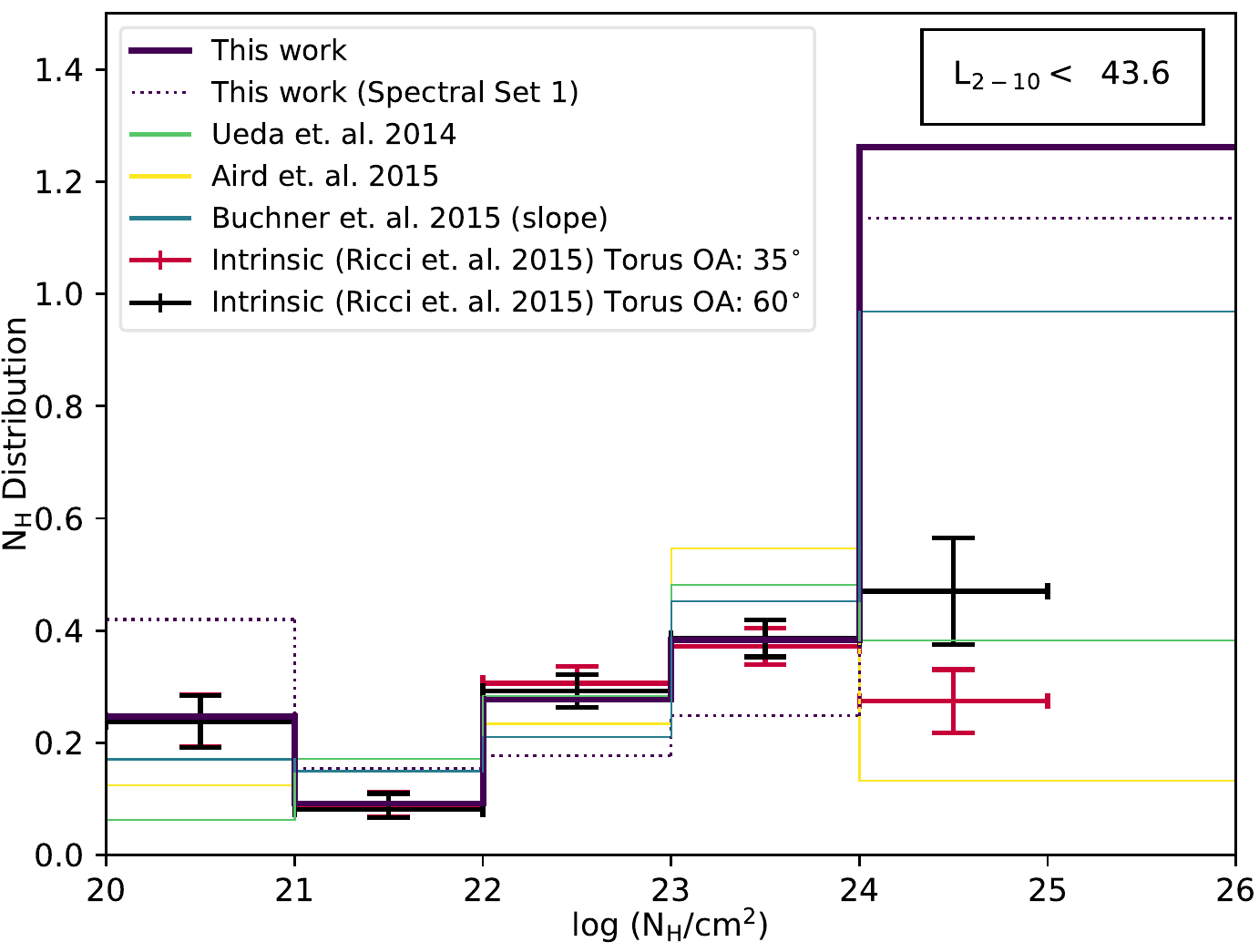}~
	\includegraphics[width=0.5\linewidth]{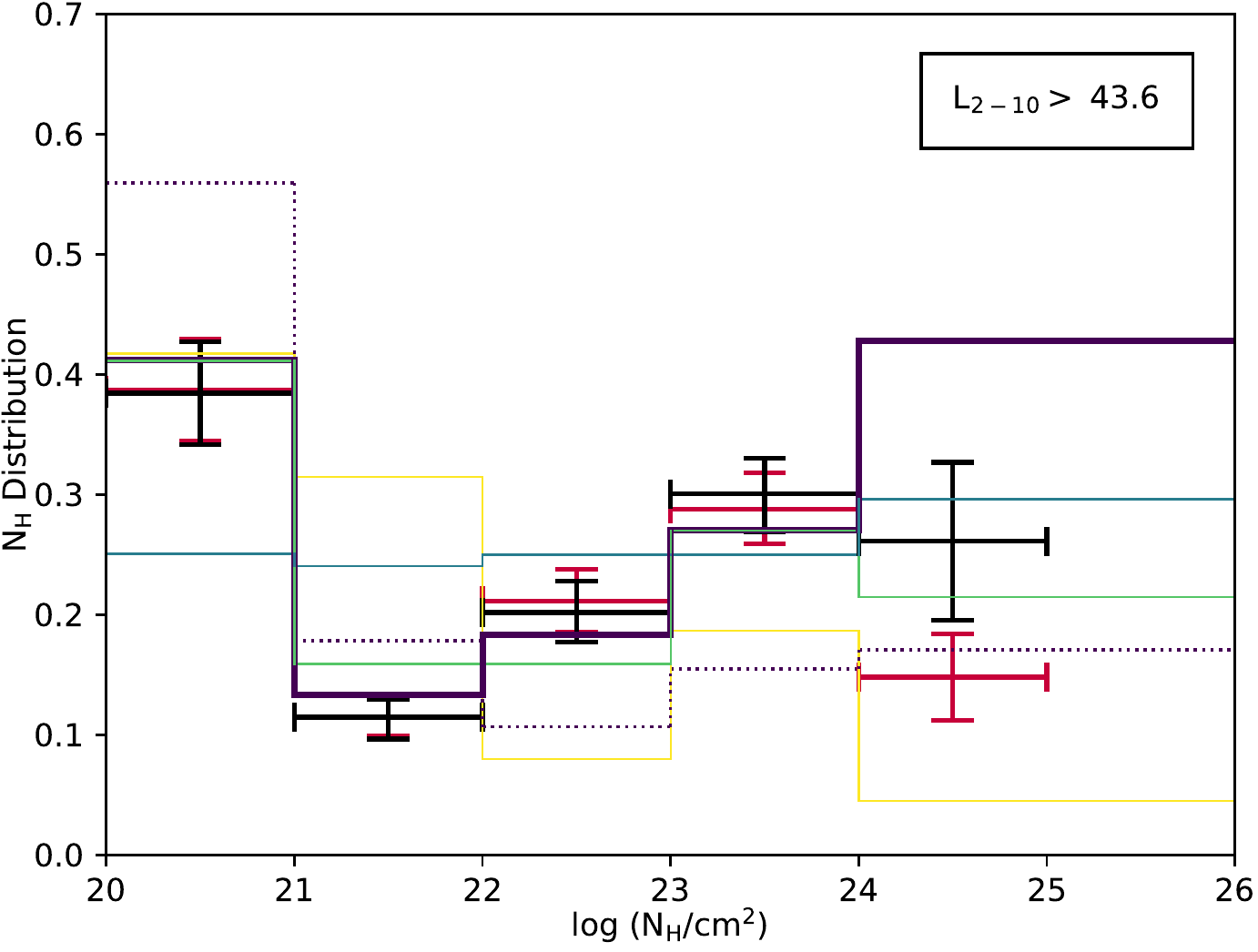}
	\includegraphics[width=0.5\linewidth]{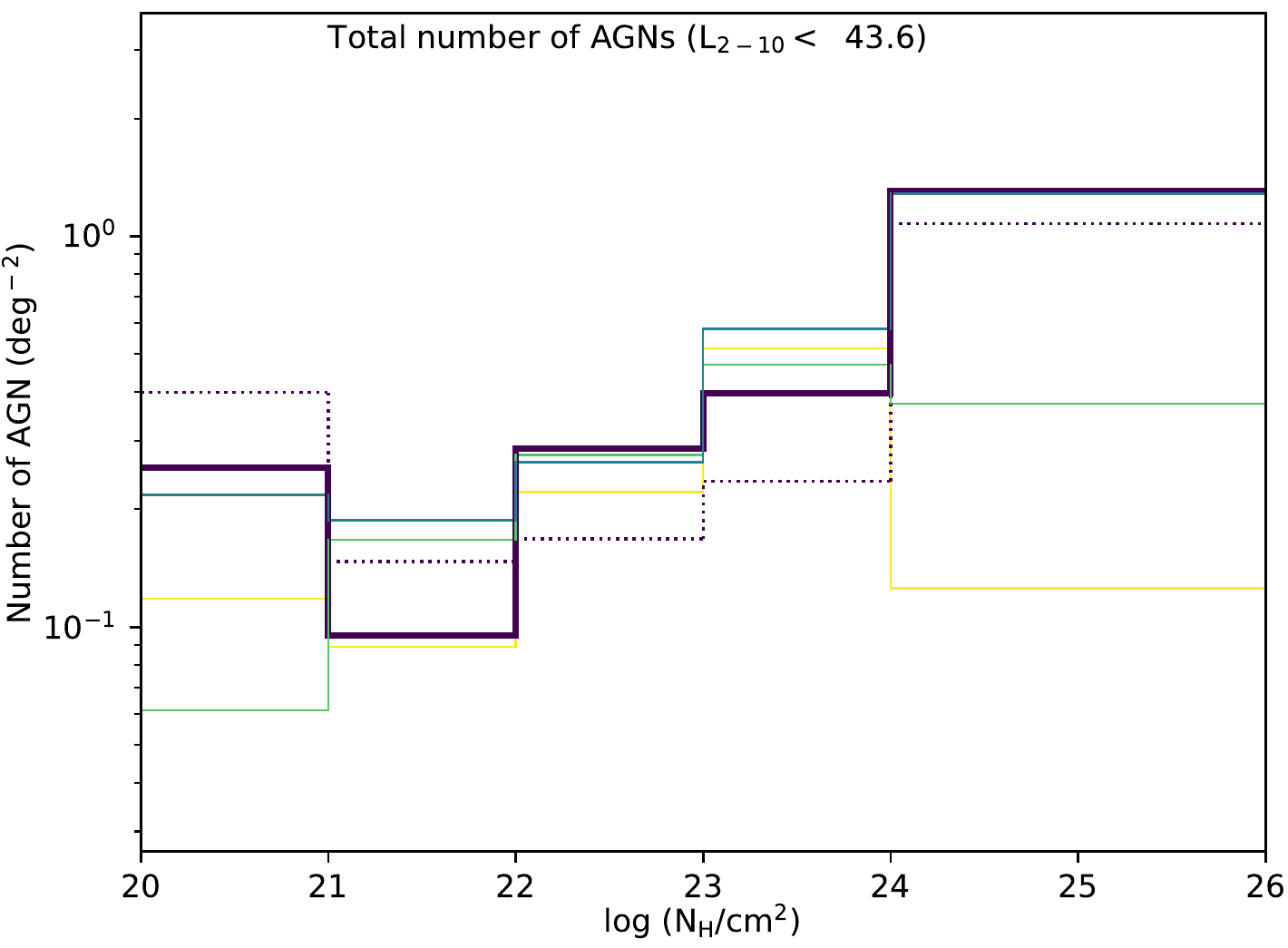}~
	\includegraphics[width=0.5\linewidth]{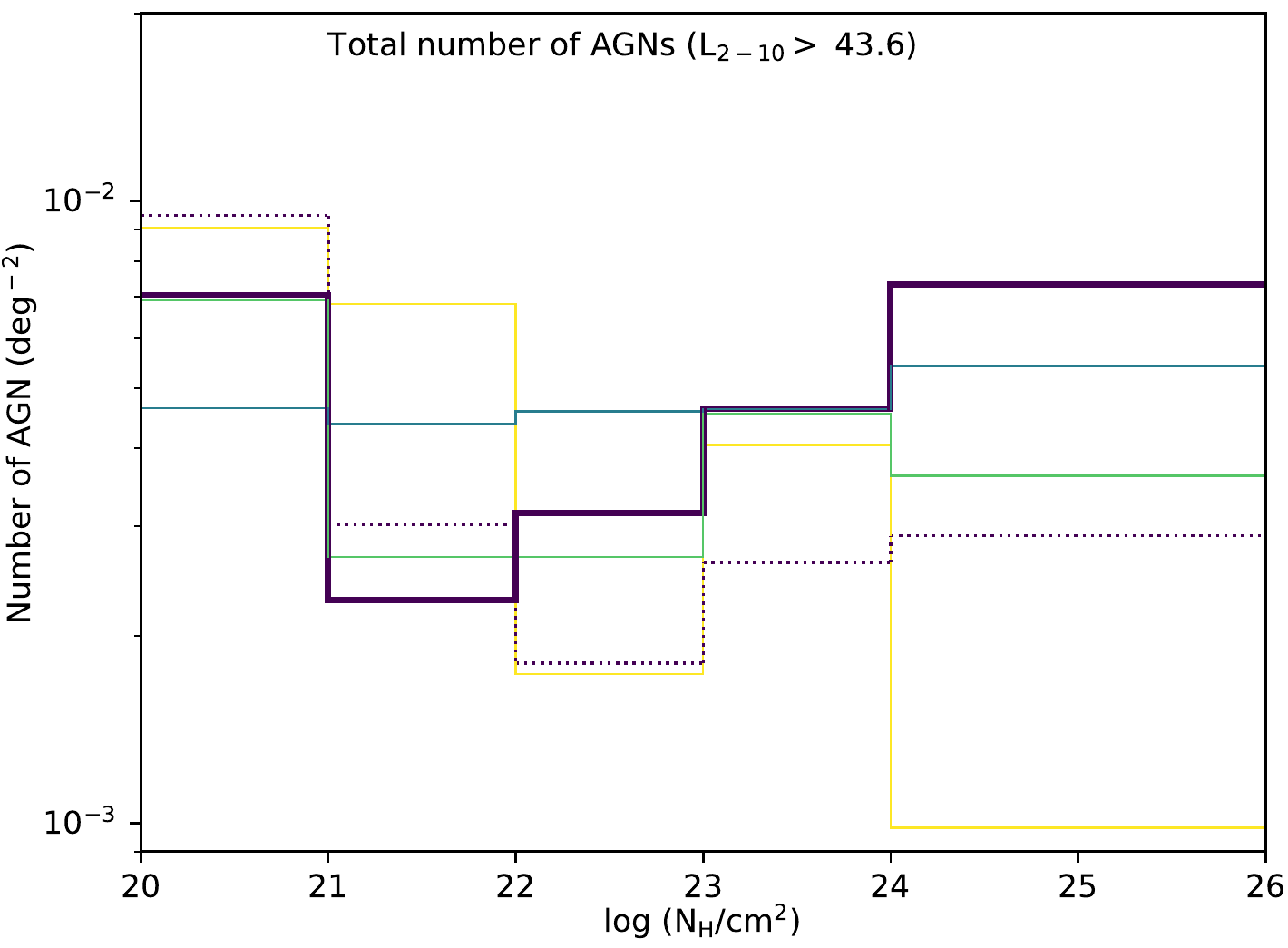}
	\caption{N$_{\rm H}$ distributions assumed for the present work and for previous population synthesis models. \textit{Top panels:} N$_{\rm H}$ distributions for various models, integrated up to z $\simeq$ 0.1 (\textit{solid lines}), including the completeness-corrected \citep{ricci2015} distributions assuming two different geometries for the torus: an opening angle of 60$^{\circ}$ (\textit{black plus sign}) and an opening angle of 35$^{\circ}$ (\textit{red plus sign}). The observations are in the $\log ({\rm N}_{\rm H}/{\rm cm}^{\rm -2}) = 20-25$ range. The models are this work (\textit{purple line}), Spectral Set 1 (\textit{dotted purple line} also from this work), U14 (\textit{light green line}), A15 (\textit{yellow line}) and B15 (median of constant slope prior; \textit{gray solid line}). All models assume that the number densities of objects in $\log ({\rm N}_{\rm H}/{\rm cm}^{\rm -2}) = 24-25$ bin are equal to those of $\log ({\rm N}_{\rm H}/{\rm cm}^{\rm -2}) = 25- 26$ bin for all redshift and luminosity, as there is very little data to constrain the $\log ({\rm N}_{\rm H}/{\rm cm}^{\rm -2}) = 25- 26$ bin. Therefore the model predictions are constant over these two absorption bins. \textit{Left panel: }The N$_{\rm H}$ distributions for the lower luminosity bin ($\log$ [L$_{\rm 2-10}/\text{erg s}^{-1}] <43.6$), and \textit{right panel:} for the higher luminosity bin  ($\log$ [L$_{\rm 2-10}/\text{erg s}^{-1}] <43.6$). Both panels are normalized in the $\log$ (N$_{\rm H}/{\rm cm}^{\rm -2}$) = 20-24 range. \textit{Bottom panels:} Total predicted number counts per square degree at each N$_{\rm H}$ bin, integrated to redshift = 0.1 in each luminosity bin.} 
	\label{fig:NH_intrinsic_distribution} 
\end{figure*}

\begin{figure*}[t]
	\centering 
	\includegraphics[width=0.8\linewidth]{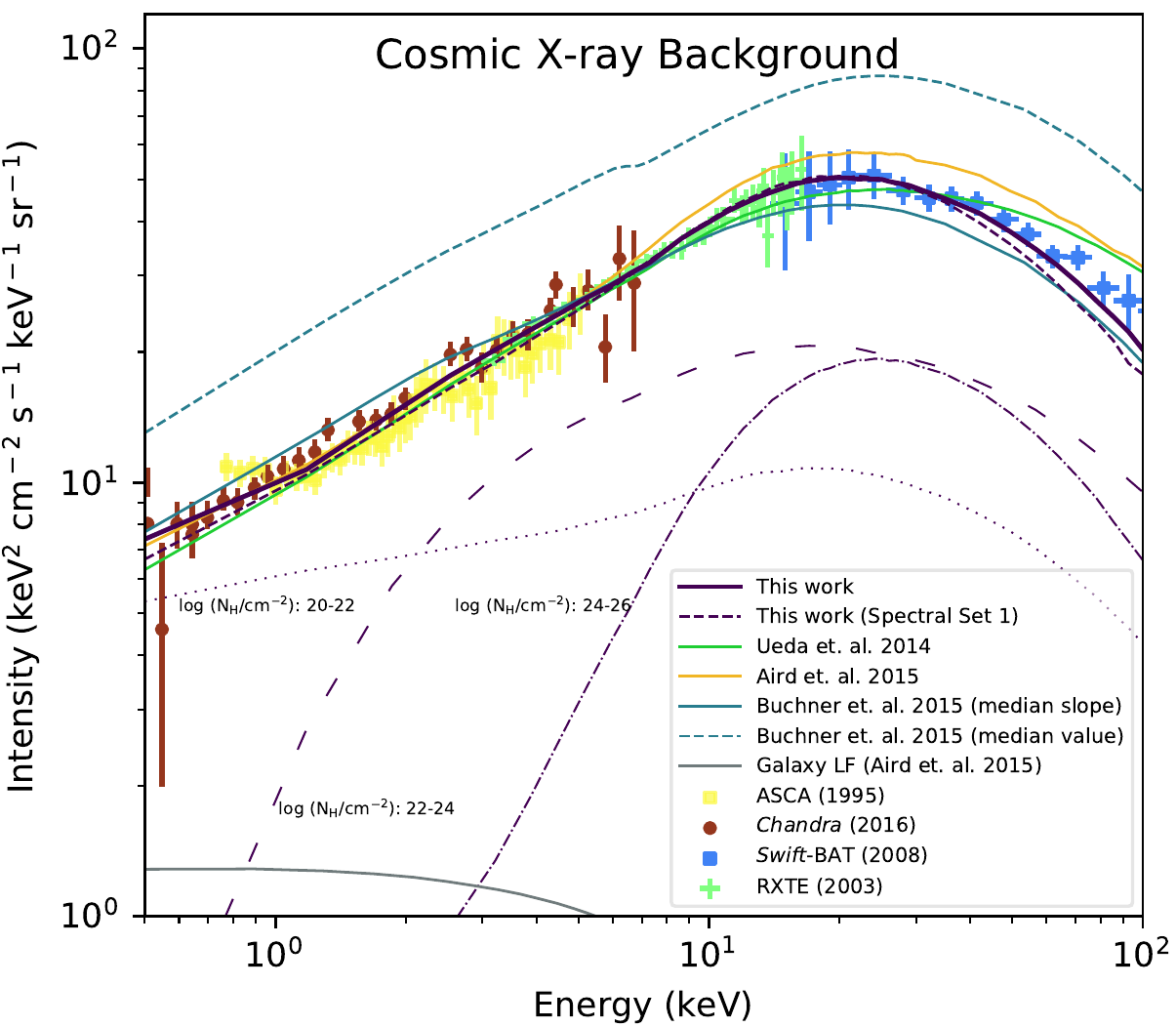}
	\caption{The empirical X-ray background (CXB) from \textit{Chandra} COSMOS (\textit{red dots}), \textit{ASCA} (\textit{yellow squares}), \textit{\textit{RXTE}} (\textit{green crosses}) and \textit{Swift}-BAT (\textit{blue squares}). One outlying data point from \textit{Chandra} COSMOS at 1.5 keV was removed due to incorrect background subtraction.  models are this work (\textit{solid purple line}), Spectral Set 1 results (\textit{dotted purple line}, also from this work), U14 (\textit{light green line}), A15 (\textit{yellow line}), B15 median of constant slope prior (\textit{gray solid line}) and B15 median of constant value prior (\textit{gray dashed line}). We added a cutoff energy of 200 keV to the B15 spectral model to bring the XLF in better agreement with CXB at higher energies. A galaxy contribution has been added to each CXB model prediction, according to the A15 galaxy luminosity function (\textit{black solid line}). Total contributions to CXB from three absorption bins for this work are also shown: $\log$ (N$_{\rm H}$/cm$^{\rm -2}$): 20-22 is shown in \textit{dotted purple line}, $\log$ (N$_{\rm H}$/cm$^{\rm -2}$): 22-24 is shown in \textit{sparsely dashed purple line and $\log$ (N$_{\rm H}$/cm$^{\rm -2}$): 24-26 is shown in \textit{dashdotted purple line}. The discrepancy in the CXB model predictions of U14 and A15 models between this plot and those published in the corresponding papers are addressed in Appendix~\ref{sec:cxb_discrepancy}.}} 
	\label{fig:xrb} 
\end{figure*}

\begin{figure*}[t]
	\centering
	\includegraphics[width=0.7\linewidth]{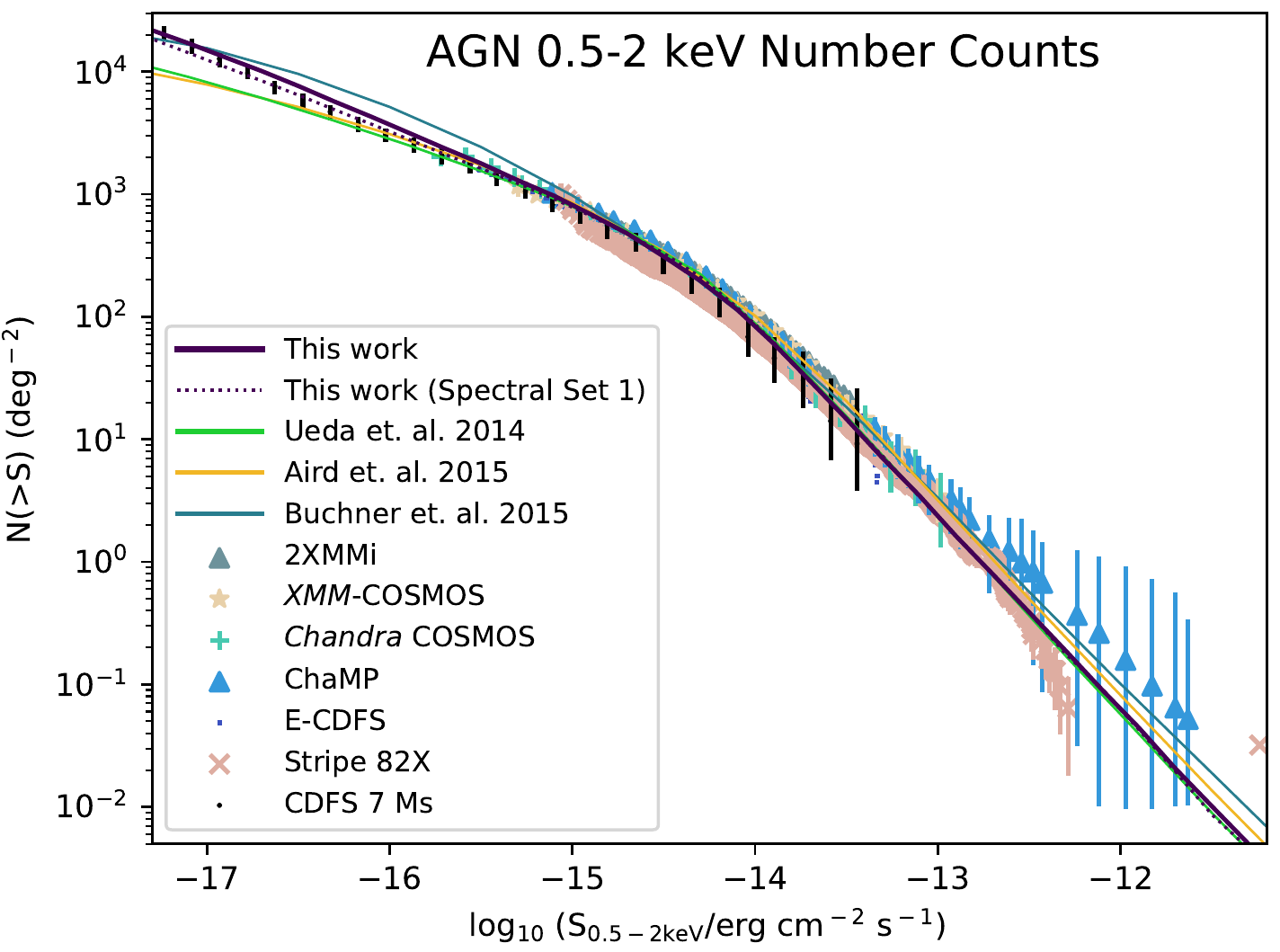}
	\includegraphics[width=0.7\linewidth]{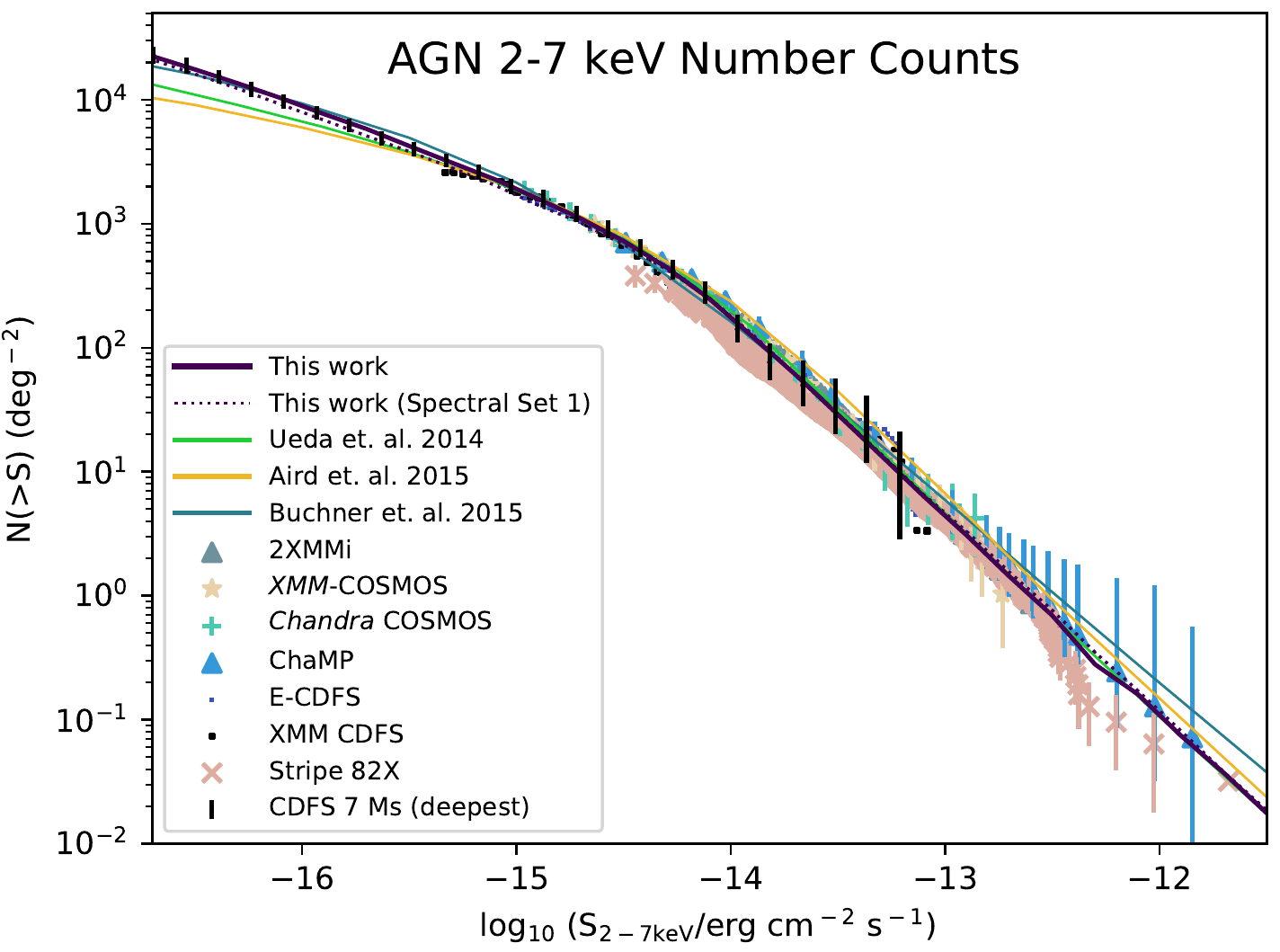}
	\caption{Number counts per  square degree of the sky versus X-ray flux, as observed in the \textit{Chandra} 7Ms (\textit{black dot + vertical error bars}), 2\textit{XMM}i (\textit{gray triangle}), \textit{XMM}-COSMOS (\textit{light pink stars}), \textit{Chandra}-COSMOS (\textit{green plus signs}), ChaMP (\textit{blue triangles}) and Stripe~82X (\textit{pink cross}), E-CDFS (\textit{deep blue dotted line}) and \textit{XMM}-CDFS (\textit{black dotted line}) surveys (all references in text). The models are this work (Spectral Set 5, \textit{solid purple line}), U14 (\textit{light green line}), A15 (\textit{yellow line}) and B15 median of constant slope prior (\textit{gray solid line}). Spectral Set 1 results are also shown in \textit{dotted purple line}. As all the data points cannot be seen in this static plot, an interactive version of this figure is available \href{https://yale.box.com/s/dzrn0zfjvi0eepc6w924yaixwgelt59m}{here}. \textit{Top panel:} 0.5$-$2 keV number counts. \textit{Bottom panel:} 2$-$7 keV band number counts. For the \textit{Chandra } 7 Ms and 4 Ms number counts in both bands, nearly all existing luminosity functions considered in this work underestimate the number counts, although the high flux number counts are generally well reproduced. B15 constant slope prior median value reproduces the hard band count for \textit{Chandra} 7 Ms very well, but overestimate the soft counts by $>$ 2$\sigma$ at $\log$ (S$_{\rm 0.5-2 keV}/ \text{erg cm}^{-2}\text{ s}^{-1})  \simeq - 16$.} 
	\label{fig:luo_counts} 
\end{figure*}

\begin{figure*}[t]
	\centering
	\includegraphics[width=0.7\linewidth]{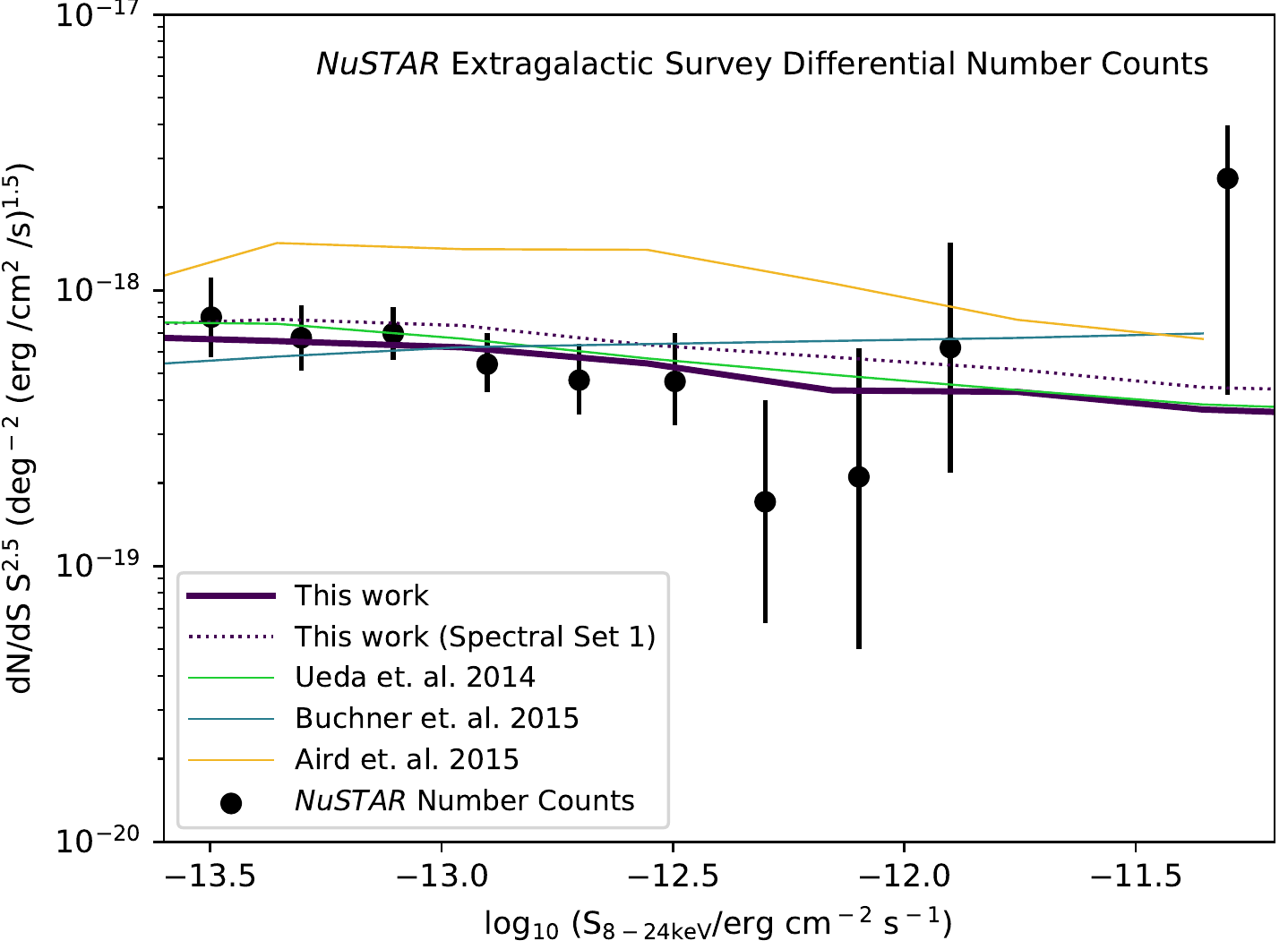}
	\caption{\textit{NuSTAR} Extragalactic Survey number counts for AGN. \textit{Black data points} are the differential number counts calculated using the \textit{NuSTAR}-COSMOS, \textit{NuSTAR} ECDFS, \textit{NuSTAR} EGS and \textit{NuSTAR} Serendipitous Surveys \citep{harrison2016}. The models are this work (Spectral Set 5, \textit{solid purple line}), U14 (\textit{light green line}), A15 (\textit{yellow line}) and B15 median of constant slope prior (\textit{gray solid line}). Spectral Set 1 results are also shown in \textit{dotted purple line}.} 
	\label{fig:nustar_numcount} 
\end{figure*}

\begin{figure*}[t]
	\centering
	\includegraphics[width=0.7\linewidth]{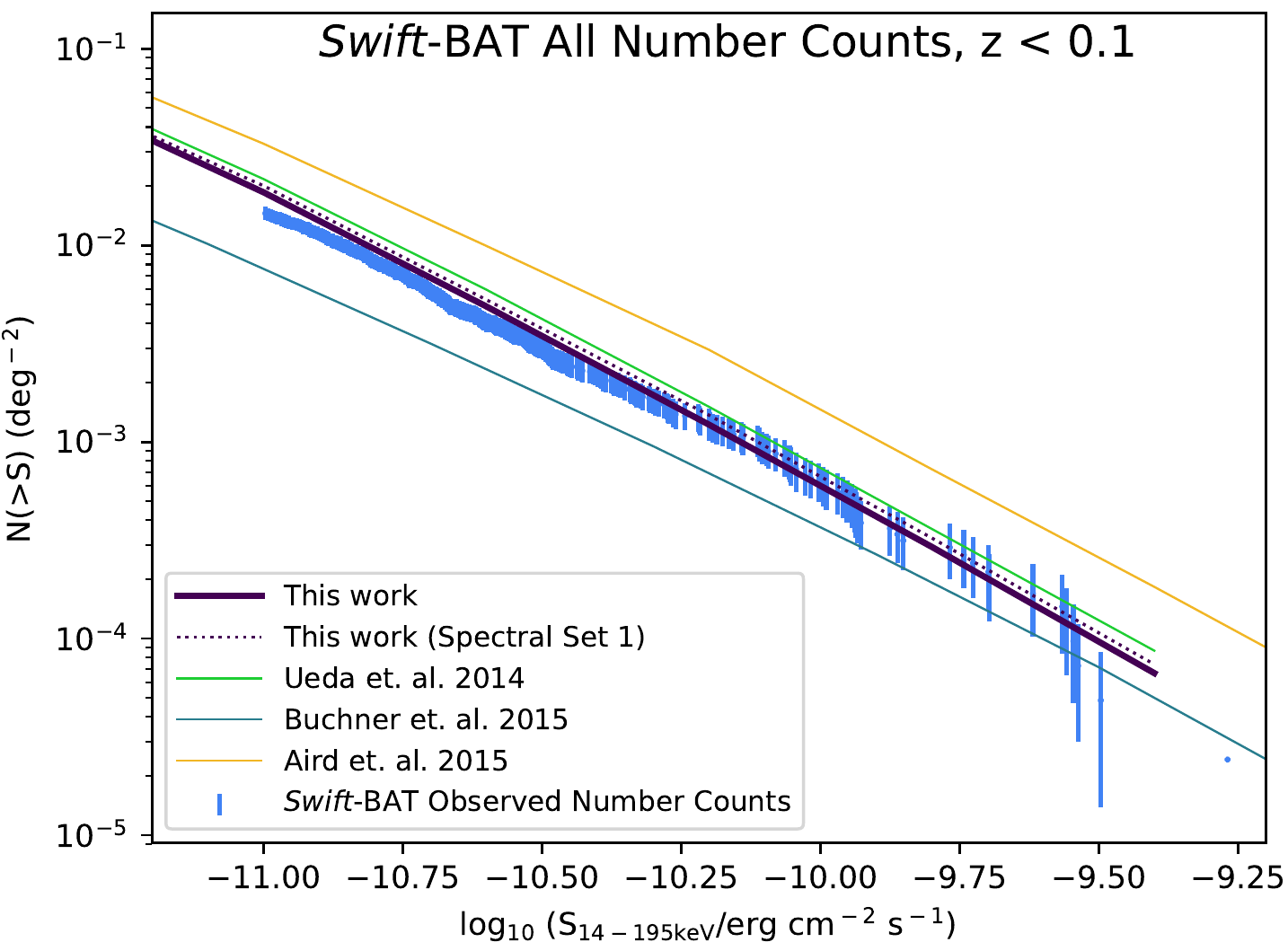}
	\caption{\textit{Swift}-BAT overall number counts (\textit{blue dots with vertical error bars}) calculated using the 70-month survey data \citep{claudio2017bat}. The models are this work (Spectral Set 5, \textit{solid purple line}), U14 (\textit{light green line}), A15 (\textit{yellow line}) and B15 median of constant slope prior (\textit{gray solid line}). Spectral Set 1 results are also shown in \textit{dotted purple line}.} 
	\label{fig:swiftbat_numcount} 
\end{figure*}

\begin{figure*}[t]
	\centering
	\includegraphics[width=0.7\linewidth]{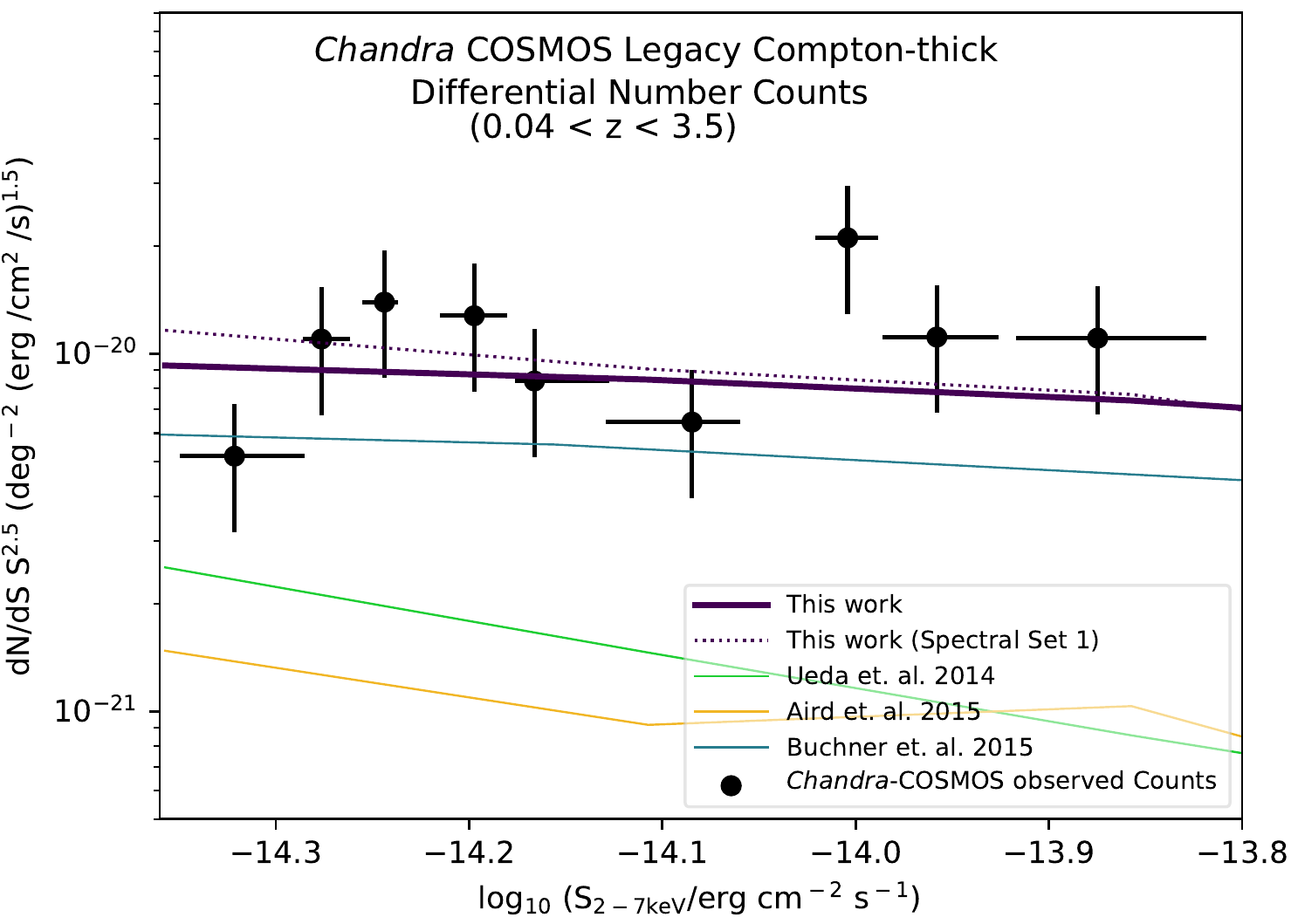}
	\caption{\textit{Chandra} COSMOS Legacy Compton-thick number counts. The data points are represented by \textit{black dots}. The models are this work (Spectral Set 5, \textit{solid purple line}), U14 (\textit{light green line}), A15 (\textit{yellow line}) and B15 median of constant slope prior (\textit{gray solid line}). Spectral Set 1 results are also shown in \textit{dotted purple line}. The counts were calculated using 41.9 objects after a careful Bayesian analysis of the spectra of each object by \citet{lanzuisi2018cosmosctk}. The flux area curve for three different redshift ranges were calculated specifically using the spectra of Compton-thick objects. We use the appropriate areas by redshift, and the fractional probability of each object of being Compton-thick to calculate number counts for the whole sample.} 
	\label{fig:cosmos_numcount} 
\end{figure*}

\begin{figure*}[t]
	\centering
	\includegraphics[width=0.7\linewidth]{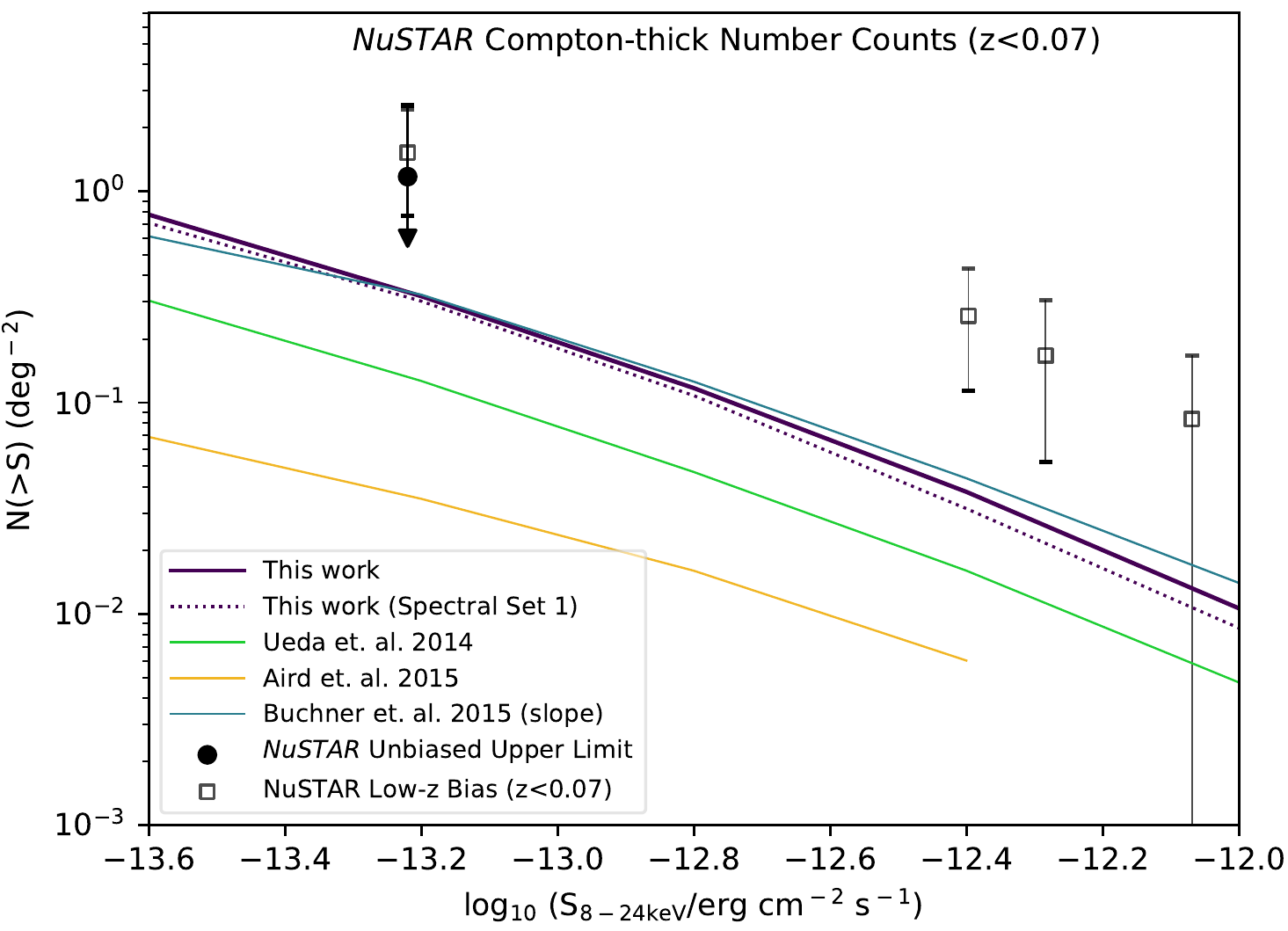}
	\caption{Compton-thick number counts in the 8-24 keV band from the \textit{NuSTAR} 40-month Serendipitous Survey, calculated by \citet{lansbury2017Ctk}. The \textit{bold solid black point} is the unbiased upper limit on Compton-thick fraction for z $<$ 0.5, and the gray number counts, calculated using four objects with extremely hard spectra, show the number counts with a low-redshift bias (z $<$ 0.07). The models are this work (Spectral Set 5, \textit{solid purple line}), U14 (\textit{light green line}), A15 (\textit{yellow line}) and B15 median of constant slope prior (\textit{gray solid line}). Spectral Set 1 results are also shown in \textit{dotted purple line}. We find that the counts with the low-redshift bias is underestimated by all the luminosity functions, but this work and B15 are within 1.5 $\sigma$, and U14 is within 2 $\sigma$. All the XLFs are consistent with the unbiased upper limit.} 
	\label{fig:nustar_counts} 
\end{figure*}

\begin{figure*}[t]
	\centering
	\includegraphics[width=0.49\linewidth]{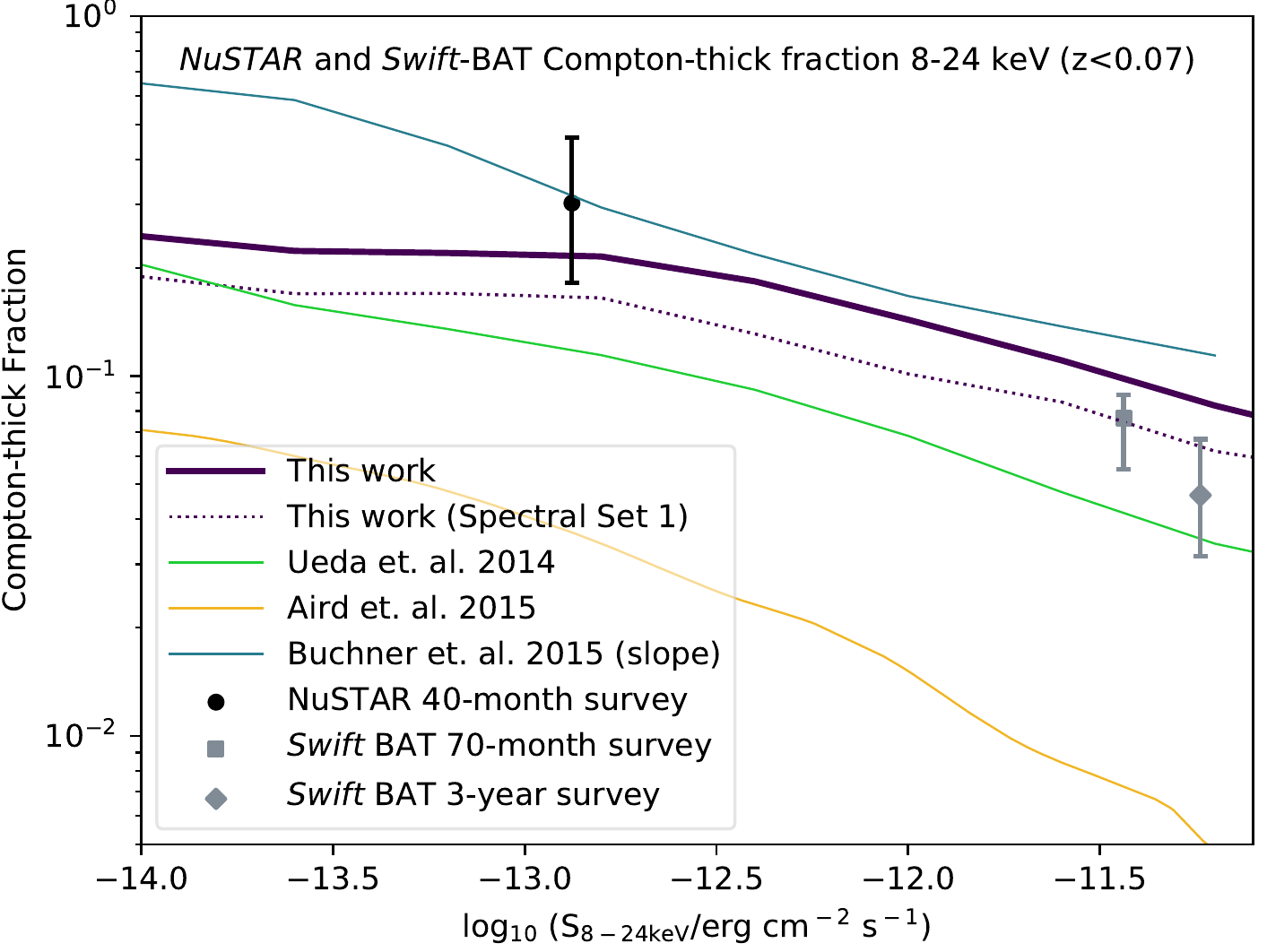}~
	\includegraphics[width=0.51\linewidth]{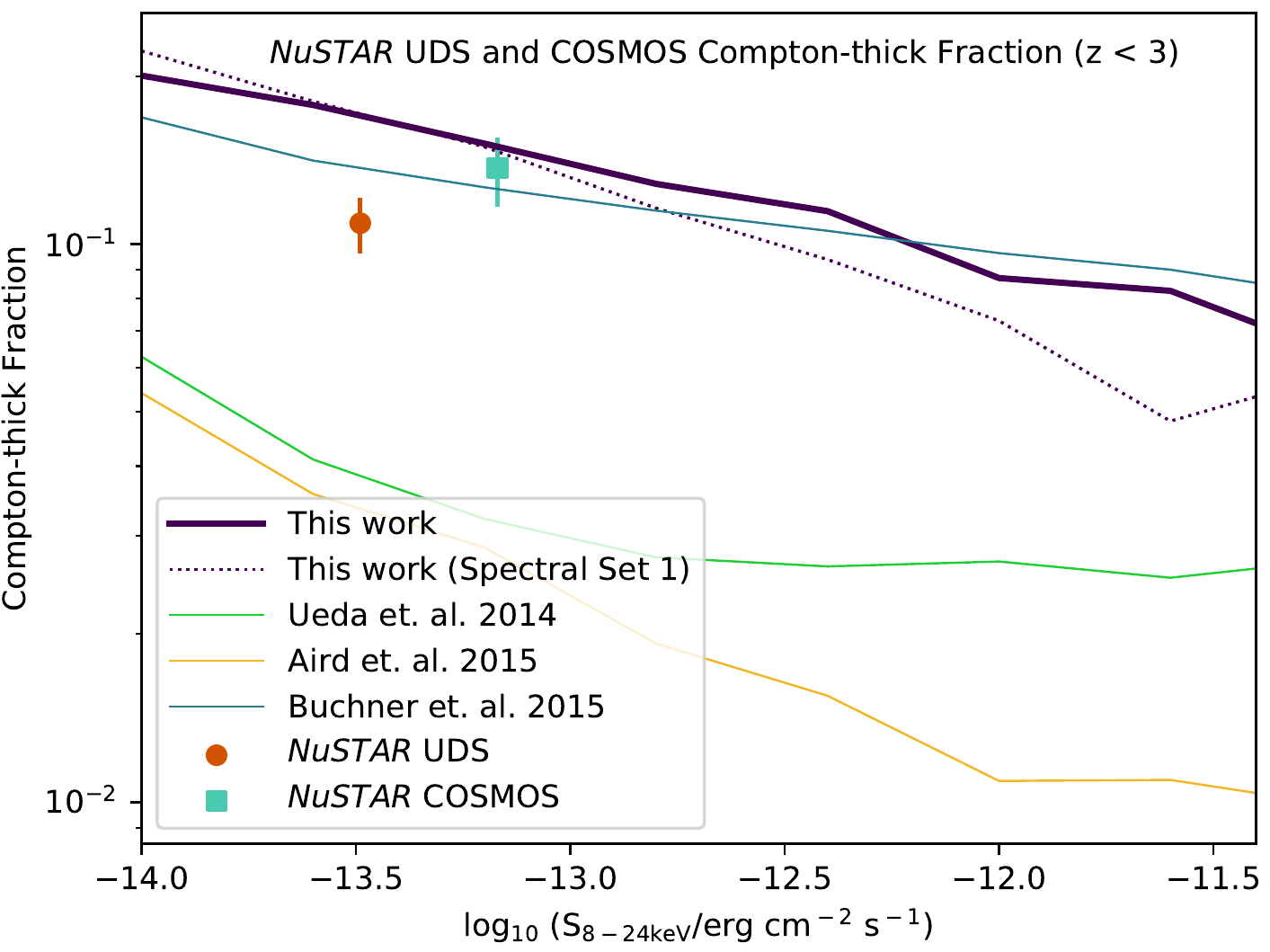}
	\caption{For both panels, the models are this work (Spectral Set 5, \textit{solid purple line}), U14 (\textit{light green line}), A15 (\textit{yellow line}) and B15 median of constant slope prior (\textit{gray solid line}). Spectral Set 1 results are also shown in \textit{dotted purple line}. \textit{Left:} Compton-thick  fraction in the 8-24 keV band from the \textit{NuSTAR} 40-month Serendipitous Survey, \textit{Swift} BAT 70-month and 3-year surveys. B15 fits the \textit{NuSTAR} Compton-thick fraction well, but overestimates the \textit{Swift}-BAT fractions. Spectrum 5 is within 1$\sigma$ of the \textit{NuSTAR} data and also fits the \textit{Swift}-BAT data well. Spectrum 1 and U14 is 1 $\sigma$ below \textit{NuSTAR} Compton-thick fraction, but fit the \textit{Swift}-BAT data well. \textit{Right:} Compton-thick  fraction in the 8-24 keV band from the \textit{NuSTAR} UDS \citep{masini2018} and COSMOS surveys \citep{civano2015nustarcosmos}. This results span a higher redshift range (0 $<$ z $<$ 3), and is therefore plotted separately for clarity.} 
	\label{fig:Ctk_fraction} 
\end{figure*}

The final results of our analysis are shown in Figures~\ref{fig:dphi}, \ref{fig:NH_intrinsic_distribution},  \ref{fig:xrb}, \ref{fig:luo_counts}, \ref{fig:nustar_numcount}, \ref{fig:swiftbat_numcount}, \ref{fig:cosmos_numcount},  \ref{fig:nustar_counts}, and \ref{fig:Ctk_fraction}, where we fit all the population synthesis models to observations. For clarity, we only show solutions for our best fit results for Spectral Set 5 (solid purple line in all figures), and include the Spectral Set 1 results (dotted purple line) for comparison. Our final population synthesis model is comprised of Spectral Set 5 and the XLF associated with it, which fits all the observed constraints listed in Table~\ref{tab:observed_constraints}. Figure~\ref{fig:dphi} shows that the XLF produced in this work has a shape similar to U14, monotonous and smooth, but with a somewhat wider bend than a double power-law. At $\log$ (L$_{\rm 2-10}$/erg s$^{\rm -1}$) $<$ 44, the normalization is closer to B15 constant slope prior median prediction. We plot our final result  in the $\log$ (N$_{\rm H}/{\rm cm}^{\rm -2}$) = 20-24 range with U14 Figure 10 observed data points in the same absorption bins, at four different redshifts. Figure~\ref{fig:dphi} shows that the U14 binned data points are reproduced by this work as well.

The upper panels of Figure~\ref{fig:NH_intrinsic_distribution} shows the absorption functions for U14, B15 and A15 integrated up to z $\sim$ 0.1. The upper panels show only the fraction of objects in each bin, normalized in the  $\log$ (N$_{\rm H}/{\rm cm}^{\rm -2}$) = 20-24 region, while the lower panels show absolute number of objects per deg$^{\rm 2}$.
The obscuration bias corrected absorption function for \textit{Swift}-BAT 70-month catalog, calculated by R15, is also plotted on the upper panels, for both of the assumed torus opening angles. The low and high luminosity bins are in left and right panels, respectively. This work is in agreement with N$_{\rm H}$ distribution derived by R15 at $\log (\text{N}_{\rm H}/\text{cm}^{-2}) < 24$, but predicts much higher fractions of Compton-thick objects, similar to B15.  This figure sheds light on the discrepancy between this work and some of the previous works. The space densities of unabsorbed and Compton-thin objects in this work is comparable to previous works, but the number densities of Compton-thick objects is much higher than U14 and A15 in both bins, and equal to B15 in the lower luminosity bin and higher in the higher luminosity bin, as shown in the lower panels of Figure~\ref{fig:NH_intrinsic_distribution}.

We take the A15 galaxy contribution to the CXB into account to avoid overestimating AGN space densities. The A15 galaxy contribution is calculated using a simple  power-law, with $\Gamma=1.9\pm0.2$, that results in a constant (with respect to energy) contribution of $\simeq$ 1.7 keV$^2$ cm$^{\rm -2}$ s$^{\rm -1}$ keV$^{\rm -1}$ sr$^{\rm -1}$ to the CXB at E $<$ 100 keV. Observations show that starburst galaxy spectra drops off very quickly above 10 keV \citep{wik2014,lehmer2015,yukita2016}, and a cutoff power-law is more appropriate. Therefore, we introduce a uniform distribution of cutoff energies between 20$-$30 keV to the spectra \citep{persic2002,persic2003,treister2010,wik2014,lehmer2015,yukita2016}, and recalculate contribution from galaxies using A15 galaxy luminosity function. The resulting galaxy X-ray background drops off rapidly at E $>$ 7 keV. We add this contribution to the CXB predictions from all models. 

 \begin{deluxetable*}{lccccc}[th]
	\tablewidth{0pt}
	\tablecaption{\label{tab:xrb_chisquare} \textsc{Statistical Significance of the match to X-ray background.}}
	\tablehead{\colhead{\textsc{CXB Constraint}} &\colhead{\textsc{Number of Datapoints}} & \colhead{\textsc{This work ($\chi^2$)}}&\colhead{\textsc{\citealp{ueda2014}} ($\chi^2$)}&\colhead{\textsc{\citealp{aird2015xlf} ($\chi^2$)}} & \colhead{\textsc{\citealp{buchner2015} ($\chi^2$)\tablenotemark{1} }}\\
     & & \colhead{\textsc{\textbf{Spec5}~Spec1}}& & & \colhead{\textsc{No E$_{\rm C}$~~~~E$_{\rm C}$ = 200 keV}}}
	\startdata
\textit{Chandra} COSMOS & 25 & \textbf{21.75}~~58.3 & 70.84 & 17.36 & 68.43~~~~~~~~~~~~55.91 \\
\textit{RXTE} & 34 & \textbf{17.98}~~45.74 & 46.5 & 152.95 & 203.06~~~~~~~~~~~~76.21 \\
\textit{Swift}-BAT & 15 & \textbf{11.66}~~46.02 & 36.5 & 137.88 & 1099~~~~~~~~~~~~69.79 \\
Total & 74 & \textbf{51.4}~~150.05 & 153.84 & 308.2 & 1370~~~~~~~~~~~~201.9 \\
Reduced $\chi^2$ & & \textbf{0.87}~~2.54 & 2.08 & 4.16 & 18.52~~~~~~~~~~~~2.73 \\
\enddata		
\tablenotetext{1}{For B15, results for only median constant slope prior are shown here.} 
\end{deluxetable*}	

\subsection{CXB}\label{sec:cxb}

Figure~\ref{fig:xrb} and Table~\ref{tab:xrb_chisquare} summarizes the fits to the CXB by prior XLFs and this work. The CXB is an important assessment of X-ray population synthesis models: individual surveys and number counts from these surveys can be affected by cosmic variance, but data for the CXB comes from a number of different experiments, in general averaged over very large areas, and converge on a similar shape (we show the latest data in the plot for better presentation). 

The CXB plotted in Figure~\ref{fig:xrb} for each model is our computation using the prescriptions in each work, and not taken directly from the results plotted in those papers. The discrepancy in the CXB model predictions of U14 and A15 models between this plot and those published in the corresponding papers is addressed in Appendix~\ref{sec:cxb_discrepancy}. We find a better fit to the CXB than previous XLFs, as shown Figure~\ref{fig:xrb} and Table~\ref{tab:xrb_chisquare}. Of the previous XLFs, U14 produces the best match to the CXB. Originally, B15 did not assume any cutoff energy, which leads to overproduction of CXB at high energies. We found that a cutoff energy of 200 keV greatly improves B15 fits, especially for the median constant slope prior. 
For B15, the range spanned between constant value prior and constant slope prior results illustrate the variety of CXB predictions possible from fitting XLF with no shape imposed. The constant slope prior fits the CXB results more closely, so we ignore the constant value prior results in the remaining analysis.

\subsection{Overall Number Counts}

Figure~\ref{fig:dphi} shows that the XLF presented in this work roughly follows a bending power-law shape similar to U14 - which is expected as U14+R15 is the basis of our function. We fit overall counts from \textit{XMM}, \textit{Chandra}, \textit{Swift}-BAT and \textit{NuSTAR} using these XLFs to verify that our population synthesis model reproduces these observations.

The XLF formulated using our neural network reproduces the CDFS number counts in both soft and hard bands down to the faintest fluxes, as shown in Figure~\ref{fig:luo_counts}. The figure contains all of the soft (0.5$-$2 keV) and hard (2$-$7 keV) band number counts for \textit{XMM}-Newton and \textit{Chandra} surveys. U14 and A15 fits these number counts down to flux levels of 10$^{\rm -15.5}$ $\rm erg\,cm^{-2}\,s^{-1}$, but at fainter levels starts to underestimate counts by several $\sigma$. B15 overestimates counts in 0.5$-$2 keV band at flux limits fainter than 10$^{\rm -14.5}$ $\rm erg\,cm^{-2}\,s^{-1}$, but fits the 2$-$7 keV band well.  As there are many datasets plotted on these two plots, we made interactive versions of these plots available \href{https://yale.box.com/s/dzrn0zfjvi0eepc6w924yaixwgelt59m}{here}. The older 4 Ms counts of CDFS \citep{lehmer2012} are also included in that plot.

The \textit{NuSTAR} observed overall counts in the 8$-$24 keV band \citep{harrison2016} are shown in Figure~\ref{fig:nustar_numcount}. They are fitted well by this work, B15 and U14. A15 overestimates the counts in the lower fluxes.

We calculated \textit{Swift}-BAT number counts using \citet{claudio2017bat} data in the 14$-$195 keV region, as shown in Figure~\ref{fig:swiftbat_numcount}. This work provides the best fit to these number counts. U14 slightly overestimates these counts by $1-2 \sigma$. A15 overestimates the counts by $ 2-4 \sigma$ at all fluxes, whereas B15 underestimates these counts by $ 1- 4\sigma$, with the discrepancy increasing at lower fluxes.

\subsection{Compton-Thick Number Counts and Fractions}

The \textit{Chandra} COSMOS Legacy Compton-thick counts span the highest redshift space, and is a result of the Bayesian analysis presented in \citet{lanzuisi2018cosmosctk} that considers the total probability of being Compton-thick for each object in their sample. They use this probability (fraction) to calculate number counts. The model fits to these data are shown in Figure~\ref{fig:cosmos_numcount}. Our models, both Spectral Set 1 and 5, fit these data better than prior XLFs. B15 also produces a close fit. U14 underestimates the counts, as does A15.

\citet{lansbury2017Ctk} calculated \textit{NuSTAR} Compton-thick number counts with low-redshift bias, and an upper limit on Compton-thick counts without the bias, using four objects with extremely hard X-ray spectra from the \textit{NuSTAR} Serendipitous Survey data. These counts are shown in Figure~\ref{fig:nustar_counts}. Our model fit lies 2$\sigma$ below  the counts with low-redshift bias. B15 comes slightly closer to these Compton-thick number counts, and U14 underestimates these counts by 2$-$3 $\sigma$. The A15 Compton-thick counts is much lower than the observations. As explained in \citet{lansbury2017Ctk}, the sample is biased because the three lowest redshift, highest flux objects show evidence of being weakly associated with \textit{Swift}-BAT AGN targets of \textit{NuSTAR} observations, which have a higher tendency of galaxy clustering. The bias may be the cause of the discrepancy between the models and these number counts.
The unbiased upper limit is calculated using only the highest redshift object, which is consistent with prior XLFs and our results.

The Compton-thick fraction for the \citet{lansbury2017Ctk} \textit{NuSTAR} sample is shown in  the left panel of Figure~\ref{fig:Ctk_fraction}, along with the \textit{Swift}-BAT observed 70-month and 3-year Compton-thick fractions (these fractions were calculated in 8$-$24 keV band by \citealp{lansbury2017Ctk}). Our model fits all the fractions within 1-1.5 $\sigma$; B15 fits the \textit{NuSTAR} fractions properly, but overestimates the \textit{Swift}-BAT fractions by 2$\sigma$. U14 underestimates the \textit{NuSTAR} Compton-thick fraction by 2 $\sigma$, but fits the \textit{Swift}-BAT 3-year survey fraction at high fluxes well. In the right panel of Figure~\ref{fig:Ctk_fraction}, we show that our model's predicted Compton-thick fraction of 18\% is higher than the \textit{NuSTAR} UDS \citep{masini2018} observed Compton-thick fraction of 11.5\%, but fits the \textit{NuSTAR}-COSMOS Compton-thick fraction \citep{civano2015nustarcosmos} well. B15 fits the \textit{NuSTAR}-COSMOS Compton-thick fraction well but slightly overestimates the UDS fraction, while both U14 and A15 underestimate these fractions.

U14 is within 2$\sigma$ of most of these counts and does fit the \textit{Swift}-BAT Compton-thick fractions in Figure~\ref{fig:Ctk_fraction}. A15 has much smaller Compton-thick number counts and fractions than the observed \textit{NuSTAR} and \textit{Swift}-BAT values. The reason behind this discrepancy is shown in Figure~\ref{fig:NH_intrinsic_distribution} lower panels --- A15 has lower Compton-thick space densities compared to other models in both luminosity bins. This discrepancy might be caused by the statistical approach A15 used to derive the absorption function. Instead of calculating N$_{\rm H}$ of individual sources through X-ray spectral fittings as done by U14 and B15, they compare the 0.5$-$2 keV selected sample with the 2$-$7 keV sample, taking into account the shift in AGN spectra with redshift. A15 discusses the limitation in this approach in distinguishing between Compton-thick sources from heavily absorbed Compton-thin sources. These uncertainties may have contributed to underestimation of Compton-thick number density.

The Compton-thick fractions found in this work and in B15, as shown in Figure~\ref{fig:NH_intrinsic_distribution}, are much higher than U14 and A15; this work fits the R15 fractions at $\log$ (N$_{\rm H}/{\rm cm}^{\rm -2}$) $<$ 24 (as shown in the figure), but the Compton-thick fraction is much higher than that derived by R15. This plot demonstrates why the U14+R15 update failed to reproduce the Compton-thick number counts. Our new model fits the \textit{NuSTAR} and \textit{Chandra} COSMOS Legacy Compton-thick number counts significantly better because the Compton-thick number counts need to be higher that R15 corrections in both luminosity bins. U14, \citet{ricci2015} and Balokovic (in prep) found that even \textit{Swift}-BAT is not completely unbiased towards Compton-thick sources beyond $\log$ (N$_{\rm H}/{\rm cm}^{\rm -2}$) $\sim$ 23, which could be the cause of the discrepancy between our model and the fractions derived by R15.

\section{Conclusions}\label{sec:conclusion}

The most noteworthy aspect of our best-fit luminosity function is the high intrinsic Compton-thick fraction predicted by this model. Of the overall AGN population  integrated up to z $\simeq$ 0.1 (1.0), 50$\pm$9\% (56$\pm$7\%)  is predicted to be Compton-thick by this model. This intrinsic Compton-thick fraction is consistent with observed number counts and fractions when flux, redshift limits and bandwidths in different surveys are taken into account, as shown in \S~\ref{sec:results}. 

This work generally has higher space densities of Compton-thick objects compared to the three prior XLFs, as shown in  Figure~\ref{fig:NH_intrinsic_distribution}. In the lower panels of that figure, there is a comparison between integrated space densities from local Universe, up to z = 0.1, in two luminosity bins. In the low-luminosity bin, B15 has $\simeq$ 2\% lower, U14 has $\simeq$ 72\% lower and A15 has $\simeq$ 91\% lower Compton-thick space densities than this work.
In the high-luminosity bin, this work predicts the highest space densities: B15 has 26\% lower, U14 has $\simeq$ 51\% lower and A15 has $\simeq$ 87\% lower space densities of Compton-thick objects than this work. 

The three nearest AGN ($\simeq$ 4 Mpc), Circinus, NGC 4945 and Centaurus A are all heavily obscured, which can either be due to a Compton-thick bias in the local Universe or it can be representative of the true AGN population. Obscured sources tend to be bright in IR due to reprocessed emission. \citet{matt2000} explored the IR and X-ray emission from some of the closest heavily obscured AGN, and find that the IR LF is 20 times the XLF in the local Universe. \citet{gandhi2003} also predicts obscured to unobscured ratio of 5:1 by formulating a population synthesis model where obscured AGN are assumed to follow the same distribution as luminous IR galaxies. \citet{fiore2008} selected sources with very high mid-IR to optical ratio in CDFS field, and found that 80$\pm$15\% of these objects have no direct X-ray detection in the hard X-ray band, and are likely to be Compton-thick. 
\citet{chen2015} show that obscured fraction is 30-70\% in objects with high far-IR luminosities (4 $\times 10^{12}$ L$_{\rm sol}$), however it is not clear what fraction of these obscured objects are Compton-thick. Ultra-luminous IR galaxies, which tend to be gas rich mergers \citep{clements1996}, have Compton-thick fractions as high as 65\%(\citealp{ricci2017ulirg}), much higher than the \textit{Swift}-BAT selected sample.

B15 and this work is closer to the high Compton-thick number counts and fractions observed by \textit{Chandra}, \textit{NuSTAR} and \textit{Swift}-BAT surveys. Previous works have stated that the most efficient way to find Compton-thick objects is using high-energy X-ray surveys (E $>$ 10 keV;  \citealp{gilli2007,treister2009,ballantyne2011}), and the results of these surveys, particularly \textit{NuSTAR} surveys, indicate higher Compton-thick space densities than those predicted by prior models. It has been suggested that a smaller Compton-thick population, and a very large reflection component (R $\simeq$ 2) at all luminosities can also reproduce the CXB (\citealp{aird2015b,akylas2012}). \citet{claudio2017bat} \textit{Swift}-BAT 70-month sample has a median biased R value (biased towards high R) of 1.3, and a bias corrected median R value of 0.53. \citet{zappacosta2018} analyzed \textit{NuSTAR} spectra of 63 sources and found $\langle R \rangle$ = 0.43, and an inter-quartile range of 0.06$-$1.5. Therefore, observed R values are smaller than the value required to support small Compton-thick number densities. Therefore, these densities may indeed be very high. In a future work, we will explore the consequences of these space densities on SMBH mass function.

This work and U14 fit overall AGN number counts better than A15 and B15, which serves as a cross-validation for the higher proportion of Compton-thick objects. Currently, our model is the only XLF which consistently fits all existing constraints.

The second most important result from our work is that we demonstrate, with examples, that the parameter space of AGN spectra that can reasonably reproduce the CXB is limited. Spectral Sets 2 and 3 are examples of combination of spectral parameters that do not consistently reproduce all parts of the CXB for any underlying XLF, and Spectral Set 4 is an example where the CXB can be reproduced reasonably, but number counts, in this case Compton-thick number counts, remain largely underestimated. 


The results for Spectral Set 1 and Spectral Set 5 are very similar, but the two spectral sets are not, as shown in Figure~\ref{fig:spectral_combinations}, and therefore the XLFs that produces the results are also different. We find that Spectral Set 5 is a better fit to the CXB (Figure~\ref{fig:xrb} and Table~\ref{tab:xrb_chisquare}) and to the R15 intrinsic absorption function (Figure~\ref{fig:NH_intrinsic_distribution}). 
The similarity between Spectral Sets 1 and 5, as shown in Figure~\ref{fig:spectral_combinations}, is that despite the different $\Gamma$ and E$_{\rm C}$ values, they are both considerably lower at E $>$ 20 keV than Spectral Sets 2, 3 and 4. This steep decline of the intrinsic power-law seems to be a necessary condition to reproduce the CXB spectra, and can be caused by a higher photon index or a lower cutoff energy. Observed AGN spectra indicates that cutoff energies are ubiquitous \citep{claudio2017bat}, and in our analysis, we find that a high photon index cannot be used to completely replace cutoff energies; B15 XLF originally did not have a cutoff energy but had a high photon index (1.95). However, this approach makes CXB constant from E $\simeq$ 30 keV for all 1000 XLF predictions, consequently heavily overestimating CXB at E $>$ 30 keV (results of that fit are shown in Table~\ref{tab:xrb_chisquare}). 

Spectral Set 1 slightly underestimates CXB at E $>$ 60 keV. Spectral Set 5 produces a better fit, and can possibly be further improved with a less steep spectra, i.e. with lower $\Gamma$ value than 1.96. The best distribution of $\Gamma$, assuming the bias-corrected cutoff energy from \textit{Swift}-BAT 70-month sample (200 keV) and reflection scaling factors, should have a $\langle \Gamma \rangle$ between 1.8 and 1.96. 

Finally, it must be noted that the upcoming \textit{Swift}-BAT 105-month spectral measurements, and the increasing \textit{NuSTAR} data, will contribute to better constraints on AGN spectra.

\section{Summary}\label{sec:summary}

We find that the most recent population synthesis models do not fit all the current X-ray observational constraints. We generated a comprehensive population synthesis model for black hole growth, consisting of AGN number densities as a function of luminosity, redshift and absorbing column density, which simultaneously accounts for the number counts and Compton-thick fractions in X-ray surveys spanning a range of depths and areas (corresponding to a wide range in luminosity and redshifts), and integrated spectrum of the Cosmic X-ray background. Specifically, given a set of input AGN spectra, we employed a neural network to find space densities that fit the X-ray background, then identified the best-fit model according to fits to the observed number counts and Compton-thick fractions. We took observational uncertainties in AGN spectra into account.
 
We find that the new population synthesis model 
predicts a much higher space density of Compton-thick objects, especially at high luminosities, than prior luminosity functions. This population matches observed Compton-thick number counts and fractions from \textit{XMM}-Newton, \textit{Chandra}, \textit{Swift}-BAT and \textit{NuSTAR} surveys, and predicts that intrinsically 50$\pm$9\% (56$\pm$7\%) of all AGN within z $\simeq$ 0.1 (1.0) are Compton-thick. We also show that AGN spectral assumptions affect the shape of the predicted X-ray background in population synthesis models, and certain spectral combinations do not suitably reproduce it for any space densities of AGN.

Our XLF is available as a 3D {\tt numpy} array, with instructions on calculating space densities using a 3D grid interpolator. It can be downloaded by cloning \href{https://bitbucket.org/tonimatas/xlf-final-result/downloads/}{this} repository.

This material is based upon work supported by the National Science Foundation under Grant No.  AST-1715512, and Yale University. ET acknowledges support from FONDECYT Regular 1160999, CONICYT PIA ACT172033 and Basal-CATA PFB-06/2007 and AFB170002 grants. TA wishes to thank her parents, M. A. Quayum and Shamim
Ara Begum, her husband, Mehrab Bakhtiar, and her sisters Arnita Tasnim and Raysa Tasnim for their
support.  TA also wishes to thank Dr. Trey Ashton Belew for his help with multiprocessing.

\textit{Software:} {\tt numpy} \citep{numpy2011}, {\tt Astropy} \citep{astropy2018},
{\tt Matplotlib} \citep{matplotlib}, {\tt Topcat} \citep{taylor2005}, \textsc{xspec} and \textsc{pyxpsec} \citep{xspec}, and {\tt Vegas} \citep{vegas}. Parts of the neural network code was adapted from \citet{nielsen2015}.

	\appendix

\section{Initial Approach: Updating N$_{\rm H}$ function}\label{sec:ricci}
  
Here we provide detailed background to the U14 absorption function and how it was updated using R15 results. The first step in formulating the U14 N$_{\rm H}$ distribution was to find the fraction of Compton-thin objects ($\psi$) --- $\log$ (N$_{\rm H}$/cm$^{\rm -2}$) = 22-24 --- among all objects with $\log$ (N$_{\rm H}$/cm$^{\rm -2}$) $<$  24. The function itself was normalized within  $\log$ (N$_{\rm H}$/cm$^{\rm -2}$) = 20-24 as there were too few Compton-thick objects securely identified in the three fields to adequately formulate the $\log$ (N$_{\rm H}$/cm$^{\rm -2}$) $>$ 24 region. As a result, the number/fraction of Compton-thick sources was essentially a free parameter. The fraction of Compton-thin objects is dependent on luminosity and redshift, $\psi (L_X, z)$, and is best constrained at $\log$ (L$_X $/erg s$^{-1}$) = 43.75 in the local Universe using \textit{Swift}-BAT data. Then, on the basis of \citet{treister2006} and \citet{hasinger2008} and independent U14 analysis, a redshift dependence is added to $\psi$ (L$_X$ = 43.75, z=0) as follows:

\begin{equation}\label{eqn:z_dep_of_psi_u14}
\begin{cases}
(1+z)^{0.48} & z < 2.0\\
(1+2)^{0.48} & z \geq 2.0
\end{cases}
\end{equation}
\nolinebreak Therefore, the complete luminosity and redshift dependent absorbed fraction is: $\psi(L_X, z) = \psi (L_X = 43.75, z) - 0.24 \times (\log L_X - 43.75)$ with forced upper and lower bounds at 0.84 and 0.2, respectively. These limits are imposed based on U14 analysis of \textit{Swift}-BAT data ($\psi_{\rm max}$=0.84) and \citealp{burlon2011} ($\psi_{\rm min}$ = 0.2). 

The absorption function is described in detail in \S~3.1 of U14. The data allowed the unabsorbed and Compton-thin bins to be much more robustly constrained than Compton-thick bins. Therefore,  U14 normalized the absorption function in the log (N$_{\rm H}/{\rm cm}^{\rm -2}$) = 20-24 bins, and assumed the same number of Compton-thick sources (at each luminosity and redshift) as the total number of Compton-thin sources, uniformly spread over log (N$_{\rm H}/{\rm cm}^{\rm -2}$) = 24-26 bins.  We replaced the local absorption function using R15 absorption function as it is based on \textit{Swift}-BAT 70-month survey, which updated the older 9-month survey.

The R15 intrinsic N$_{\rm H}$ function is normalized to one between $\log$ (N$_{\rm H}/{\rm cm}^{\rm -2}$)  = 20-25 in two 14$-$195 keV luminosity bins, for a z $<$ 0.3 sample with median z $\sim$ 0.055. These luminosity boundaries translate to different 2$-$10 keV luminosities, depending on the spectrum we assume. We convert $\log$  (${\rm L}_{\rm 14-195}$/$\text{erg s}^{-1}$) = 43.7 to $\log$ (${\rm L}_{\rm 2-10}$/$\text{erg s}^{-1}$)  = 43.58 assuming $\Gamma$ = 1.76 and cutoff energy of 60 keV, which are median spectral parameters for Swift-BAT 70-month sample \citep{claudio2017bat}. 

In our initial attempt, we renormalized R15 in the $\log$ (N$_{\rm H}/{\rm cm}^{\rm -2}$)  = 20-24 bins. We kept the total number of U14 objects in $\log$ N$_{\rm H}$/cm$^{\rm 2} < 24$ unchanged, and only redistributed objects within adjacent bins - which is sufficient to reproduce the R15 absorption function (as shown in Figure~\ref{fig:NH_intrinsic_distribution}). However, we add more objects in the Compton-thick bins, according to the R15 fraction. Then, we incorporated luminosity dependence into $\psi(L_x, z)$, by taking the R15 fractions as flat for the two luminosity bins ($\log$  (${\rm L}_{\rm 2-10}$/$\text{erg s}^{-1}$) $<$ 43.58  and $\log$  (${\rm L}_{\rm 2-10}$/$\text{erg s}^{-1}$) $\ge$ 43.58).  

The luminosity and redshift dependence of the absorbed fraction follows (z increases to a maximum value of 2):

\begin{equation}\label{eqn:lum_z_dep_of_psi}
	\psi(L_x, z) = \begin{cases}
	(0.68 \pm 0.04)(1+z)^{0.48 \pm 0.05}, & L_{\rm 2- 10} < 43.58\\
	 (0.50 \pm 0.04)(1+z)^{0.48 \pm 0.05}, & L_{\rm 2- 10} \ge 43.58
	\end{cases}
\end{equation}
Following U14, we use a maximum and minimum $\psi$ of 0.84 and 0.2 respectively. The updated N$_{\rm H}$ function is normalized in the log (N$_{\rm H}/{\rm cm}^{\rm -2}$) = 20-24 region. The ratio of number of objects in the log (N$_{\rm H}/{\rm cm}^{\rm -2}$) = 23-24 bin to log (N$_{\rm H}/{\rm cm}^{\rm -2}$) =  22-23 bin, the $\epsilon$ parameter from U14, is slightly lower in the R15 model than in U14, but it is within the range reported in literature - between 1.3 and 1.7 \citep{risaliti1999,tueller2008,vasudevan2013}. In R15 $\log$ (${\rm L}_{\rm 2-10}$/$\text{erg s}^{-1}$) $<$ 43.6 bin, $\epsilon =$ 1.3205 $\pm$ 0.17469, and in the L$_{\rm 2 - 10} \ge 43.6$ bin, $\epsilon =$ 1.4885 $\pm$ 0.24079. U14 uses a fixed value for $\epsilon$. 
Overall, the absorption function is as follows.

For $\log$ (${\rm L}_{\rm 2-10}$/$\text{erg s}^{-1}$) $<$ 43.58 (in the following equations, $\psi$ = $\psi(L_x,z)$ as shown in Equation~\ref{eqn:lum_z_dep_of_psi}):
\begin{equation}\label{eq:frac_low}
f(L_x,z; N_H) = \begin{cases}
(0.74 \pm 0.06) \times (1-\psi) & [20 \leq \log N_H < 21] \\
(0.26 \pm 0.06) \times (1-\psi) & [21 \leq \log N_H < 22] \\
\frac{1}{1+\epsilon} \psi & [22 \leq \log N_H < 23] \\
\frac{\epsilon}{1+\epsilon} \psi & [23 \leq \log N_H < 24] \\
(0.69 \pm 0.15) \times \psi & [24 \leq \log N_H < 26] \\
\end{cases}
\end{equation}

For $\log$  [${\rm L}_{\rm 2-10}$/$\text{erg s}^{-1}$] $\geq 43.58$:
\begin{equation}\label{eqn:frac_high}
	f(L_x,z; N_H) = \begin{cases}
		(0.77 \pm 0.05) \times (1-\psi) & [20 \leq \log N_H < 21] \\
		(0.23 \pm 0.05) \times (1-\psi) & [21 \leq \log N_H < 22] \\
		\frac{1}{1+\epsilon} \psi & [22 \leq \log N_H < 23] \\
		 \frac{\epsilon}{1+\epsilon} \psi & [23 \leq \log N_H < 24] \\
		 (0.52 \pm 0.08) \times \psi & [24 \leq \log N_H < 26] \\
	\end{cases}
\end{equation}

Similar to U14, we assumed the Compton-thick absorption function to be flat over the log (N$_{\rm H}/{\rm cm}^{\rm -2}$) = 24-26 range.  After editing the U14 XLF with R15 absorption function (U14+R15), we find that number counts in the 0.5$-$2 keV and 2$-$7 keV bands from CDFS 7Ms catalog are still underestimated by $\simeq$ 40\% at the faintest flux ends, and the Compton-thick number counts remain underestimated.  We also tried adding the luminosity dependence of $\psi$ found in \citet{barger2005}, where it linearly decreases from 0.8 to 0.2 between $\log$ (L$_{\rm 2-10} / \text{erg s}^{-1}$) = 42.0 to 46.0. This relationship can be incorporated without violating R15, if we relax the linear fraction to vary from 0.7 to 0.3 between $\log$ (L$_{\rm 2-10} / \text{erg s}^{-1}$) of 42.0 to 46.0. This made the absorption function more complicated, but did not improve results, so we neglected it. We used the U14 XLF  with modified absorption function, as described in Equations~\ref{eqn:lum_z_dep_of_psi}, \ref{eq:frac_low} and \ref{eqn:frac_high}, as input into the neural network.

\section{Back Propagation Algorithm}\label{sec:backpropagation}
 Here we explain how a back-propagation algorithm is used to determine our best-fit population synthesis model. In neural networks, a layer of input neurons receive input broken down into chunks; for instance, a handwriting recognition network would take information about black-to-white ratio of pixels for different parts of the image for different input neurons. This input vector of ratio would be called $\vec{x}$ - where $x_j$ is the ratio of black-to-white ink in the j-th input neuron. The input from all the layers sum up and reach each neuron in a new layer of neurons. Each of these neurons in the new layer have different weights associated with the neurons in the previous layer, such that, for neuron i in layer 2, the output ``activation" value $a_i$ is:
 \begin{equation}
 a_i = \begin{cases}
 0, & \vec{w} \cdot \vec{x} < {\rm threshold}_i \\
 1, & \vec{w} \cdot \vec{x} >= {\rm threshold}_i \\
 \end{cases}
 \end{equation}
 Here, $\vec{w}$ is the vector of weights associated with each neuron in the input layer, and threshold$_i$ is the bias associated with neuron i in the new layer (mathematically, bias $= -$threshold). If this threshold$_i$ value is exceeded by the dot product of the input and weights, then this neuron is activated and sends an input signal of 1 to the next layer, and if it falls below the threshold, this neuron sends a 0 to the next layer. Usually a more sophisticated function is used to calculate $a_i$ which outputs a range of values between 0 and 1, rather than just the binary values. A sigmoid function ($\sigma (z) = \frac{1}{1 + e^{-z}}$) is a commonly used activation function because its derivatives have nice properties, but we cannot do an analytic differentiation with CXB so the sigmoid function is not used in our neural network.
 
 A normal neural network, such as a handwriting recognition network, is comprised of a series of layers of neurons, each neuron with an array of weights associated with each neuron in the previous layer, and a bias. All these weights and biases are readjusted according to each input in the training dataset, so that the final set of weights and biases can predict which handwritten letter is seen using the ratio of black and white pixels in different parts of the image. The readjustment process is done by optimizing a cost function, which is:
 \begin{equation}\label{eqn:cost}
 C(w) \propto  \Sigma_x\vert\vert y (x) - a(w) \vert\vert^2 ~~,
 \end{equation}
 where $w$ is the collection of all weights in the network, $y$ is final output and $a$ is the activation. The weights are updated using gradient descent (see comprehensive explanation by \citealt{nielsen2015}). In this work, the activation function is the predicted X-ray background, and the cost function is the difference between the observed X-ray background and the model predictions, as shown in Figure~\ref{fig:neuralnetwork}:
 
 \begin{equation}\label{eqn:cost_here}
 C(w) = \frac{1}{2n} \Sigma_E \vert\vert CXB_{\rm obs} (E) - CXB_{\rm model} (E, w) \vert\vert^2 ~~,
 \end{equation}
 
Where E is the energy for each CXB data point, and $CXB_{\rm model}$ is evaluated using the following integral:
 
\begin{equation}
CXB_{\rm model} (E, w) \propto \int^{z_{\rm max}}_{z_{\rm min}} \int^{L_{\rm 2-10, max}}_{L_{\rm 2-10, min}} \int^{\log N_{\rm H, max}}_{\log N_{\rm H, min}} XLF(z, L_{\rm 2-10}, \log N_{\rm H} \vert w_1,... ,w_{15}) \times {\rm Spectra} (E(1+z))~dz~dL_{\rm 2-10}~d \log N_{\rm H} 
\end{equation}
 
After calculating cost C, the derivative of C, $\frac{\partial C(w_i)}{\partial w_i}$, is calculated numerically for each weight:

\begin{equation}
 	\frac{\partial C(w_i)}{\partial w_i} = \frac{C(w_{1},..., w_i + \partial w_i,...) - C(w_{1},..., w_i,...)}{\partial w_i}
\end{equation}

where $C(w_{1},..., w_i,...)$ is the cost for the current step. The weights are then updated so that the cost function is minimized:
 
 \begin{equation}
 w_i \mapsto w_i^{\prime} = w_i - \eta \frac{\partial C(w_i)}{\partial w_i} ~~,
 \end{equation}

where $\eta$ is the step size.
 
  Using this method, we readjust weights in such a way that the fit to the CXB improves with each iteration. We present our best solution as a new XLF in \S~\ref{sec:results}.
  
\section{Discrepancy In CXB Plot}\label{sec:cxb_discrepancy}

\begin{figure*}[h]
	\centering 
	\includegraphics[width=0.8\linewidth]{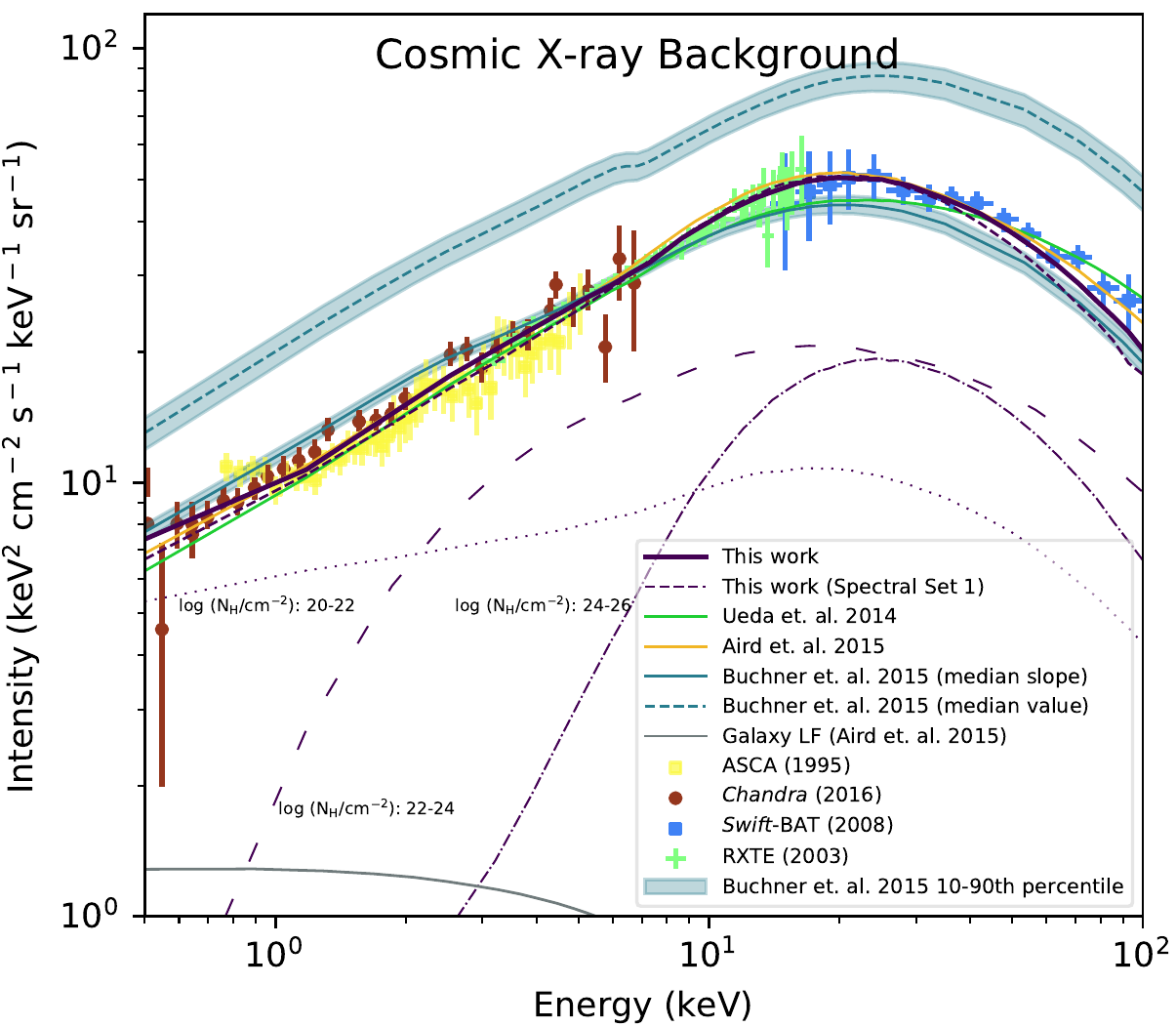}
	\caption{The empirical X-ray background (CXB) from \textit{Chandra} COSMOS (\textit{red dots}), \textit{ASCA} (\textit{yellow squares}), \textit{\textit{RXTE}} (\textit{green crosses}) and \textit{Swift}-BAT (\textit{blue squares}). The models are this work (\textit{solid purple line}), Spectral Set 1 results (\textit{dotted purple line}, also from this work), U14 (\textit{light green line}), A15 (\textit{yellow line}), B15 median of constant slope prior (\textit{gray solid line}) and B15 median of constant value prior (\textit{gray dashed line}). We added a cutoff energy of 200 keV to the B15 spectral model to bring the XLF in better agreement with CXB at higher energies. A galaxy contribution has been added to each CXB model prediction, according to the A15 galaxy luminosity function (\textit{black solid line}). Total contributions to CXB from three absorption bins for this work are also shown: $\log$ (N$_{\rm H}$/cm$^{\rm -2}$): 20-22 is shown in \textit{dotted purple line}, $\log$ (N$_{\rm H}$/cm$^{\rm -2}$): 22-24 is shown in \textit{sparsely dashed purple line} and $\log$ (N$_{\rm H}$/cm$^{\rm -2}$): 24-26 is shown in \textit{dashdotted purple line}. In this plot, we remove the dispersion in photon indices for the U14 and A15 models, which brings our calculations in agreement with published results in U14 and A15. We also show the 10-90th percentile CXB predictions (\textit{gray shaded region}) for B15 constant slope and constant value assumptions. We ignore the constant value assumption in the number count plots as the constant slope prediction is a closer match to X-ray background data.} 
	\label{fig:xrb_figure_consistent} 
\end{figure*}

In U14, the photon index for absorbed objects is $\langle \Gamma \rangle = 1.84$ with a dispersion of $\Gamma_{\mu} = 0.15$, and the photon index of unabsorbed objects is $\langle \Gamma \rangle = 1.94$ with a dispersion of $\Gamma_{\mu} = 0.09$. For A15, the photon index for all AGN is $\langle \Gamma \rangle = 1.9$ with a dispersion of $\Gamma_{\mu} = 0.2$. We find that using a Gaussian distribution of photon indices with these dispersions produces the CXB shown in Figure~\ref{fig:xrb}. However, if we keep the photon index constant at $\langle \Gamma \rangle$ values instead of using a distribution, we recover the CXB published in U14 and A15. The result of not using a distribution for A15 and U14 is shown in Figure~\ref{fig:xrb_figure_consistent}. In B15, no CXB was plotted for the model, so we use a prescribed distribution of photon index in B15 in both Figures~\ref{fig:xrb} and \ref{fig:xrb_figure_consistent}. Here, we show the 10$-$90th percentile predictions of the constant slope and constant value predictions by B15.

\begin{thebibliography}{}
	\bibitem[Aird et al.(2010)]{aird2010} Aird, J., Nandra, K., Laird, E.~S., et al.\ 2010, \mnras, 401, 2531 
	\bibitem[Aird et al.(2015)]{aird2015xlf} Aird, J., Coil, A.~L., Georgakakis, A., et al.\ 2015, \mnras, 451, 1892 
	\bibitem[Aird et al.(2015)]{aird2015b} Aird, J., Alexander, D.~M., Ballantyne, D.~R., et al.\ 2015, \apj, 815, 66 
	\bibitem[Ajello et al.(2008)]{ajello2008} Ajello, M., Greiner, J., Sato, G., et al.\ 2008, \apj, 689, 666-677 
	\bibitem[Alexander et al.(2013)]{alexander2013} Alexander, D.~M., Stern, D., Del Moro, A., et al.\ 2013, \apj, 773, 125 
	\bibitem[Akiyama et al.(2000)]{akiyama2000} Akiyama, M., Ohta, K., Yamada, T., et al.\ 2000, \apj, 532, 700 
	\bibitem[Akiyama et al.(2003)]{akiyama2003} Akiyama, M., Ueda, Y., Ohta, K., Takahashi, T., \& Yamada, T.\ 2003, \apjs, 148, 275 
    \bibitem[Akylas et al.(2012)]{akylas2012} Akylas, A., Georgakakis, A., Georgantopoulos, I., Brightman, M., \& Nandra, K.\ 2012, \aap, 546, A98 
	\bibitem[Alexander et al.(2003)]{alexander2003cdfs} Alexander, D.~M., Bauer, F.~E., Brandt, W.~N., et al.\ 2003, \aj, 126, 539 
    \bibitem[Alexander et al.(2011)]{alexander2011} Alexander, D.~M., Bauer, F.~E., Brandt, W.~N., et al.\ 2011, \apj, 738, 44 
	\bibitem[Ananna et al.(2017)]{ananna2017} Ananna, T.~T., Salvato, M., LaMassa, S., et al.\ 2017, \apj, 850, 66
	\bibitem[Arnaud(1996)]{xspec} Arnaud, K.~A.\ 1996, Astronomical Data Analysis Software and Systems V, 101, 17 
	\bibitem[Assef et al.(2013)]{assef2013} Assef, R.~J., Stern, D., Kochanek, C.~S., et al.\ 2013, \apj, 772, 26 
    \bibitem[The Astropy Collaboration et al.(2018)]{astropy2018} The Astropy Collaboration, Price-Whelan, A.~M., Sip{\H o}cz, B.~M., et al.\ 2018, arXiv:1801.02634 
	\bibitem[Ballantyne et al.(2011)]{ballantyne2011} Ballantyne, D.~R., Draper, A.~R., Madsen, K.~K., Rigby, J.~R., \& Treister, E.\ 2011, \apj, 736, 56
\bibitem[Balokovi{\'c} et al.(2018)]{mislav2018} Balokovi{\'c}, M., Brightman, M., Harrison, F.~A., et al.\ 2018, \apj, 854, 42 
\bibitem[Barger et al.(2003)]{barger2003} Barger, A.~J., Cowie, L.~L., Capak, P., et al.\ 2003, \aj, 126, 632 
	\bibitem[Barger et al.(2005)]{barger2005} Barger, A.~J., Cowie, L.~L., Mushotzky, R.~F., et al.\ 2005, \aj, 129, 578 
	\bibitem[Barthelmy et al.(2005)]{barthelmy2005} Barthelmy, S.~D., Barbier, L.~M., Cummings, J.~R., et al.\ 2005, \ssr, 120, 143 
	\bibitem[Baumgartner et al.(2013)]{baumgartner2013} Baumgartner, W.~H., Tueller, J., Markwardt, C.~B., et al.\ 2013, \apjs, 207, 19 
	\bibitem[Bottacini et al.(2012)]{bottacini2012} Bottacini, E., Ajello, M., \& Greiner, J.\ 2012, \apjs, 201, 34 
	\bibitem[Boyle et al.(1993)]{boyle1993} Boyle, B.~J., Griffiths, R.~E., Shanks, T., Stewart, G.~C., \& Georgantopoulos, I.\ 1993, \mnras, 260, 49 
	\bibitem[Boyle \& Terlevich(1998)]{boyle1998} Boyle, B.~J., \& Terlevich, R.~J.\ 1998, \mnras, 293, L49 
	\bibitem[Brandt \& Hasinger(2005)]{brandthasinger2005} Brandt, W.~N., \& Hasinger, G.\ 2005, \araa, 43, 827 
	\bibitem[Brightman \& Nandra(2011)]{bntorus2011} Brightman, M., \& Nandra, K.\ 2011, \mnras, 413, 1206 
	\bibitem[Brunner et al.(2008)]{brunner2008xmmlh} Brunner, H., Cappelluti, N., Hasinger, G., et al.\ 2008, \aap, 479, 283 
	\bibitem[Buchner et al.(2014)]{buchner2014spectra} Buchner, J., Georgakakis, A., Nandra, K., et al.\ 2014, \aap, 564, A125 
	\bibitem[Buchner et al.(2015)]{buchner2015} Buchner, J., Georgakakis, A., Nandra, K., et al.\ 2015, \apj, 802, 89 
	\bibitem[Burlon et al.(2011)]{burlon2011} Burlon, D., Ajello, M., Greiner, J., et al.\ 2011, \apj, 728, 58 
	\bibitem[Burrows et al.(2005)]{swiftxrt} Burrows, D.~N., Hill, J.~E., Nousek, J.~A., et al.\ 2005, \ssr, 120, 165 
	\bibitem[Cappelluti et al.(2007)]{nicocosmos2007} Cappelluti, N., Hasinger, G., Brusa, M., et al.\ 2007, \apjs, 172, 341 
	\bibitem[Cappelluti et al.(2017)]{nico2017} Cappelluti, N., Li, Y., Ricarte, A., et al.\ 2017, \apj, 837, 19 
	\bibitem[Cardamone et al.(2007)]{Cardamone2007} Cardamone, C.~N., Urry, C., Damen, M., et al.\ 2007, Bulletin of the American Astronomical Society, 39, 46.06 
	\bibitem[Cardamone et al.(2008)]{Cardamone2008} Cardamone, C.~N., Urry, C.~M., Damen, M., et al.\ 2008, \apj, 680, 130-142 
	\bibitem[Cavaliere \& Padovani(1989)]{cavaliere1989} Cavaliere, A., \& Padovani, P.\ 1989, \apjl, 340, L5 
	\bibitem[Civano et al.(2015)]{civano2015nustarcosmos} Civano, F., Hickox, R.~C., Puccetti, S., et al.\ 2015, \apj, 808, 185 
	\bibitem[Civano et al.(2016)]{civanocosmos2016} Civano, F., Marchesi, S., Comastri, A., et al.\ 2016, \apj, 819, 62 
    \bibitem[Chen et al.(2015)]{chen2015} Chen, C.-T.~J., Hickox, R.~C., Alberts, S., et al.\ 2015, \apj, 802, 50 
	\bibitem[Chokshi \& Turner(1992)]{chokshiandturner1992} Chokshi, A., \& Turner, E.~L.\ 1992, \mnras, 259, 421 
	\bibitem[Churazov et al.(2007)]{churazov2007} Churazov, E., Sunyaev, R., Revnivtsev, M., et al.\ 2007, \aap, 467, 529 
    \bibitem[Clements et al.(1996)]{clements1996} Clements, D.~L., Sutherland, W.~J., McMahon, R.~G., \& Saunders, W.\ 1996, \mnras, 279, 477 
    \bibitem[Comastri et al.(1995)]{comastri1995} Comastri, A., Setti, G., Zamorani, G., \& Hasinger, G.\ 1995, \aap, 296, 1 
	\bibitem[Cowie et al.(2003)]{cowie2003} Cowie, L.~L., Barger, A.~J., Bautz, M.~W., Brandt, W.~N., \& Garmire, G.~P.\ 2003, \apjl, 584, L57
	\bibitem[Della Ceca et al.(2004)]{dellaceca2004} Della Ceca, R., Maccacaro, T., Caccianiga, A., et al.\ 2004, \aap, 428, 383 
	\bibitem[Della Ceca et al.(2008)]{dellaceca2008} Della Ceca, R., Caccianiga, A., Severgnini, P., et al.\ 2008, \aap, 487, 119 
	\bibitem[Del Moro et al.(2016)]{delmoro2016} Del Moro, A., Alexander, D.~M., Bauer, F.~E., et al.\ 2016, \mnras, 456, 2105 
	\bibitem[De Luca \& Molendi(2004)]{xmmxrb2004} De Luca, A., \& Molendi, S.\ 2004, \aap, 419, 837 
	\bibitem[Di Matteo et al.(2005)]{matteo2005} Di Matteo, T., Springel, V., \& Hernquist, L.\ 2005, \nat, 433, 604
	\bibitem[Donley et al.(2010)]{donley2010} Donley, J.~L., Rieke, G.~H., Alexander, D.~M., Egami, E., \& P{\'e}rez-Gonz{\'a}lez, P.~G.\ 2010, \apj, 719, 1393  
	\bibitem[Donley et al.(2012)]{donley2012} Donley, J.~L., Koekemoer, A.~M., Brusa, M., et al.\ 2012, \apj, 748, 142 
	\bibitem[Elvis et al.(2009)]{elvis2009ccosmos} Elvis, M., Civano, F., Vignali, C., et al.\ 2009, \apjs, 184, 158 
	\bibitem[Ferrarese \& Merritt(2000)]{ferrarese2000} Ferrarese, L., \& Merritt, D.\ 2000, \apjl, 539, L9 
	\bibitem[Ferrarese \& Ford(2005)]{ferrarese2005} Ferrarese, L., \& Ford, H.\ 2005, \ssr, 116, 523 
	\bibitem[Fiore et al.(2003)]{fiore2003hella2xmm} Fiore, F., Brusa, M., Cocchia, F., et al.\ 2003, \aap, 409, 79 
    \bibitem[Fiore et al.(2008)]{fiore2008} Fiore, F., Grazian, A., Santini, P., et al.\ 2008, \apj, 672, 94 
    \bibitem[Fiore et al.(2009)]{fiore2009} Fiore, F., Puccetti, S., Brusa, M., et al.\ 2009, \apj, 693, 447 
    \bibitem[Fiore et al.(2012)]{fiore2012} Fiore, F., Puccetti, S., Grazian, A., et al.\ 2012, \aap, 537, A16 
	\bibitem[Fiore et al.(2017)]{fiore2017} Fiore, F., Feruglio, C., Shankar, F., et al.\ 2017, \aap, 601, A143 
	\bibitem[Fischer et al.(1998)]{fischer1998} Fischer, J.-U., Hasinger, G., Schwope, A.~D., et al.\ 1998, Astronomische Nachrichten, 319, 347 
	\bibitem[Gebhardt et al.(2000)]{gebhardt2000} Gebhardt, K., Bender, R., Bower, G., et al.\ 2000, \apjl, 539, L13 
	\bibitem[Gehrels et al.(2004)]{gehrels2004} Gehrels, N., Chincarini, G., Giommi, P., et al.\ 2004, \apj, 611, 1005 
	\bibitem[Gendreau et al.(1995)]{ascasisxrb} Gendreau, K.~C., Mushotzky, R., Fabian, A.~C., et al.\ 1995, \pasj, 47, L5 
    \bibitem[Lepage (1995)]{vegas} Lepage, G.~Peter \ 1980, CLNS-80/447 
     \bibitem[Gandhi \& Fabian(2003)]{gandhi2003} Gandhi, P., \& Fabian, A.~C.\ 2003, \mnras, 339, 1095 
	\bibitem[Giacconi et al.(2002)]{cdfs2002} Giacconi, R., Zirm, A., Wang, J., et al.\ 2002, \apjs, 139, 369 
    \bibitem[Gilli et al.(2001)]{gilli2001} Gilli, R., Salvati, M., \& Hasinger, G.\ 2001, \aap, 366, 407 
	\bibitem[Gilli et al.(2007)]{gilli2007} Gilli, R., Comastri, A., \& Hasinger, G.\ 2007, \aap, 463, 79 
	\bibitem[Gruber et al.(1999)]{gruber1999hea0a4} Gruber, D.~E., Matteson, J.~L., Peterson, L.~E., \& Jung, G.~V.\ 1999, \apj, 520, 124 
	\bibitem[Hasinger et al.(2005)]{hasinger2005} Hasinger, G., Miyaji, T., \& Schmidt, M.\ 2005, \aap, 441, 417 
	\bibitem[Harrison et al.(2013)]{harrison2013} Harrison, F.~A., Craig, W.~W., Christensen, F.~E., et al.\ 2013, \apj, 770, 103 
	\bibitem[Harrison et al.(2016)]{harrison2016} Harrison, F.~A., Aird, J., Civano, F., et al.\ 2016, \apj, 831, 185
	\bibitem[Hasinger(2008)]{hasinger2008} Hasinger, G.\ 2008, \aap, 490, 905 
	\bibitem[Hasinger et al.(2001)]{hasinger2001xmmlh} Hasinger, G., Altieri, B., Arnaud, M., et al.\ 2001, \aap, 365, L45 
	\bibitem[Hickox \& Markevitch(2006)]{handm2006} Hickox, R.~C., \& Markevitch, M.\ 2006, \apj, 645, 95 
    \bibitem[Hunter(2007)]{matplotlib} Hunter, J.~D.\ 2007, Computing in Science and Engineering, 9, 90 
    \bibitem[Jim{\'e}nez-Vicente et al.(2014)]{accretionsize} Jim{\'e}nez-Vicente, J., Mediavilla, E., Kochanek, C.~S., et al.\ 2014, \apj, 783, 47 
	\bibitem[Jones et al.(1997)]{jones1997} Jones, L.~R., McHardy, I.~M., Merrifield, M.~R., et al.\ 1997, \mnras, 285, 547 
	\bibitem[Hiroi et al.(2011)]{hiroi2011} Hiroi, K., Ueda, Y., Isobe, N., et al.\ 2011, \pasj, 63, S677 
	\bibitem[Hopkins et al.(2006)]{hopkins2006} Hopkins, P.~F., Hernquist, L., Cox, T.~J., et al.\ 2006, \apjs, 163, 1 
	\bibitem[Ichikawa et al.(2012)]{ichikawa2012} Ichikawa, K., Ueda, Y., Terashima, Y., et al.\ 2012, \apj, 754, 45 
	\bibitem[Jansen et al.(2001)]{jansen2001} Jansen, F., Lumb, D., Altieri, B., et al.\ 2001, \aap, 365, L1 
	\bibitem[Kim et al.(2007)]{kim2007champ} Kim, M., Kim, D.-W., Wilkes, B.~J., et al.\ 2007, \apjs, 169, 401 
	\bibitem[Kinzer et al.(1997)]{kinzer1997} Kinzer, R.~L., Jung, G.~V., Gruber, D.~E., et al.\ 1997, \apj, 475, 361
	\bibitem[Kormendy \& Gebhardt(2001)]{kormendy2001} Kormendy, J., \& Gebhardt, K.\ 2001, 20th Texas Symposium on relativistic astrophysics, 586, 363
	\bibitem[Kormendy \& Ho(2013)]{kormendyho2013} Kormendy, J., \& Ho, L.~C.\ 2013, \araa, 51, 511 
	\bibitem[Krivonos et al.(2007)]{krivonos2007} Krivonos, R., Revnivtsev, M., Lutovinov, A., et al.\ 2007, \aap, 475, 775 
	\bibitem[Lacy et al.(2004)]{lacy2004} Lacy, M., Storrie-Lombardi, L.~J., Sajina, A., et al.\ 2004, \apjs, 154, 166 
	\bibitem[Laird \& Aegis-X Team(2009)]{laird2009egs} Laird, E., \& Aegis-X Team 2009, Chandra's First Decade of Discovery, 14
	\bibitem[LaMassa et al.(2013a)]{lamassa2013a} LaMassa, S.~M., Urry, C.~M., Glikman, E., et al.\ 2013, \mnras, 432, 1351 
    \bibitem[LaMassa et al.(2013b)]{lamassa2013b} LaMassa, S.~M., Urry, C.~M., Cappelluti, N., et al.\ 2013, \mnras, 436, 3581 
	\bibitem[LaMassa et al.(2016)]{lamassa2016} LaMassa, S.~M., Urry, C.~M., Cappelluti, N., et al.\ 2016, \apj, 817, 172 
	\bibitem[Lansbury et al.(2017)]{lansbury2017} Lansbury, G.~B., Stern, D., Aird, J., et al.\ 2017, \apj, 836, 99 
	\bibitem[Lansbury et al.(2017)]{lansbury2017Ctk} Lansbury, G.~B., Alexander, D.~M., Aird, J., et al.\ 2017, \apj, 846, 20
	\bibitem[Lanzuisi et al.(2018)]{lanzuisi2018cosmosctk} Lanzuisi, G., Civano, F., Marchesi, S., et al.\ 2018, arXiv:1803.08547 
	\bibitem[Lehmer et al.(2005)]{lehmer2005} Lehmer, B.~D., Brandt, W.~N., Alexander, D.~M., et al.\ 2005, \apjs, 161, 21 
	\bibitem[Lehmer et al.(2012)]{lehmer2012} Lehmer, B.~D., Xue, Y.~Q., Brandt, W.~N., et al.\ 2012, \apj, 752, 46 
    \bibitem[Lehmer et al.(2015)]{lehmer2015} Lehmer, B.~D., Tyler, J.~B., Hornschemeier, A.~E., et al.\ 2015, \apj, 806, 126 
	\bibitem[Luo et al.(2017)]{luo2017} Luo, B., Brandt, W.~N., Xue, Y.~Q., et al.\ 2017, \apjs, 228, 2 
	\bibitem[Kim et al.(2007)]{kim2007} Kim, M., Wilkes, B.~J., Kim, D.-W., et al.\ 2007, \apj, 659, 29 
	\bibitem[Kirkpatrick et al.(2013)]{kirkpatrick2013} Kirkpatrick, A., Pope, A., Charmandaris, V., et al.\ 2013, \apj, 763, 123 
	\bibitem[Kirkpatrick et al.(2015)]{kirkpatrick2015} Kirkpatrick, A., Pope, A., Sajina, A., et al.\ 2015, \apj, 814, 9 
\bibitem[Matt et al.(2000)]{matt2000} Matt, G., Fabian, A.~C., Guainazzi, M., et al.\ 2000, \mnras, 318, 173 
	\bibitem[Maccacaro et al.(1991)]{maccacaro1991} Maccacaro, T., della Ceca, R., Gioia, I.~M., et al.\ 1991, \apj, 374, 117 
    \bibitem[Maiolino \& Rieke(1995)]{mandr1995} Maiolino, R., \& Rieke, G.~H.\ 1995, \apj, 454, 95 
    \bibitem[Malizia et al.(2014)]{malizia2014} Malizia, A., Molina, M., Bassani, L., et al.\ 2014, \apjl, 782, L25 
	\bibitem[Marchesi et al.(2016)]{marchesicosmos2016} Marchesi, S., Lanzuisi, G., Civano, F., et al.\ 2016, \apj, 830, 100 
	\bibitem[Magdziarz \& Zdziarski(1995)]{pexrav} Magdziarz, P., \& Zdziarski, A.~A.\ 1995, \mnras, 273, 837 
	\bibitem[Magorrian et al.(1998)]{magorrian1998} Magorrian, J., Tremaine, S., Richstone, D., et al.\ 1998, \aj, 115, 2285 
	\bibitem[Mart{\'{\i}}n-Navarro et al.(2018)]{martinnavarro2018} Mart{\'{\i}}n-Navarro, I., Brodie, J.~P., Romanowsky, A.~J., Ruiz-Lara, T., \& van de Ven, G.\ 2018, \nat, 553, 307
	\bibitem[Marshall et al.(1980)]{marshall1980} Marshall, F.~E., Boldt, E.~A., Holt, S.~S., et al.\ 1980, \apj, 235, 4 
    \bibitem[Masini et al.(2016)]{masini2016} Masini, A., Comastri, A., Balokovi{\'c}, M., et al.\ 2016, \aap, 589, A59 
	\bibitem[Masini et al.(2018)]{masini2018} Masini, A., Civano, F., Comastri, A., et al.\ 2018, \apjs, 235, 17 
	\bibitem[Mateos et al.(2008)]{mateos2008} Mateos, S., Warwick, R.~S., Carrera, F.~J., et al.\ 2008, \aap, 492, 51 
    \bibitem[Mateos et al.(2017)]{mateos2017} Mateos, S., Carrera, F.~J., Barcons, X., et al.\ 2017, \apjl, 841, L18 
    \bibitem[Matt et al.(2000)]{matt2000} Matt, G., Fabian, A.~C., Guainazzi, M., et al.\ 2000, \mnras, 318, 173 
    \bibitem[Matt(2001)]{matt2001} Matt, G.\ 2001, X-ray Astronomy: Stellar Endpoints, AGN, and the Diffuse X-ray Background, AIPC, 599, 209 
	\bibitem[Mendez et al.(2013)]{mendez2013} Mendez, A.~J., Coil, A.~L., Aird, J., et al.\ 2013, \apj, 770, 40 
	\bibitem[Merloni et al.(2010)]{merloni2010} Merloni, A., Bongiorno, A., Bolzonella, M., et al.\ 2010, \apj, 708, 137 
	\bibitem[Merritt \& Ferrarese(2001)]{merrit2001} Merritt, D., \& Ferrarese, L.\ 2001, \apj, 547, 140 
	\bibitem[Miyaji et al.(2000)]{miyaji2000} Miyaji, T., Hasinger, G., \& Schmidt, M.\ 2000, \aap, 353, 25 
	\bibitem[Moretti(2009)]{swiftxrt2} Moretti, A.\ 2009, American Institute of Physics Conference Series, 1126, 223 
	\bibitem[Mullaney et al.(2015)]{mullaney2015} Mullaney, J.~R., Del-Moro, A., Aird, J., et al.\ 2015, \apj, 808, 184 
	\bibitem[Murray et al.(2005)]{murray2005} Murray, S.~S., Kenter, A., Forman, W.~R., et al.\ 2005, \apjs, 161, 1 
	\bibitem[Murphy \& Yaqoob(2009)]{mytorus} Murphy, K.~D., \& Yaqoob, T.\ 2009, \mnras, 397, 1549 
	\bibitem[Nandra \& Pounds(1994)]{nandra1994} Nandra, K., \& Pounds, K.~A.\ 1994, \mnras, 268, 405 
	\bibitem[Nandra(2006)]{nandra2006} Nandra, K.\ 2006, \mnras, 368, L62 
	\bibitem[Nandra et al.(2007)]{pexmon} Nandra, K., O'Neill, P.~M., George, I.~M., \& Reeves, J.~N.\ 2007, \mnras, 382, 194 
	\bibitem[Nandra et al.(2015)]{nandra2015} Nandra, K., Laird, E.~S., Aird, J.~A., et al.\ 2015, \apjs, 220, 10 
	\bibitem[Nenkova et al.(2002)]{nenkova2002} Nenkova, M., Ivezi{\'c}, {\v Z}., \& Elitzur, M.\ 2002, \apjl, 570, L9 
	\bibitem[Nielsen (2015)]{nielsen2015}Michael A. Nielsen, "Neural Networks and Deep Learning", Determination Press, 2015.
	\bibitem[Page et al.(1997)]{page1997} Page, M.~J., Mason, K.~O., McHardy, I.~M., Jones, L.~R., \& Carrera, F.~J.\ 1997, \mnras, 291, 324 
    \bibitem[Persic \& Rephaeli(2002)]{persic2002} Persic, M., \& Rephaeli, Y.\ 2002, \aap, 382, 843 
    \bibitem[Persic \& Rephaeli(2003)]{persic2003} Persic, M., \& Rephaeli, Y.\ 2003, \aap, 399, 9 
    \bibitem[Petrucci et al.(2001)]{petrucci2001} Petrucci, P.~O., Haardt, F., Maraschi, L., et al.\ 2001, \apj, 556, 716 
	\bibitem[Pierre et al.(2016)]{pierre2016} Pierre, M., Pacaud, F., Adami, C., et al.\ 2016, \aap, 592, A1 
	\bibitem[Ranalli et al.(2013)]{ranalli2013} Ranalli, P., Comastri, A., Vignali, C., et al.\ 2013, \aap, 555, A42 
	\bibitem[Revnivtsev et al.(2003)]{rxte} Revnivtsev, M., Gilfanov, M., Sunyaev, R., Jahoda, K., \& Markwardt, C.\ 2003, \aap, 411, 329 
	\bibitem[Ricci et al.(2015)]{ricci2015} Ricci, C., Ueda, Y., Koss, M.~J., et al.\ 2015, \apjl, 815, L13 
	\bibitem[Ricci et al.(2017)]{claudio2017bat} Ricci, C., Trakhtenbrot, B., Koss, M.~J., et al.\ 2017, \apjs, 233, 17 
    \bibitem[Ricci et al.(2017)]{ricci2017ulirg} Ricci, C., Bauer, F.~E., Treister, E., et al.\ 2017, \mnras, 468, 1273 
	\bibitem[Richstone et al.(1998)]{richstone1998} Richstone, D., Ajhar, E.~A., Bender, R., et al.\ 1998, \nat, 395, A14 
	\bibitem[Risaliti et al.(1999)]{risaliti1999} Risaliti, G., Maiolino, R., \& Salvati, M.\ 1999, \apj, 522, 157 
	\bibitem[Salvatier et al.(2016)]{pymc32016} Salvatier, J., Wiecki, T.~V., \& Fonnesbeck, C.\ 2016, Astrophysics Source Code Library, ascl:1610.016 
	\bibitem[Schwope et al.(2000)]{schwope2000} Schwope, A., Hasinger, G., Lehmann, I., et al.\ 2000, Astronomische Nachrichten, 321, 1 
	\bibitem[Soltan(1982)]{soltan1982} Soltan, A.\ 1982, \mnras, 200, 115 
	\bibitem[Steffen et al.(2004)]{steffen2004} Steffen, A.~T., Barger, A.~J., Capak, P., et al.\ 2004, \aj, 128, 1483 
	\bibitem[Stern et al.(2012)]{stern2012} Stern, D., Assef, R.~J., Benford, D.~J., et al.\ 2012, \apj, 753, 30 
	\bibitem[Taylor(2005)]{taylor2005} Taylor, M.~B.\ 2005, Astronomical Data Analysis Software and Systems XIV, 347, 29 
	\bibitem[Treister et al.(2004)]{treister2004} Treister, E., Urry, C.~M., Chatzichristou, E., et al.\ 2004, \apj, 616, 123 
	\bibitem[Treister \& Urry(2006)]{treister2006} Treister, E., \& Urry, C.~M.\ 2006, \apjl, 652, L79 
	\bibitem[Treister et al.(2009)]{treister2009} Treister, E., Urry, C.~M., \& Virani, S.\ 2009, \apj, 696, 110 
    \bibitem[Treister et al.(2010)]{treister2010} Treister, E., Urry, C.~M., Schawinski, K., Cardamone, C.~N., \& Sanders, D.~B.\ 2010, \apjl, 722, L238 
	\bibitem[Trouille et al.(2008)]{trouille2008clasxsclanscdfn} Trouille, L., Barger, A.~J., Cowie, L.~L., Yang, Y., \& Mushotzky, R.~F.\ 2008, \apjs, 179, 1-18 
	\bibitem[Trouille et al.(2009)]{trouille2009clans} Trouille, L., Barger, A.~J., Cowie, L.~L., Yang, Y., \& Mushotzky, R.~F.\ 2009, \apj, 703, 2160 
	\bibitem[Tueller et al.(2008)]{tueller2008} Tueller, J., Mushotzky, R.~F., Barthelmy, S., et al.\ 2008, \apj, 681, 113-127
	\bibitem[Ueda et al.(1999)]{ueda1999} Ueda, Y., Takahashi, T., Ishisaki, Y., Ohashi, T., \& Makishima, K.\ 1999, \apjl, 524, L11 
	\bibitem[Ueda et al.(1999)]{ueda1999alss} Ueda, Y., Takahashi, T., Inoue, H., et al.\ 1999, \apj, 518, 656 
	\bibitem[Ueda et al.(2001)]{ueda2001amss} Ueda, Y., Ishisaki, Y., Takahashi, T., Makishima, K., \& Ohashi, T.\ 2001, \apjs, 133, 1 
	\bibitem[Ueda et al.(2003)]{ueda2003} Ueda, Y., Akiyama, M., Ohta, K., \& Miyaji, T.\ 2003, \apj, 598, 886 
	\bibitem[Ueda et al.(2008)]{ueda2008sxds} Ueda, Y., Watson, M.~G., Stewart, I.~M., et al.\ 2008, \apjs, 179, 124-141 
	\bibitem[Ueda et al.(2011)]{ueda2011} Ueda, Y., Hiroi, K., Isobe, N., et al.\ 2011, \pasj, 63, S937 
	\bibitem[Ueda et al.(2014)]{ueda2014} Ueda, Y., Akiyama, M., Hasinger, G., Miyaji, T., \& Watson, M.~G.\ 2014, \apj, 786, 104
	\bibitem[Van Der Walt et al.(2011)]{numpy2011} Van Der Walt, S., Colbert, S.~C., \& Varoquaux, G.\ 2011, arXiv:1102.1523 
	\bibitem[Vasudevan et al.(2013)]{vasudevan2013} Vasudevan, R.~V., Brandt, W.~N., Mushotzky, R.~F., et al.\ 2013, \apj, 763, 111 
	\bibitem[Weisskopf et al.(2002)]{weisskopf2002} Weisskopf, M.~C., Brinkman, B., Canizares, C., et al.\ 2002, \pasp, 114, 1 
    \bibitem[Wik et al.(2014)]{wik2014} Wik, D.~R., Lehmer, B.~D., Hornschemeier, A.~E., et al.\ 2014, \apj, 797, 79 
	\bibitem[Worsley et al.(2005)]{worsley2005} Worsley, M.~A., Fabian, A.~C., Bauer, F.~E., et al.\ 2005, \mnras, 357, 1281  
	\bibitem[Xue et al.(2012)]{xue2012} Xue, Y.~Q., Wang, S.~X., Brandt, W.~N., et al.\ 2012, \apj, 758, 129 
	\bibitem[Yang et al.(2004)]{yang2004clasxs} Yang, Y., Mushotzky, R.~F., Steffen, A.~T., Barger, A.~J., \& Cowie, L.~L.\ 2004, \aj, 128, 1501 
    \bibitem[Yukita et al.(2016)]{yukita2016} Yukita, M., Hornschemeier, A.~E., Lehmer, B.~D., et al.\ 2016, \apj, 824, 107 
	\bibitem[Zappacosta et al.(2018)]{zappacosta2018} Zappacosta, L., Comastri, A., Civano, F., et al.\ 2018, arXiv:1801.04280 
    \bibitem[Zdziarski et al.(1999)]{z99} Zdziarski, A.~A., Lubi{\'n}ski, P., \& Smith, D.~A.\ 1999, \mnras, 303, L11 
    \bibitem[Zdziarski et al.(2000)]{z2000} Zdziarski, A.~A., Poutanen, J., \& Johnson, W.~N.\ 2000, \apj, 542, 703 
\end{thebibliography}
\end{document}